\documentclass[12pt]{article}

\usepackage{verbatim,color,amssymb}
\usepackage{amsmath}					
\usepackage{amsthm}					
\usepackage{natbib}
\usepackage{multirow}
\usepackage{setspace}
\usepackage[mathscr]{euscript}
\usepackage{fancyhdr}
\usepackage{enumitem}
\usepackage{graphicx}
\usepackage{geometry}
\usepackage{xcolor}

\usepackage{multibib}
\newcites{latex}{References}

\usepackage{tabularx, booktabs}
\newcolumntype{Y}{>{\centering\arraybackslash}X}
\usepackage{lscape}

\usepackage{subfig}
\usepackage{float}
\usepackage{lineno}

\def\eE{\mathbb{E}}

\def\H{{\cal H}}

\def\X{{\cal X}}
\def\Z{{\cal Z}}
\def\Y{{\cal Y}}

\def\wh{\widehat}

\def\var{\hbox{var}}

\def\Dir{\hbox{Dir}}

\def\Ga{\hbox{Ga}}

\def\HC{\hbox{C}^{+}}

\def\MVN{\hbox{MVN}}

\def\Normal{\hbox{Normal}}

\def\Unif{\hbox{Unif}}
\def\Mult{\hbox{Mult}}

\def\P_25_ICML{{\it Proceedings of the 25th international conference on Machine learning}}

\def\bse{\begin{eqnarray*}}
\def\ese{\end{eqnarray*}}
\def\be{\begin{eqnarray}}
\def\ee{\end{eqnarray}}
\def\bq{\begin{equation}}
\def\eq{\end{equation}}

\def\wh{\widehat}

\def\trans{^{\rm T}}

\def\th{^{th}}

\def\bB{{\mathbf B}}

\def\bD{{\mathbf D}}
\def\b1e{{\mathbf e}}
\def\b1f{{\mathbf f}}

\def\bI{{\mathbf I}}

\def\bP{{\mathbf P}}

\def\bs{{\mathbf s}}

\def\bv{{\mathbf v}}

\def\by{{\mathbf y}}

\def\bz{{\mathbf z}}

\def\bzero{{\mathbf 0}}

\newcommand{\bmu}{\mbox{\boldmath $\mu$}}

\newcommand{\bpi}{\mbox{\boldmath $\pi$}}

\newcommand{\btheta}{\mbox{\boldmath $\theta$}}

\newcommand{\bbeta}{\mbox{\boldmath $\beta$}}

\newcommand{\bzeta}{\mbox{\boldmath $\zeta$}}

\newcommand{\bSigma}{\mbox{\boldmath $\Sigma$}}

\newcommand{\abs}[1]{\left\vert#1\right\vert}

\renewcommand\footnoterule{\kern-3pt \hrule \textwidth 2in \kern 2.6pt}

\def\boxit#1{\vbox{\hrule\hbox{\vrule\kern6pt \vbox{\kern6pt \textcolor{blue}{#1}\kern6pt}\kern6pt\vrule}\hrule}}

\def\authorfootnote#1{{\let\thefootnote\relax\footnotetext{#1}}}

\allowdisplaybreaks

\begin{document}
\thispagestyle{empty}
\baselineskip=28pt

\begin{center}
{\LARGE{\bf Bayesian Semiparametric Longitudinal\\ 
\vskip -8pt Drift-Diffusion Mixed Models\\ 
for Tone Learning in Adults
}}
\end{center}
\baselineskip=12pt
\vskip 20pt

\begin{center}
Giorgio Paulon$^{1}$ (giorgio.paulon@utexas.edu)\\
Fernando Llanos$^{2,3}$ (f.llanos@pitt.edu)\\
Bharath Chandrasekaran$^{3}$(b.chandra@pitt.edu)\\
Abhra Sarkar$^{1}$ (abhra.sarkar@utexas.edu)\\

\vskip 7mm
$^{1}$Department of Statistics and Data Sciences, \\
University of Texas at Austin,\\ 2317 Speedway D9800, Austin, TX 78712-1823, USA\\
\vskip 8pt 
$^{2}$Department of Linguistics, \\
University of Texas at Austin,\\ 305 East 23rd Street B5100, Austin, TX 78712, USA\\
\vskip 8pt 
$^{3}$Department of Communication Science and Disorders,\\ 
University of Pittsburgh,\\
4028 Forbes Tower, Pittsburgh, PA 15260, USA
\end{center}

\vskip 20pt 
\begin{center}
{\Large{\bf Abstract}} 
\end{center}
\baselineskip=12pt

Understanding how adult humans learn non-native speech categories such as tone information has shed novel insights into the mechanisms underlying experience-dependent brain plasticity. 
Scientists have traditionally examined these questions using longitudinal learning experiments under a multi-category decision making paradigm. 
Drift-diffusion processes are popular in such contexts for their ability to mimic underlying neural mechanisms. 
Motivated by these problems, we develop a novel Bayesian semiparametric inverse Gaussian drift-diffusion mixed model for multi-alternative decision making in longitudinal settings. 
We design a Markov chain Monte Carlo algorithm for posterior computation. 
We evaluate the method's empirical performances through synthetic experiments. 
Applied to our motivating longitudinal tone learning study, 
the method provides novel insights into  
how the biologically interpretable model parameters evolve with learning, 
differ between input-response tone combinations,  
and differ between well and poorly performing adults.  

\vskip 20pt 
\baselineskip=12pt
\noindent\underline{\bf Key Words}: 
Auditory category/tone learning, 
Auditory neuroscience, 
B-splines, 
Drift-diffusion models, 
(Factorial) hidden Markov models, 
Functional models, 
Inverse Gaussian distributions, 
Local clustering, 
Longitudinal mixed models, 
Perceptual decision making, 
Speech learning, 
Wiener processes

\par\medskip\noindent
\underline{\bf Short/Running Title}: Longitudinal Drift-Diffusion Mixed Models

\par\medskip\noindent
\underline{\bf Corresponding Author}: Abhra Sarkar (abhra.sarkar@utexas.edu)

\pagenumbering{arabic}
\setcounter{page}{0}
\newlength{\gnat}
\setlength{\gnat}{16pt}
\baselineskip=\gnat

\newpage
\section{Introduction}

Understanding the cognitive and biological mechanisms underlying our ability to learn new speech categories in adulthood constitute important questions in auditory neuroscience. 
Recent studies have demonstrated that adults are capable of learning features of a second language to a high degree of efficiency, demonstrating that age need not always constrain language learning abilities. 
The inherent dynamic complexities underlying learning in adulthood are not yet well understood but are being studied through extensive ongoing research.

The research reported here is motivated
particularly by experiments 
on the acquisition of Mandarin tones by native speakers of English. 
Native speech categories are acquired during the first year of life, within a so-called phonetic sensitivity period.
There is a greater neural commitment to native-language speech sounds, and this commitment may preclude the learning of novel speech categories in adulthood \citep{johnson1989critical, iverson2003perceptual}. 
In Mandarin Chinese, there are four tone categories that systematically change word meaning, similar to consonants and vowels in English. 
These tones are, however, linguistically  irrelevant in English. 
English native speakers thus struggle to distinguish the four tones and generalize their differences
\citep{wang1999training, chandrasekaran2010individual,maddox2014tests}. 
In laboratory settings, combining exposure to perceptually variable tones with trial-by-trial corrective feedback 
can improve tone categorization skills within a few hundred trials. 
Reaching a native like proficiency, however, may take several sessions of training \citep{xie2017stability,reetzke2018tracing}. 
The perceptual and sensory representation of Mandarin tones gets fundamentally refined over the course of this learning period \citep{feng2018role}. 
Understanding this longitudinal evolution 
is critical to assess the cognitive dynamics of speech category learning.
The statistical challenge is to make this assessment indirectly from behavioral data on tone categorization responses and response times.

To this end, 
we identify the Mandarin tone categorization problem with the broader class of problems of multi-category decision making under perceptual stimuli 
\citep{smith2004psychology, heekeren2004general, gold2007neural, schall2001neural, purcell2013neural, glimcher2013neuroeconomics}. 
In such contexts, drift-diffusion processes are popular models for behavioral accuracies and response times 
as they mimic the accumulation of sensory evidence in favor of different decision alternatives in the human brain \citep{ratcliff1978theory, ratcliff2016diffusion}. 
The existing literature on drift-diffusion models is substantive \citep{smith1988accumulator, ratcliff1998modeling, ratcliff2008diffusion}. 
These classical methods, as well as their recent adaptations using reinforcement learning based ideas \citep{fontanesi2019reinforcement, pedersen2017drift, peters2020drift}, 
are, however, heavily focused on the two category case 
with a single latent diffusion process and two boundaries, one for each of the two decision alternatives. 
This is despite the fact that humans often are required to learn more than two categories at once. 
For example, English has 14 vowels and 24 consonant phonemes; Mandarin has four tone categories, etc. 
The joint likelihood of accuracies and response times under models with a single diffusion process is mathematically complex and computationally expensive \citep{navarro2009fast, tuerlinckx2004efficient, tuerlinckx2001comparison}. 
Inference in such models is thus often based 
on approximations of the likelihood \citep{vandekerckhove2007fitting}, 
or on the conditional likelihood of the response times, conditioned on the decisions \citep{vandekerckhove2008bayesian}. 
Multi-category drift-diffusion models with separate latent processes, one for each decision category and simultaneously at play, 
have been developed to address some of the limitations \citep{usher2001time, brown2008simplest, leite2010modeling, dufau2012say, kim2017bayesian}, 
but the relevant literature remains sparse and focused only on simple static designs. 

Learning to distinguish Mandarin tones or, more generally, to make categorization decisions is, however, a dynamic process, driven by continuous and nuanced perceptual adjustments in our brain and behavior over time. 
The existing simple static models are thus severely limited in their ability to capture the true inherent complexities, including assessing the biologically relevant changes that take place over the learning period. 
Principled statistical approaches to multi-category dynamic drift-diffusion mixed effects models, 
that appropriately accommodate fixed effects of experimental factors as well as random effects due to subjects, 
are therefore highly needed but present daunting methodological and computational challenges. 

In this article, we address these challenges by developing a novel biologically interpretable flexible Bayesian semiparametric inverse Gaussian drift-diffusion mixed model for studying multi-alternative perceptual decision making processes in longitudinal settings. 


%
%

Our construction proceeds by characterizing the accumulation of evidence for different input-response tone combinations by associated independent Wiener diffusion processes, 
resulting in an inverse Gaussian distribution based joint probability model for the final response tone and the associated response time. 
To adapt this to a longitudinal mixed model setting, 
we then assume the model parameters to comprise input-response tone specific fixed effects and subject specific random effects, 
modeling them both by mixtures of locally supported B-spline bases \citep{de1978practical, eilers1996flexible} 
spanning the length of the longitudinal experiment. 
Both these effects are thus allowed to evolve flexibly as smooth functions over the training period \citep{ramsay2007applied, morris2015functional, wang2016functional} as the participants get more experience and training in their assigned decision tasks.

Dependence in the fixed effects model spline coefficients across adjacent temporal regions is induced via hidden Markov models (HMMs) \citep{mcdonald_zuchhini:1997,Rabiner:1989,fruhwirth2006finite,cappe2009inference}, 
one for each input-response tone combination but all sharing a common state space, 
as well as a novel smoothness inducing Markovian prior on the core spline coefficients. 
The HMMs, adapted in such novel ways, induce a local clustering of the fixed effects spline coefficients associated with different input-response tone combinations, 
in effect, 
allowing us to assess local similarities and differences between the 
corresponding parameter trajectories in different learning phases. 

This ability to infer local similarities and differences in the cognitive dynamics is theoretically and practically relevant for tone learning applications. 
The underlying mechanisms are expected to be very similar when the participants are first introduced to the tones; 
differences may appear as they get better at identifying the tones as some tones may be easier to identify than others in this stage; 
these differences may start to disappear again in later stages of the experiment as the participants become highly proficient in identifying all the different tones. 
As for individual heterogeneity, neural measures of sensory encoding information collected prior to the learning task show no clear individual differences, 
even though the process of learning itself results in good and poor learners \citep{reetzke2018tracing}. 


The literature on longitudinal data analysis models is enormous.
See, for example, books by \cite{diggle2002analysis, singer2003applied, fitzmaurice2008longitudinal} and the references therein. 
Bayesian methods for longitudinal data have also been extensively developed \citep[][etc.]{daniels2002bayesian, chib2002semiparametric, li2010bayesian, muller2013bayesian, quintana2016bayesian}. 
The problem of modeling locally clustered effects has, however, not garnered much attention. 
We can only mention \cite{petrone2009hybrid, nguyen2011dirichlet, nguyen2014bayesian}, 
all of which were designed primarily for normally distributed functional data with continuous covariates. 
It is not clear how these approaches can be adapted to our problem.  


Overall, our proposed method 
takes the existing state-of-the-art many significant steps forward, 
including (a) introducing a novel biologically interpretable class of multi-category inverse Gaussian drift-diffusion models for decision making,
(b) accommodating fixed effects of perceptual stimuli and random effects due to subject specific heterogeneity in such models in a statistically principled manner, 
(c) adapting these models to longitudinal study designs, studying the temporal evolution of the underlying process parameters as the subjects get trained and experienced in their assigned decision tasks, 
(d) allowing the process parameters to be locally clustered, 
enabling the assessment of their similarities and differences 
in various learning stages. 

Applied to our motivating tone learning data set, 
the proposed method provides many novel insights into the cognitive dynamics, 
allowing us to answer important scientific questions 
completely outside the scope of the previously existing literature. 
These include 
a detailed understanding of how biologically significant model parameters, that systematically relate to the underlying neural processes, evolve and interplay to enable gradual longitudinal learning in the participants, 
how similar or different these parameters are across different input and output tone combinations in different learning phases, 
how these processes differ between a good and a bad learner, 
etc.

The rest of this article is organized as follows. 
Section \ref{sec: background} provides additional background on tone learning and drift-diffusion models. 
Section \ref{sec: lddmm} details our novel locally varying longitudinal drift-diffusion mixed model. 
Section \ref{sec: post inference main} outlines computational challenges and solution strategies. 
Section \ref{sec: application} presents the results of the proposed method applied to tone learning data. 
Section \ref{sec: discussion} contains concluding remarks.
Substantive additional details, including a Markov chain Monte Carlo (MCMC) based posterior inference algorithm and results of simulation experiments, are presented in the supplementary materials.

\section{Behavioral Data and Scientific Background} \label{sec: background}

The behavioral data set that motivated our research comes from an intensive multi-day longitudinal speech category training study reported previously in \cite{reetzke2018tracing}. 
In this study,   
$n = 20$ native English-speaking adults were trained to categorize Mandarin Chinese syllables into lexical tone categories as a function of their pitch contour. 
Mandarin Chinese has four syllabic pitch contours or tones that are used to convey different lexical meanings. 
For example, in Mandarin Chinese, the syllable `ma' can be interpreted as `mother', `hemp', `horse', or `scold' depending on whether is pronounced with a high-level (T1), low-rising (T2), low-dipping (T3), or high-falling (T4) tone, respectively. 
The stimuli consisted of these tones pronounced by four native Mandarin speakers. 
The trials were administered in homogeneous blocks. 
Each block comprised $40$ categorization trials for $40$ different speech exemplars, 
corresponding to different combinations of speakers, syllables, and input tones.
Participants were trained across several days, with five blocks on each day. 
On each categorization trial, participants indicated the tone category they heard via a button press on a computer keyboard. 
Following the button press, the participants were given corrective feedback (`Correct/Incorrect') on a computer screen which was previously shown to be more effective in enhancing learning compared to full feedback (for example, `Incorrect, that was a category 2') \citep{chandrasekaran2014dual}. 
Individual categorization performance was monitored across training sessions until each participant achieved and maintained accuracy levels comparable to that of native speakers of Mandarin.

\begin{figure}[!ht]
\begin{center}
\includegraphics[width=0.47\linewidth]{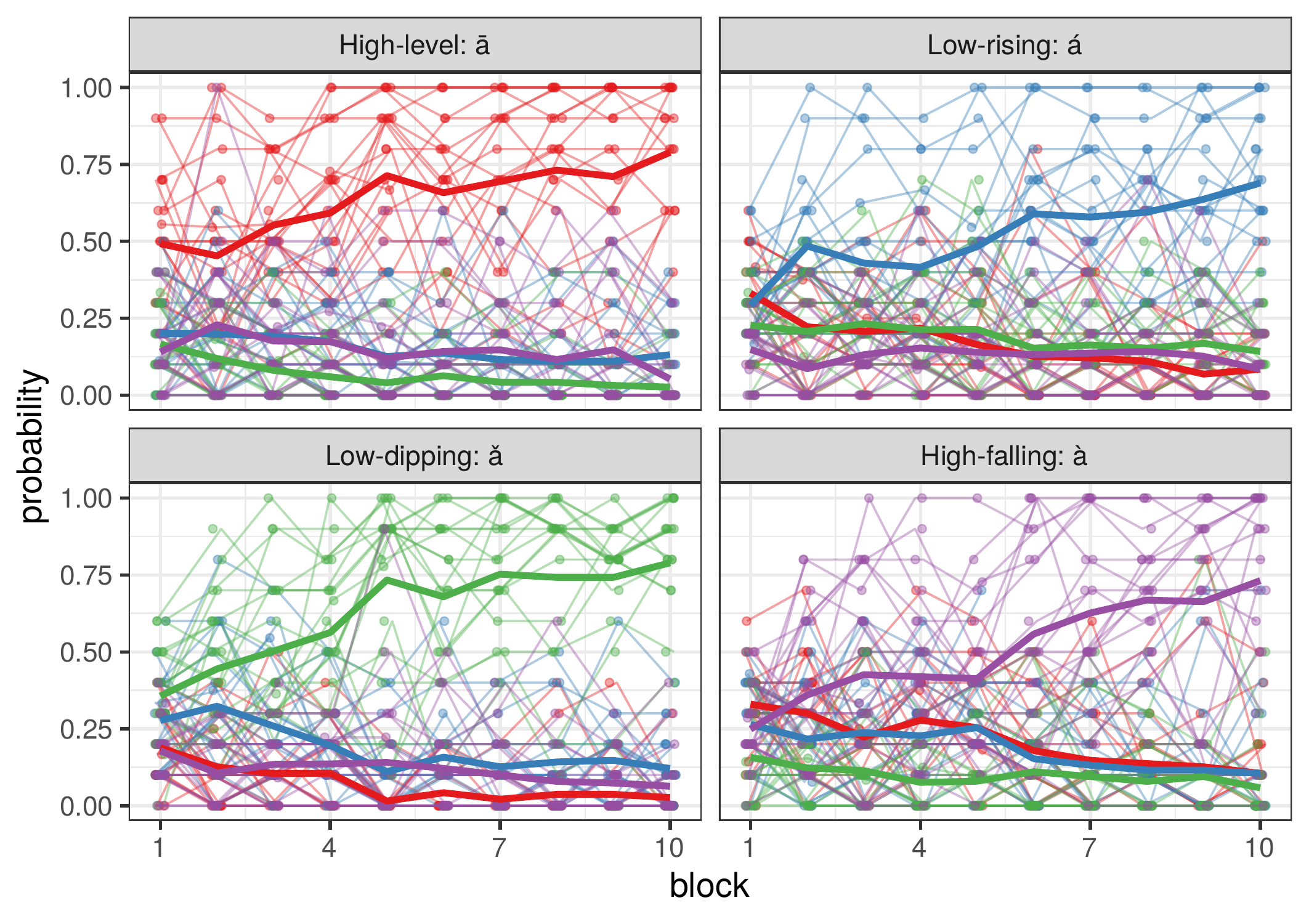} \hspace{0.5cm}
\includegraphics[width=0.47\linewidth]{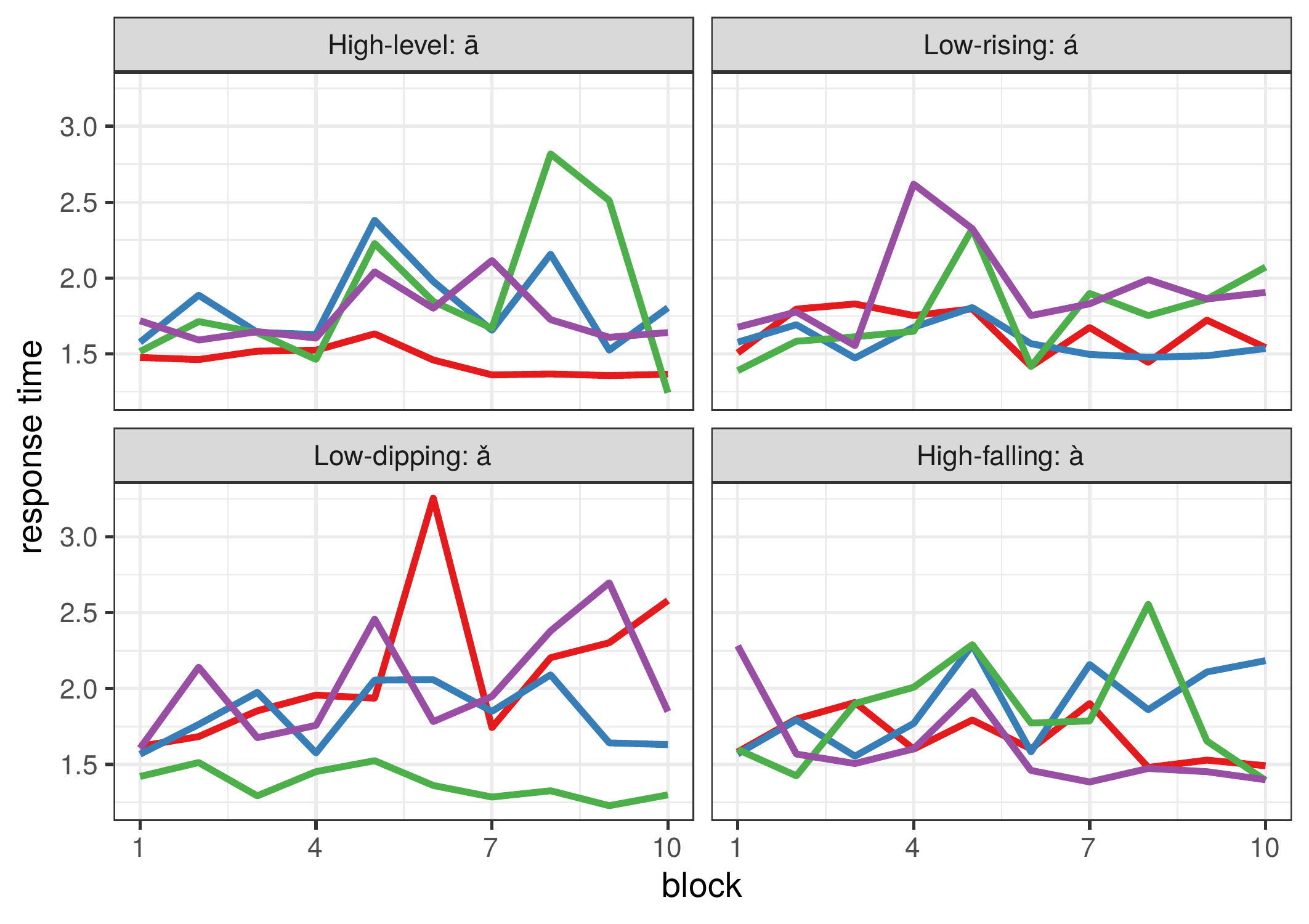}
\end{center}
\vskip -15pt
\caption{\baselineskip=10pt Left panel: Proportions of times an input tone was classified into different tone categories by different subjects. The thick line represents the average performance across subjects. 
Right panel: Associated response times averaged across subjects for clarity. 
In both panels, high-level tone responses are shown in red; low-rising in blue; low-dipping in green; and high-falling in purple.
}
\label{fig: accuracies and RTs} 
\end{figure}

The data consist of the tone responses and the associated response times for different input tones for the 20 participants. 
We focus here on the first two days of training (10 blocks in total) 
as they exhibited the steepest improvement in learning as well as the most striking individual differences relative to any other collection of blocks (Figure \ref{fig: accuracies and RTs}). 
In that sense, they provide an optimal longitudinal frame to assess the effects of learning on decision making variables.

Tone learning can be viewed from a broader perspective of multi-category decision making tasks, 
and hence can be studied using computational models developed for such tasks.  
We present here a brief nontechnical overview of how these models relate to the underlying neurobiology. 
Mathematical details and developments are deferred to Section \ref{sec: lddmm}.



In a typical multi-category decision task, the brain accumulates sensory evidence in order to make a categorical decision. This accumulation process is reflected in increasing firing rate at local neural populations associated with alternative decisions.
A decision is taken when neural activity in one of these populations crosses a particular threshold level. 
The decision category that is finally chosen is the one whose decision threshold is crossed first \citep{gold2007neural,brody2016neural}. 

Changes in evidence accumulation rates and decision thresholds can be induced by task difficulty, neurostimulation, and/or individual differences in cognitive function
\citep{cavanagh2011subthalamic,ding2013basal}. 
%
Decision-making is also regulated by demands on both speed and accuracy as a function of the task \citep{bogacz2010neural,milosavljevic2010drift}. 
The overall learning accuracies (`Correct/Incorrect' response proportions) in our data set were previously analyzed in \cite{paulon2019functional} using a binary logistic longitudinal mixed model. 
In a different context, \cite{craigmile2010hierarchical} had developed a model for response times. 
Separate models for accuracies and response times cannot, however, provide a meaningful interpretation of the speed-accuracy trade-off. 


\begin{figure}[!ht]
	\centering
	\includegraphics[width=0.7\linewidth]{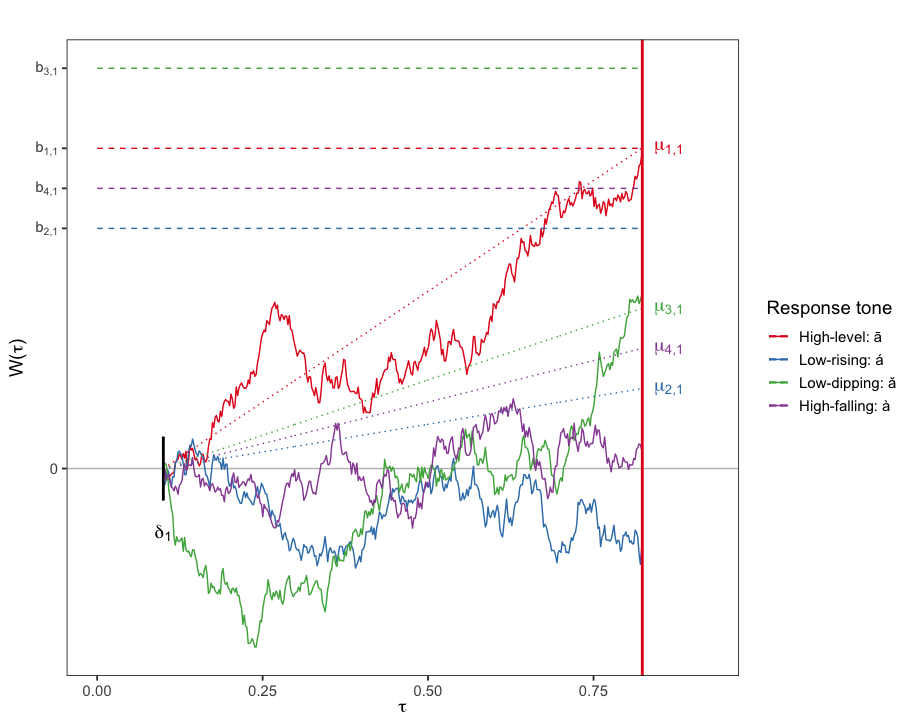}
	\caption{Drift-diffusion model for perceptual decision making.
	After an initial $\delta_{s}$ amount of time required to encode an input signal $s$,
	the evidence in favor of a response category $d$ accumulates according to a Wiener diffusion process with drift $\mu_{d,s}$. 
	The decision $d$ is eventually taken if the underlying process is the first to reach its decision boundary $b_{d,s}$.
	Here we illustrate a tone learning trial with input tone T1 ($s = 1$) that was eventually correctly identified. 
	Section \ref{sec: background} provides additional neurobiological background. 
	Section \ref{sec: lddmm} provides additional mathematical details.
	}
	\label{fig: drift diffusion}
\end{figure}

An excellent basis for jointly modeling accuracies and response times is 
obtained by imitating the underlying neural evidence accumulation mechanisms via latent drift-diffusion processes racing toward their respective boundaries, 
the process reaching its boundary first producing the final observed decision and the time taken to reach this boundary giving the associated response time (Figure \ref{fig: drift diffusion}) \citep{usher2001time}. 
The drift and the boundary parameters jointly explain the dynamics of choice, including the speed-accuracy trade-off. 
Broadly speaking, decision thresholds remaining fixed, higher drift rates 
lead to faster and more accurate responses; 
for fixed drift rates, higher decision thresholds, 
on the other hand, increase response times as well as inaccuracies. 

{In our motivating tone learning experiment, 
we are interested in understanding 
the evolution and interplay of the drift and the boundary parameters 
behind the improved tone identification performances over
training. 
Importantly, as was also discussed in the introduction, 
we are not just interested in estimating the overall trajectories of these parameters 
but also how they might differ between different input-response tone combinations locally in different longitudinal stages of the experiment. 
Additional interest lies in assessing subject level heterogeneity in these parameter trajectories, 
including particularly how they differ between good versus bad learners. 
}

\section{Longitudinal Drift-Diffusion Mixed Models} \label{sec: lddmm}
The basic Wiener diffusion process can be specified as 
$W(\tau) = \mu \tau + \sigma B(\tau)$,
where $B(\tau)$ is the standard Brownian motion, $\mu$ is the drift rate, and $\sigma$ is the diffusion coefficient \citep{cox1965theory, ross1996stochastic}. 
The process has independent normally distributed increments, that is, $\Delta W(\tau) = \{W(\tau+\Delta\tau) - W(\tau)\} \sim \Normal(\mu \Delta\tau,\sigma^{2} \Delta\tau)$, independently from $W(\tau)$. 
The first passage time of crossing a threshold $b$, $\tau = \inf \{\tau^{\prime}: W(0)=0, W(\tau^{\prime}) \geq b\}$, 
is then distributed according to an inverse Gaussian distribution \citep{whitmore1987heuristic, chhikara1988inverse, lu1995degradation} with density \vspace{-4ex}\\
\bse
\textstyle f(\tau \mid \mu,\sigma^{2},b)  =  \frac{b}{\sqrt{2\pi \sigma^{2}}} \tau^{-3/2} \exp\left\{- \frac{(b-\mu \tau)^{2}}{2\sigma^{2} \tau} \right\},~~~b>0,~~~\mu>0,~~~\sigma^{2}>0.
\ese
\vspace{-4ex}\\
With $\btheta=(\mu,\sigma,b)\trans$, we have $\eE(\tau \mid \btheta) = b / \mu$ and $\var(\tau \mid \btheta) = b \sigma^{2}/ \mu^{3}$.

Given perceptual stimuli and a set of decision choices, the neurons in the brain accumulate evidence in favor of the different alternatives. 
Modeling this behavior using Wiener processes with unit variances, 
assuming that a response is given when the decision threshold for one of the options is crossed, 
a probability model for the time $\tau_{d}$ to reach the threshold for the $d\th$ decision category under the influence of the $s\th$ stimulus is obtained as
\vspace{-4ex}\\
\be
\textstyle f(\tau_{d} \mid \delta_{s},\mu_{d,s},1, b_{d,s})  =  \frac{b_{d,s}}{ \sqrt{2\pi} } (\tau_{d}-\delta_{s})^{-3/2} \exp\left[- \frac{\{b_{d,s}-\mu_{d,s} (\tau_{d}-\delta_{s})\}^{2}}{2  (\tau_{d}-\delta_{s})} \right], \label{eq: drift-diffusion 1}
\ee
\vspace{-4ex}\\
where $\mu_{d,s}$ denotes the rate of accumulation of evidence, 
$b_{d,s}$ the decision boundaries, 
and $\delta_{s}$ an offset representing the collective time required to encode the $s\th$ signal before evidence accumulation begins, the time to press a computer key to record a response after a decision is reached, etc. 
(Figure \ref{fig: drift diffusion}). 
We now let $\btheta_{d,s}=(\delta_{s},\mu_{d,s},b_{d,s})\trans$. 
Since a decision $d$ is reached at response time $\tau$ if the corresponding threshold is crossed first, that is when $\{\tau = \tau_{d}\} \cap_{d^{\prime} \neq d} \{\tau_{d^{\prime}} > \tau_{d}\}$, we have $d = \arg \min \tau_{d^{\prime}}$. 
Assuming simultaneous accumulation of evidence for all decision categories, 
modeled by independent Wiener processes, 
and termination when the threshold for the observed decision category $d$ is reached, 
the joint distribution of $(d, \tau)$ is thus given by 
\vspace{-4ex}\\
\be
& f(d, \tau \mid s,\btheta) = g(\tau \mid \btheta_{d,s}) \prod_{d^{\prime} \neq d} \{1 - G(\tau \mid \btheta_{d^{\prime},s})\}.   \label{eq: inv-Gaussian likelihood}
\ee
\vspace{-4ex}\\
where, to distinguish from the generic notation $f$, we now use $g(\cdot \mid \btheta)$ and $G(\cdot \mid \btheta)$ to denote, 
respectively, the probability density function (pdf) and the cumulative distribution function (cdf) of an inverse Gaussian distribution, as defined in (\ref{eq: drift-diffusion 1}). 
We refer to model (\ref{eq: inv-Gaussian likelihood}) as the inverse Gaussian drift-diffusion model.

The marginal distribution of the response times $\tau$ under the influence of stimulus $s$ is then obtained as 
\vspace{-4ex}\\
\be
& f(\tau \mid s,\btheta) = \sum_{d} g(\tau \mid \btheta_{d,s}) \prod_{d^{\prime} \neq d} \left\{1 - G(\tau \mid \btheta_{d^{\prime},s})\right\}. \label{eq: marginal tau}
\ee
\vspace{-4ex}\\
The marginal probability of taking decision $d$ under the influence of stimulus $s$ is likewise obtained as 
\vspace{-4ex}\\
\be
& f(d \mid s,\btheta) = \int_{\delta_{s}}^{\infty} g(\tau \mid \btheta_{d,s}) \prod_{d^{\prime} \neq d} \left\{1 - G(\tau \mid \btheta_{d^{\prime},s})\right\} d\tau. \label{eq: multinomial inv-probit}
\ee
\vspace{-4ex}\\
Interestingly, model (\ref{eq: multinomial inv-probit}) is similar to traditional multinomial probit/logit regression models \citep{borooah2002logit,agresti2018introduction} 
except that the latent variables are now inverse Gaussian distributed as opposed to being normal or extreme-value distributed, 
and the observed category is associated with the minimum of the latent variables in contrast to being identified with the maximum of the latent variables. 

In an interesting recent work, \cite{kunkel2019bayesian} have also used an inverse Gaussian distribution based hierarchical Bayesian model for decision making, albeit in a simpler binary category case, 
focusing primarily on individual level models 
with no mechanism to assess population level effects or their dynamic complexities.

For our motivating longitudinal tone learning experiment described in Section \ref{sec: background}, 
for $i\in \{1,\dots,n=20\}, \ell\in \{1,\dots,L=40\}, t \in \{1,\dots,T=10\}$, let $s_{i,\ell,t}$ denote the input tone for the $i\th$ individual in the $\ell\th$ trial in block $t$. 
Likewise, let $d_{i,\ell,t}$ and $\tau_{i,\ell,t}$ denote, respectively, the selected Mandarin tone 
and the time taken to reach the corresponding threshold by the $i\th$ individual in the $\ell\th$ trial in block $t$. 
We now have 
\vspace{-4ex}\\
\be
& \hspace*{-0.5cm} g\{\tau_{i,\ell,t} \mid s_{i,\ell,t}=s, \btheta_{d,s}^{(i)}(t)\}  \textstyle = \frac{b_{d,s}^{(i)}(t)}{\sqrt{2\pi} (\tau_{i,\ell,t}-\delta_{s}^{(i)})^{3/2}}  \exp\left[- \frac{\{b_{d,s}^{(i)}(t)-\mu_{d,s}^{(i)}(t) (\tau_{i,\ell,t}-\delta_{s}^{(i)})\}^{2}}   {2 (\tau_{i,\ell,t}-\delta_{s}^{(i)})} \right].  \label{eq: mod1b}
\ee
\vspace{-4ex}\\
The drift rates $\mu_{d,s}^{(i)}(t)$ and the decision boundaries $b_{d,s}^{(i)}(t)$ now also vary with the blocks $t$. 
In addition, we accommodate random effects by allowing $\delta_{s}^{(i)}$, $\mu_{d,s}^{(i)}(t)$ and $b_{d,s}^{(i)}(t)$ to also depend on the subject index $i$.  
We let 
$y_{i,\ell,t} = (d_{i,\ell,t},\tau_{i,\ell,t})$, 
$\by=\{y_{i,\ell,t}\}_{i,\ell,t}$, and $d_{0} = 4$ be the number of possible decision categories (T1, T2, T3, T4). 
The likelihood function of our longitudinal drift-diffusion mixed model thus takes the form
\vspace{-4ex}\\
\bse
L(\by \mid \bs, \btheta) = \prod_{d=1}^{d_{0}} \prod_{s=1}^{d_{0}} \prod_{t=1}^{T}\prod_{i=1}^{n} \prod_{\ell=1}^{L} \left(g\{\tau_{i,\ell,t} \mid \btheta_{d,s}^{(i)}(t)\} \prod_{d^{\prime} \neq d} [1 - G\{\tau_{i,\ell,t} \mid \btheta_{d^{\prime},s}^{(i)}(t)\}]\right)^{1\{d_{i,\ell,t} = d, s_{i,\ell,t}=s\}}.
\ese
\vspace{-4ex}

\subsection{Modeling the Offsets}

The offset parameters $\delta_{s}^{(i)}$, we recall, signify 
the times spent on encoding the different input tones, 
the time to press computer keys to record the responses, etc., 
and hence are not directly relevant to the actual decision making processes. 
These parameters are thus biologically not very interesting but may still vary between individuals and have an important effect on the estimates of drift rates and boundaries \citep{teichert2016importance}. 
We thus let them vary between input stimuli and participants but assume them to remain stable across blocks as in (\ref{eq: mod1b}). 

We assign uniform priors on $\delta_{s}^{(i)} \sim \Unif(0,\delta_{s,i,\max})$, 
where $\delta_{s,i,\max}$ is the minimum of all response times under stimulus $s$ for individual $i$, that is, $\delta_{s,i,\max} = \min_{\{(\ell,t): s_{i,\ell,t}=s\}} \tau_{i,\ell,t}$.

\subsection{Modeling the Drifts and the Boundaries} \label{sec: drift and boundaries}
Our modeling efforts concentrate henceforth on flexibly characterizing  
the longitudinal evolution of the mixed effects parameters $\mu_{d,s}^{(i)}(t), b_{d,s}^{(i)}(t)$. 
Variations in these parameters over training blocks explain perceptual learning in the participants.
Variations across participants, on the other hand, explain their performance heterogeneity. 
Following the discussion in the introduction, 
of particular interest are 
the local similarities and differences 
between these parameters for different input-response tone combinations $(d,s)$ in different learning phases.

To this end, 
we propose essentially identical modeling strategies for 
$\mu_{d,s}^{(i)}(t)$ and $b_{d,s}^{(i)}(t)$. 
For ease of exposition avoiding unnecessary repetition, 
we describe below only these common strategies using simplified generic notations. 
With $x=(d,s) \in \X = \{(1,1),(1,2),\dots,(4,4)\} \equiv \{1,2, \dots, x_{\max}\}$, $x_{\max} = 4 \times 4$, succinctly representing the input-response tone combinations 
and, with some abuse, $\theta_{x}^{(i)}(t)$ being a generic for  $\mu_{d,s}^{(i)}(t)$ and $b_{d,s}^{(i)}(t)$, we let 
\vspace{-4ex}\\
\be
\begin{split}
& \theta_{x}^{(i)}(t)  =  \exp\{f_{x}(t) + u_{x}^{(i)}(t)\}, ~~~~~u_{x}^{(i)}(t) \sim f_{u}\{u_{x}^{(i)}(t)\}.
\end{split} \label{eq: function 1}
\ee
\vspace{-4ex}\\
The exponentiation in (\ref{eq: function 1}) enforces positivity constraints; 
$f_{x}(t)$ and $u_{x}^{(i)}(t)$ denote, respectively, 
additive fixed and random effects components in the exponential scale; $f_{u}$ denotes the underlying random effects distribution. 
When needed, the fixed and random effects components for the drifts and the boundaries, as well as associated parameters and hyper-parameters, will be distinguished by reintroducing the subscripts as $f_{\mu,x}(t),f_{b,x}(t)$, $u_{\mu,x}^{(i)}(t), u_{b,x}^{(i)}(t)$ etc. 
To further simplify notation, generic data recording experimental blocks in $\{1,\dots, T\}$ as well as other generic time points in $[1,T]$ will both be denoted by $t$.  
Likewise, generic input-response tone combinations as well as their particular values will both be denoted by $x$ and so forth. 

We model the components $f_{x}(t)$ and $u_{x}^{(i)}(t)$, and hence $\theta_{x}^{(i)}(t)$, to all be smoothly varying functions over $t \in [1,T]$. 
A functional approach is not strictly necessary if inference is restricted only to the $T$ data recording blocks blocks $t \in \{1,\dots, T\}$. 
Learning may, however, be viewed as a continuous process - 
the brain synthesizes information from relevant past experiences even when not being actively engaged in actual decision making.
A functional approach to modeling $f_{x}(t)$ and $u_{x}^{(i)}(t)$ for any $t \in [1,T]$, 
not just the experimental blocks $t \in \{1,\dots, T\}$, thus facilitates parameter interpretability. 
A functional approach is also practically convenient in characterizing smoothly varying longitudinal parameter trajectories. 

In modeling the fixed effects components $f_{x}(t)$, 
we are not only interested in characterizing their overall trajectories over time $t$ for different input-response combinations $x=(d,s)$ 
but also how they might vary locally between different values of $x$ in different learning stages. 
Compared to the fixed effects, we have to, however, rely on much less data to estimate the random effects $u_{x}^{(i)}(t)$ for different $x=(d,s)$ and different participant $i$, 
especially for $d \neq s$ toward later stages of the experiment when most participants identify the input tones with high accuracies. 
Our models and inferential goals for the random effects $u_{x}^{(i)}(t)$ will therefore be relatively modest. 

\subsubsection{Locally Varying Functional Fixed Effects} \label{sec: fixed effects}
\begin{figure}[!ht]
	\centering
	\includegraphics[width=.95\linewidth]{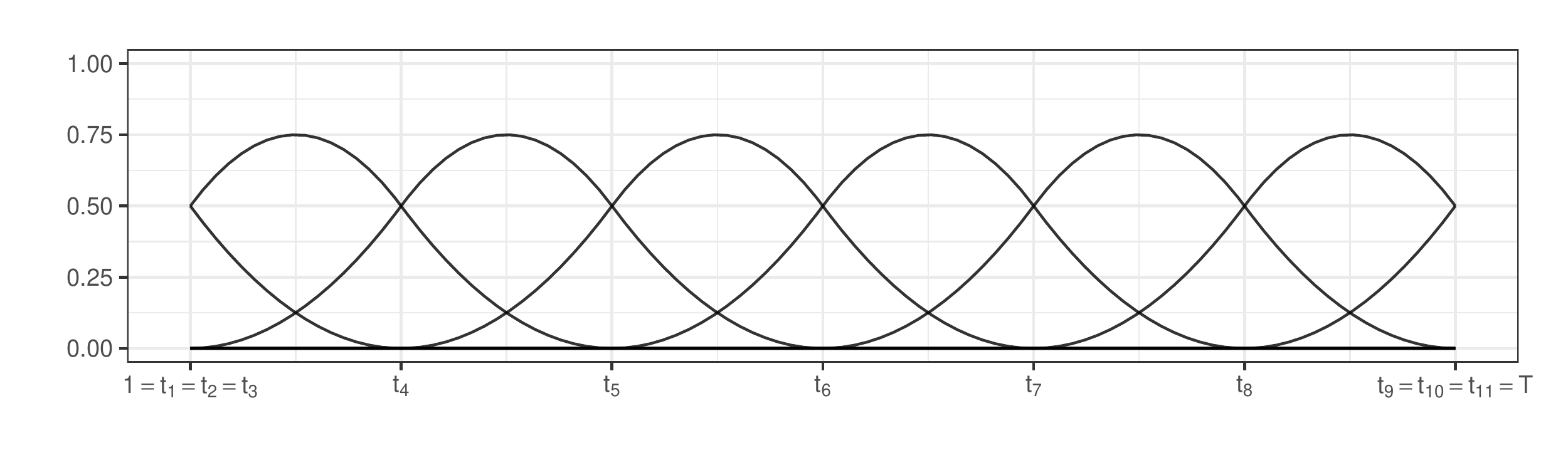}
	\caption{\baselineskip=10pt 
	Plot of 8 quadratic B-splines on an interval $[1,T]$ defined by $11$ knot points that divide $[1,T]$ into $K=6$ equal subintervals.
	}
	\label{fig: b-splines}
\end{figure}

We now propose a novel approach to modeling the latent functions $f_{x}(t)$ using basis decomposition methods that allow them to smoothly vary with the blocks $t$ 
while also depending locally on the indexing variable $x$. 
To begin with, we let 
\vspace{-4ex}\\
\be
\textstyle f_{x}(t) = \sum_{k=1}^{K} \beta_{k}^{(x)} B_{k}(t), \label{eq: fixed effects function}
\ee 
\vspace{-4ex}\\
where $\bB(t) = \{B_{1}(t),\dots, B_{K}(t)\}\trans$ are a set of known locally supported basis functions spanning $[1,T]$, 
$\bbeta^{(x)} = (\beta_{1}^{(x)},\dots,\beta_{K}^{(x)})\trans$ are associated unknown coefficients to be estimated from the data. 
In this article, we use quadratic B-spline bases with knot points coinciding with the block locations. 
B-splines are non-negative, continuous and have desirable local supports (Figure \ref{fig: b-splines}). 
Mixtures of B-splines are highly flexible \citep{de1978practical}. 
Allowing the $\beta_{k}^{(x)}$'s to flexibly vary with $x$, 
the model can accommodate widely different shapes for different input-response tone combinations.

It is difficult to assess how similar or different these functions are using such unstructured models. 
One potential solution is to cluster the spline coefficients $\bbeta^{(x)}$ associated with different input-response tone combinations $x$.  
If, for example, $\bbeta^{(x_{1})} = \bbeta^{(x_{2})}$ for two combinations $x_{1}$ and $x_{2}$, 
then we have $f_{x_{1}}(t)=f_{x_{2}}(t)$ for all $t$. 

Such global clustering of all elements of $\bbeta^{(x)}$ together does not, however, allow us to straightforwardly assess the local similarities and differences between these functions in different learning phases. 
To induce a desirable local cluster inducing mechanism, we introduce a set of latent variables $z_{k}^{(x)}$ for each input-response tone combination $x$ 
with a shared state space $\X$, and associated core coefficients $\beta_{k,z}^{\star}$ and let 
\vspace{-4ex}\\
\be
\begin{split}
& \textstyle (\beta_{k}^{(x)} \mid z_{k}^{(x)}=z_{k}) = \beta_{k,z_{k}}^{\star}, ~~\text{implying}~~\{f_{x}(t) \mid z_{k}^{(x)}=z_{k}, k=1,\dots,K\} = \sum_{k=1}^{K} \beta_{k,z_{k}}^{\star} B_{k}(t). 
\end{split} \label{eq: function 2}
\ee 
\vspace{-4ex}\\
%
The set of B-spline coefficients to be estimated at the $k\th$ location now comprises the $\beta_{k,z_{k}}^{\star}$'s 
that are indexed by $z_{k}^{(x)} = z_{k}$ at that location $k$. 
When $z_{k}^{(x_{1})} = z_{k}^{(x_{2})}$ for two different levels $x_{1}$ and $x_{2}$ of $x$, 
we have $\beta_{k}^{(x_{1})}=\beta_{k}^{(x_{2})}$ 
and the implied functions $f_{x_{1}}(t)$ and $f_{x_{2}}(t)$ will tend to be similar at location $k$. 
Indeed, 
for quadratic B-splines with knots at the blocks $\{1,\dots,T\}$, 
$f_{x_{1}}(t)$ and $f_{x_{2}}(t)$ will be exactly equal at block $t$ when $z_{t}^{(x_{1})} = z_{t}^{(x_{2})}$ and $z_{t+1}^{(x_{1})} = z_{t+1}^{(x_{2})}$.

In theory, we could use B-splines of other small degrees as they all enjoy local support properties. 
With linear splines, however, smoothness becomes harder to control, and with cubic splines, three latent variables would be needed to determine
the cluster configuration at each block $t$. 
We found quadratic B-splines to be a good compromise between the two for modeling smoothly varying curves while also maintaining easy interpretability of the latent variables.

Letting $\Z_{k} = \{z_{k}: z_{k}^{(x)}  = z_{k}~\text{for some}~x \in \X\}$, 
the case $\abs{\Z_{k}}=1$ then characterizes the scenario when the the spline coefficients for all input-response tone combinations $x$ are the same at location $k$.  
On the other end, when $\abs{\Z_{k}}= x_{\max} = 4 \times 4$, 
the spline coefficients are all different for different $x$ at location $k$. 
In our tone learning application, $\abs{\Z_{k}}$  tend to be much smaller than $x_{\max}$ uniformly for all $k$ and the restricted support $z_{k}^{(x)} \in \{1,\dots,z_{\max}\} \subset \X$ with $z_{\max} = 8 < x_{max}=16$ will suffice. 

\begin{figure}[ht!]
	\centering
	\hspace*{-0.00cm}\includegraphics[width=0.4\linewidth, trim=2cm 1.15cm 1cm 1.15cm,clip=true]{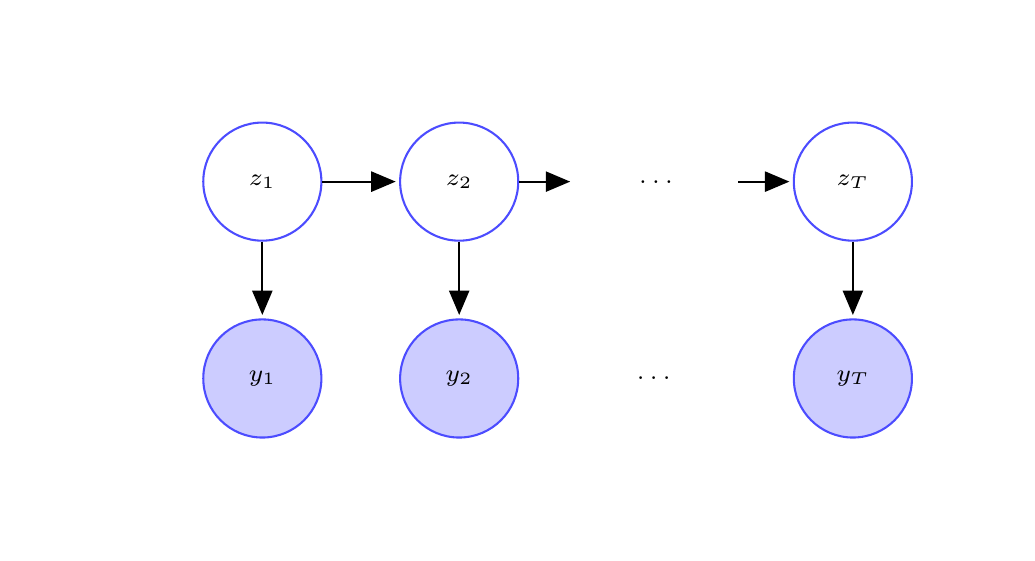} \quad
	\includegraphics[width=0.55\linewidth, trim=2cm 1.1cm 1cm 1.1cm,clip=true]{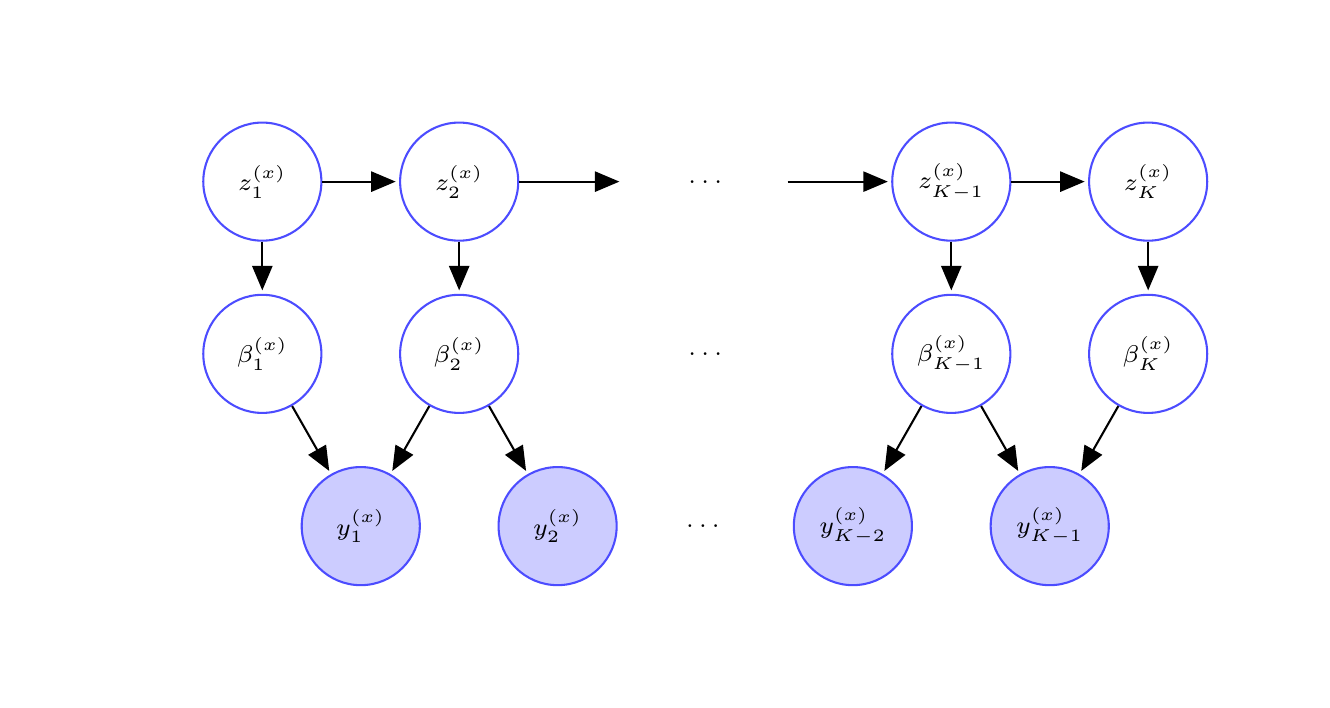}
	\caption{\baselineskip=10pt 
	Left panel: Graph of a conventional HMM. 
	Right panel: Graph of our proposed functional HMM model (\ref{eq: function 2}) with quadratic B-splines (Figure \ref{fig: b-splines}) with knots points coinciding with the data recording time blocks ($T = K - 1$). 
	}
	\label{fig: dag funHMM}
\end{figure}

We model the temporal evolution of the latent local cluster indicators $z_{k}^{(x)}, k =1,\dots,K$, using hidden Markov models (HMMs) (Figure \ref{fig: dag funHMM}). 
We consider two types of dynamics for the latent states corresponding to correct (C) and incorrect (I) identification of the tones. 
That is, 
\vspace{-4ex}\\
\bse
(z_{k}^{(d,s)} \mid z_{k-1}^{(d,s)} = z_{k-1}) \sim \Mult(\pi_{z_{k-1},1}^{(C)},\dots,\pi_{z_{k-1},z_{\max}}^{(C)})~~~~\text{when}~d=s,\\
(z_{k}^{(d,s)} \mid z_{k-1}^{(d,s)} = z_{k-1}) \sim \Mult(\pi_{z_{k-1},1}^{(I)},\dots,\pi_{z_{k-1},z_{\max}}^{(I)})~~~~\text{when}~d \neq s.
\ese
\vspace{-4ex}\\
The latent cluster inducing variables $z_{k}^{(x)}$'s are shared between $f_{\mu,x}(t)$ and $f_{b,x}(t)$, reducing computational complexities while also facilitating model interpretability. 
We assign Dirichlet priors on the transition probabilities 
\vspace{-4ex}\\
\bse
&\bpi_{z}^{(C)} = (\pi_{z,1}^{(C)},\dots,\pi_{z,z_{\max}}^{(C)})\trans \sim \Dir(\alpha^{(C)}/z_{\max},\dots,\alpha^{(C)}/z_{\max})~~~\text{with}~~~\alpha^{(C)} \sim \Ga(a_{\alpha},b_{\alpha}),\\
&\bpi_{z}^{(I)} = (\pi_{z,1}^{(I)},\dots,\pi_{z,z_{\max}}^{(I)})\trans \sim \Dir(\alpha^{(I)}/z_{\max},\dots,\alpha^{(I)}/z_{\max})~~~\text{with}~~~\alpha^{(I)} \sim \Ga(a_{\alpha},b_{\alpha}).
\ese
\vspace{-4ex}

\begin{figure}[ht!]
	\centering
	\hspace*{-0.00cm}\includegraphics[width=0.47\linewidth, trim=2cm 1.25cm 1cm 1.25cm]{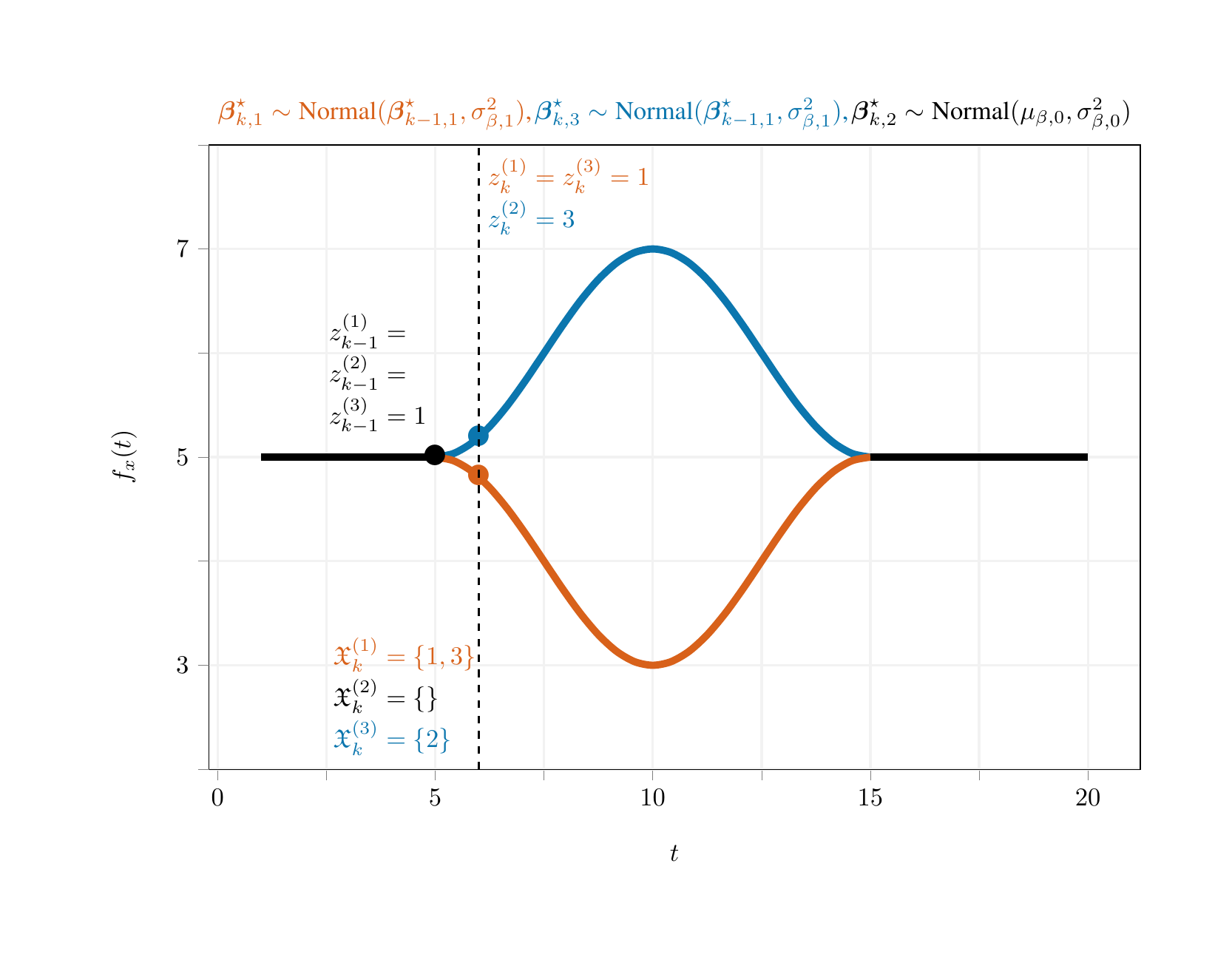} \hspace*{1cm}
	\hspace*{-0.40cm}\includegraphics[width=0.47\linewidth, trim=2cm 1.25cm 1cm 1.25cm]{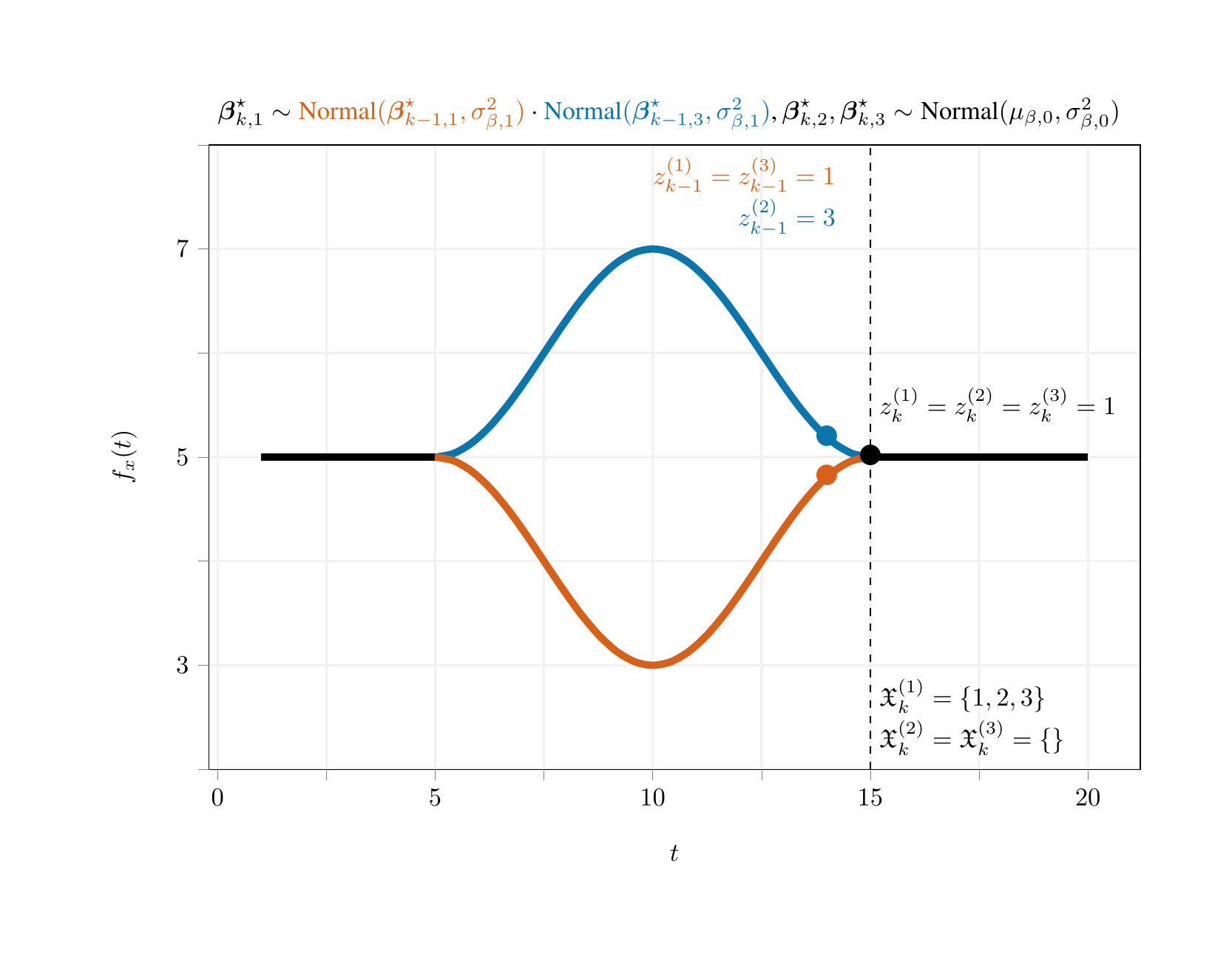}
	\caption{\baselineskip=10pt 
	An illustration of the prior on the spline core coefficients $\beta_{k,z_{k}}^{\star}$ at location $k$ (marked by the dashed vertical lines) in 
	the fixed effects model developed in Section \ref{sec: fixed effects} 
	for a synthetic scenario with $x \in \{1,2,3\}$, 
	where the curves corresponding to the three levels of $x$ are initially equal, the curves for $x=1,3$ (in red) and $x=2$ (in blue) then diverge at $t=6$, merging back again at $t=15$. 
	}
	\label{fig: prior core beta}
\end{figure}

We next consider priors for the atoms $\beta_{k,z_{k}}^{\star}$. 
Conditional on the $z_{k}^{(x)}$'s and the coefficients at the previous locations, for $k = 2, \dots, K$, we construct the priors sequentially as 
		\vspace{-4ex}\\
		\be
		\beta_{k,z_{k}}^{\star} \sim 
		\left\{\begin{array}{*2{>{\displaystyle}l}}
			\prod_{\substack{\{ z_{k-1}^{(x)} : ~ x \in \mathfrak{X}_{k}^{(z_{k})}\}}}   \hspace*{-20pt} \Normal\left( \beta^{\star}_{k-1,z_{k-1}^{(x)}}, \sigma_{\beta,1}^{2}\right) &\text{if}~|\mathfrak{X}_{k}^{(z_{k})}| > 0,
			\\
			\Normal(\mu_{\beta,0}, \sigma_{\beta,0}^{2}) &\text{otherwise},
			\end{array}\right.   \label{eq: prior on beta star}
		\ee
		\vspace{-1ex}\\
where $\mathfrak{X}_{k}^{(z_{k})} = \{x : z_{k}^{(x)} = z_{k}\}$ is the set of values of $x$ that, at the location $k$, are assigned the label $z_{k}$. 
In constructing the prior in this manner, 
we center the core coefficients around the ones that are `expressed' at the previous location (Figure \ref{fig: prior core beta}), penalizing their first order differences.
The coefficients that are not associated with any levels of $x$ are assigned a normal prior with a large variance $\sigma_{\beta,0}^{2}$. 
The initial coefficients are assigned non-informative flat priors as $\beta_{1,z_{k}}^{\star} \sim 1$. 
Additional illustrations on these smoothness inducing priors on the core coefficients can be found in Section \ref{sec: prior_details} of the supplementary materials.

The smoothness of the curves is controlled by the parameter $\sigma_{\beta,1}^{2}$ and is assigned a prior, allowing it to be informed by the data. 
We let 
\vspace{-4ex}\\
\bse
\sigma_{\beta,1}^{2} \sim \HC(0,1),
\ese  
\vspace{-4ex}\\ 
where $\HC(a,b)$ denotes a half-Cauchy distribution \citep{gelman2006prior, polson2012half} with location parameter $a$ and scale parameter $b$. 
The half-Cauchy distribution, which attains its mode at zero, is capable of capturing strong smoothness, while also having heavy tails, thus being capable of capturing wiggly functions. 
The choice of the scale hyper-parameter is discussed in Section \ref{sec: prior hyper-parameters} in the supplementary materials.


\begin{figure}[ht]
	\centering
	\includegraphics[width=0.75\linewidth, trim=2cm 1.25cm 1cm 1.25cm]{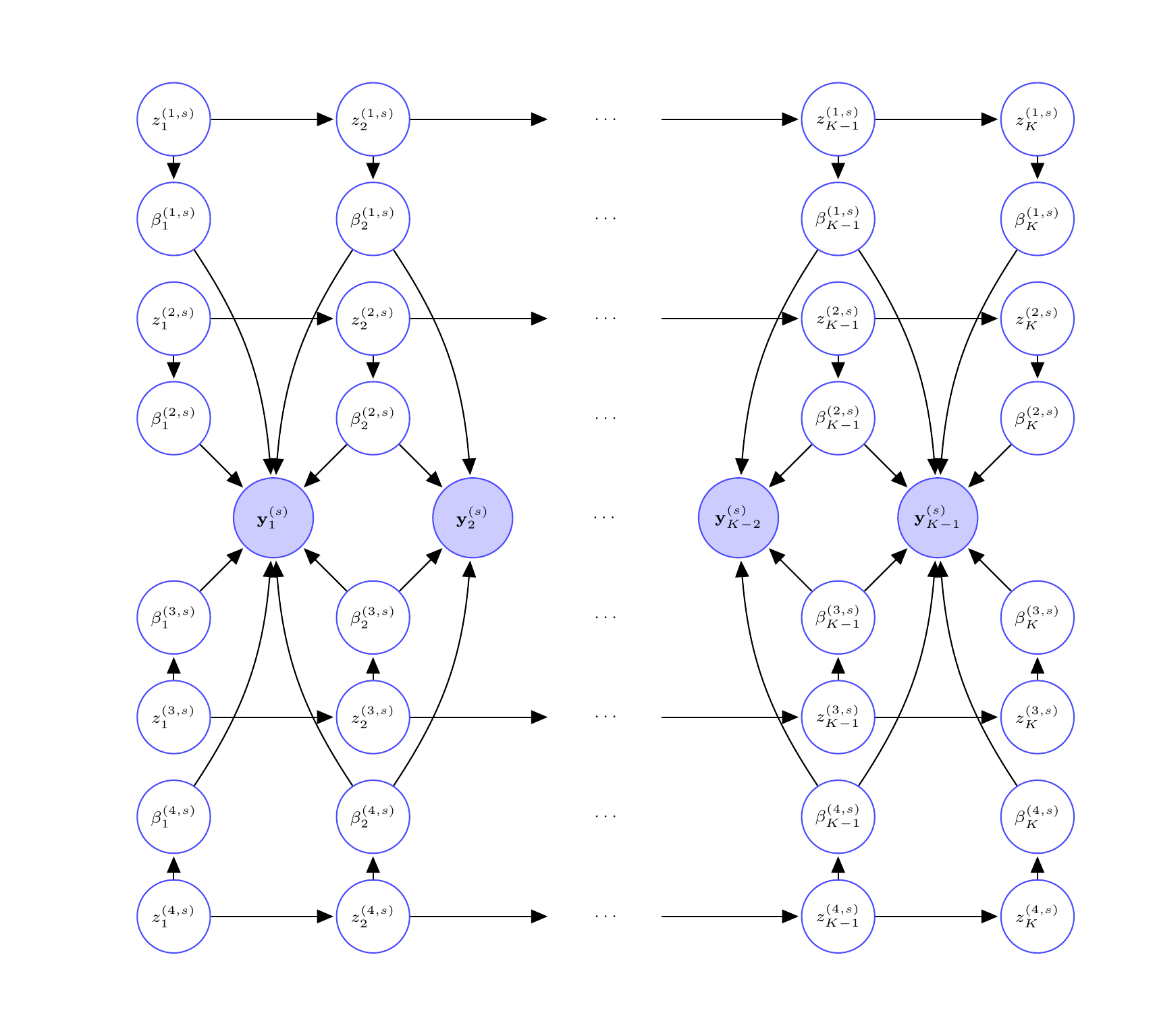} 
	\caption{\baselineskip=10pt 
	Graph of the proposed fixed effects model for tone learning. 
	}
	\label{fig: dag fHMM}
\end{figure}

Importantly, although our basic building blocks for the fixed effects components comprise conventional HMMs, 
one for each input-response tone combination $x=(d,s)$, 
for any input tone $s$, 
all four latent variables $z_{k}^{(1,s)}, z_{k}^{(2,s)}, z_{k}^{(3,s)}, z_{k}^{(4,s)}$ 
simultaneously appear in equation (\ref{eq: inv-Gaussian likelihood}).  
For each input tone, the graph for our tone learning model (Figure \ref{fig: dag fHMM} and Figure \ref{fig: dag fHMM supp} in the supplementary materials) thus resembles a factorial HMM \citep[fHMM]{ghahramani1996factorial} with four hidden layers. 
In the posterior, a latent state $z_{k}^{(d,s)}$ is thus informed by all responses generated under the tone $s$, not just the subset corresponding to $x=(d,s)$. 
This has important consequences for posterior inference, as we discuss in Section \ref{sec: post inference main}.

\subsubsection{Locally Varying Functional Random Effects} \label{sec: random effects}
We now focus on flexibly modeling the functional random effects components. 
For reasons outlined before Section \ref{sec: fixed effects},  estimating $u_{x}^{(i)}(t)$ for each different $x$ is a challenging task. 
For any participant, the random effects for correct and incorrect identification of the tones may, however, be expected to be on the opposite sides of the corresponding population level curves. 
Taking a middle path, we thus 
allow different random effects $u_{C}^{(i)}(t)$ and $u_{I}^{(i)}(t)$ for correct (C) and incorrect (I) identifications, respectively, as 
\vspace{-4ex}\\
\bse
& u_{d,s}^{(i)}(t) = u_{C}^{(i)}(t)~~~~\text{when}~d=s,
~~~~~u_{d,s}^{(i)}(t) = u_{I}^{(i)}(t)~~~~\text{when}~d \neq s.
\ese 
\vspace{-4ex} 

\begin{figure}[!ht]
	\centering
	\begin{center}
	\includegraphics[width=0.65\linewidth, trim=0cm 0cm 0cm 0cm, clip=true]{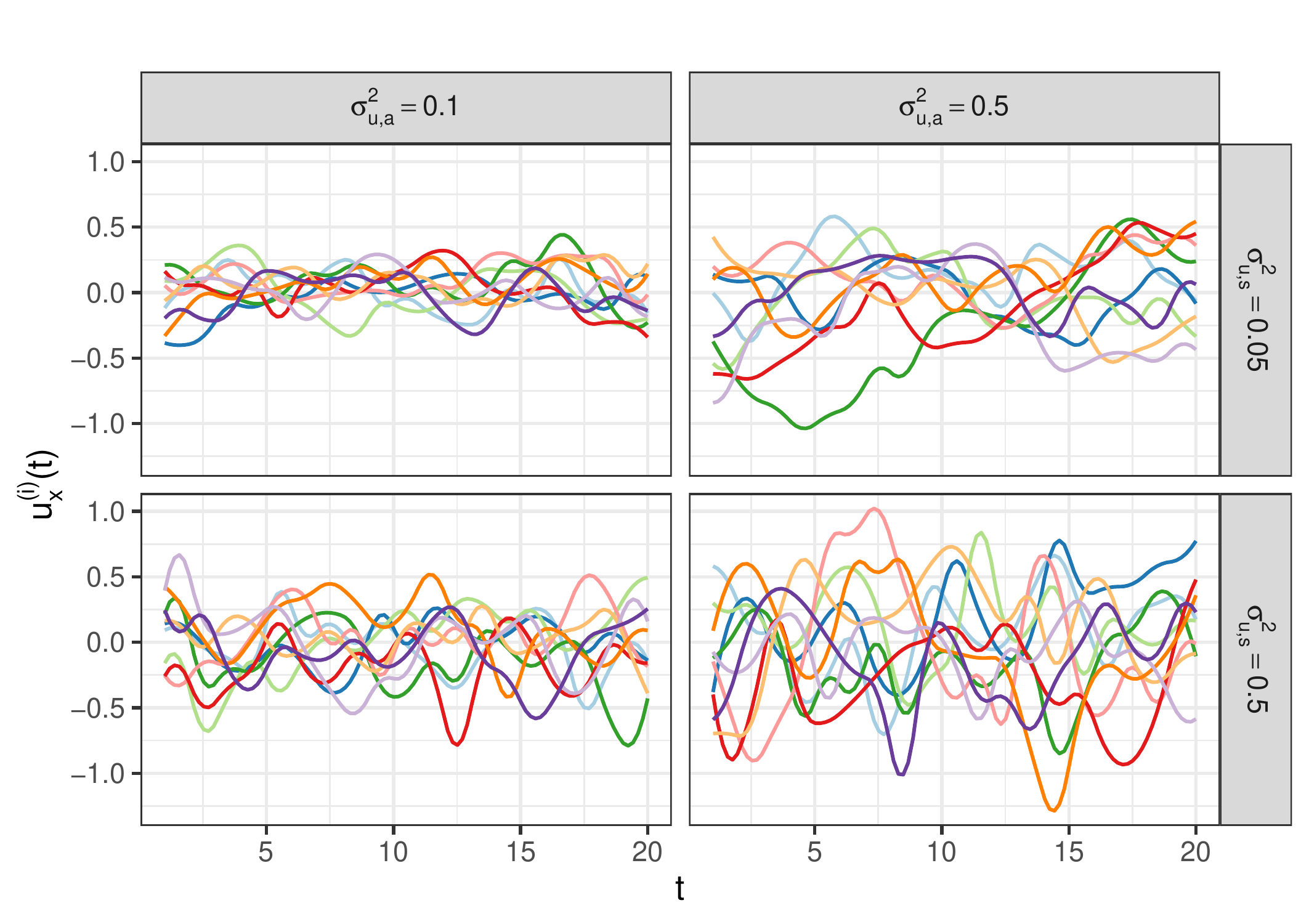}
	\end{center}
	\vskip -15pt
	\caption{An illustration of the functional random effects model proposed in Section \ref{sec: random effects}. 
	Each panel shows a collection of $10$ random draws from the random effects distribution for a combination of values of $(\sigma_{u,s}^{2}, \sigma_{u,a}^{2})$.  
	}
	\label{fig: random effects distribution illustrations}
\end{figure}

We adopt a common strategy to model both $u_{C}^{(i)}(t)$ and $u_{I}^{(i)}(t)$. 
Suppressing the subscripts to simplify notation and avoid repetition, 
we model the time-varying random effects components $u^{(i)}(t)$ as 
\vspace{-4ex}\\
\be
\begin{split}
& \textstyle u^{(i)}(t) = \sum_{k=1}^{K} \beta_{k,u}^{(i)} B_{k}(t), ~~~~~\\
& \bbeta_{u}^{(i)} \sim 
\MVN_{K}\{\bzero,(\sigma_{u,a}^{-2}\bI_{K}+\sigma_{u,s}^{-2}\bP_{u})^{-1}\},~~~~~\\
\end{split} \label{eq: random effects}
\ee
\vspace{-2ex}\\
where $\bbeta_{u}^{(i)} = (\beta_{1,u}^{(i)},\dots,\beta_{K,u}^{(i)})\trans$ are subject-specific spline coefficients, 
$\MVN_{K}(\bmu,\bSigma)$ denotes a $K$ dimensional multivariate normal distribution with mean $\bmu$ and covariance $\bSigma$.
We choose $\bP_{u} = \bD_{u}\trans \bD_{u}$, where the $(K-1) \times K$ matrix $\bD_{u}$ is such that $\bD_{u} \bbeta_{u}^{(i)}$ computes the first order differences in $\bbeta_{u}^{(i)}$.
The model thus penalizes $\sum_{k=1}^K (\nabla \beta_{k,u}^{(i)})^{2} = \bbeta_{u}^{(i) \rm{T}} \bP_{u} \bbeta_{u}^{(i)}$, the sum of squares of first order differences in $\bbeta_{u}^{(i)}$ \citep{eilers1996flexible}. 
The random effects variance parameter $\sigma_{u,s}^{2}$ models the smoothness of the random effects curves, smaller $\sigma_{u,s}^{2}$ inducing smoother $u^{(i)}(t)$'s. 
Additional variations from the constant zero curve are explained by $\sigma_{u,a}^{2}$ (Figure \ref{fig: random effects distribution illustrations}). 
The absence of random effects is signified by the limiting case $\sigma_{u,s}^{2} = \sigma_{u,a}^{2} = 0$.  
We assign half-Cauchy priors on the variance parameters as
\vspace{-4ex}\\
\bse
& \sigma_{u,s}^{2} \sim \HC (0, 1),~~~~~\sigma_{u,a}^{2} \sim \HC (0, 1).
\ese
\vspace{-4ex}

\vskip -0pt

Modeled in the same space of quadratic B-splines, 
the fixed and the random effects curves thus share similar smoothness properties. 
Having different smoothness controlling parameters, they are, however, allowed to have different smoothness levels. 
A similar approach, but with additional assumptions on the covariance matrix of the random effects, has previously been developed in \cite{guo2002functional}.
To our knowledge, model (\ref{eq: random effects}) for the random effects is thus also novel to the literature.

Integrating out the random effects, the corresponding population level parameters $\theta_{x}(t)$ are obtained as 
\vspace{-4ex}\\
\bse
\textstyle \theta_{x}(t) = \int \exp\{f_{x}(t) + u_{x}^{(i)}(t)\} f_{u}\{u_{x}^{(i)}(t)\} du_{x}^{(i)}(t) = \exp\left[f_{x}(t) + \frac{\var\{u_{x}^{(i)}(t)\}}{2}\right].
\ese
\vspace{-4ex}

\section{Posterior Inference} \label{sec: post inference main}
Posterior inference for conventional HMMs
can generally be based on samples drawn from the posterior using dynamic message passing MCMC algorithms \citep{Rabiner:1989,Scott:2002}. 
The nonstandard inverse Gaussian likelihood and the fHMM type model structure of our proposed longitudinal drift-diffusion mixed model, however, 
bring in significant additional complexities. 
We adapt recent advances in MCMC algorithms for discrete spaces \citep{neal2003slice,van2008beam,titsias2016hamming,zanella2019informed} in novel non-trivial ways, 
designing locally informative slice sampling moves that carefully exploits the conditional independence relationships encoded in the model 
to overcome the computational challenges. 
Due to space constraints, the details are deferred to Section \ref{sec: post inference sm} in the supplementary materials.

\section{Application to Tone Categorization Data} \label{sec: application}
In this section, we discuss the results produced by our method applied to the tone category learning data described in Section \ref{sec: background}. 
Our primary inference goals, we recall, 
include understanding systematic longitudinal variations in perceptual categorization decision
as the participants get better at identifying the four Mandarin tones  
with there being some additional interests in assessing individual specific trajectories, especially how they differ between good and bad learners.

\begin{figure}[ht!]
	\centering
	\includegraphics[width=6.5cm, trim=2cm 0.25cm 1cm 0.25cm]{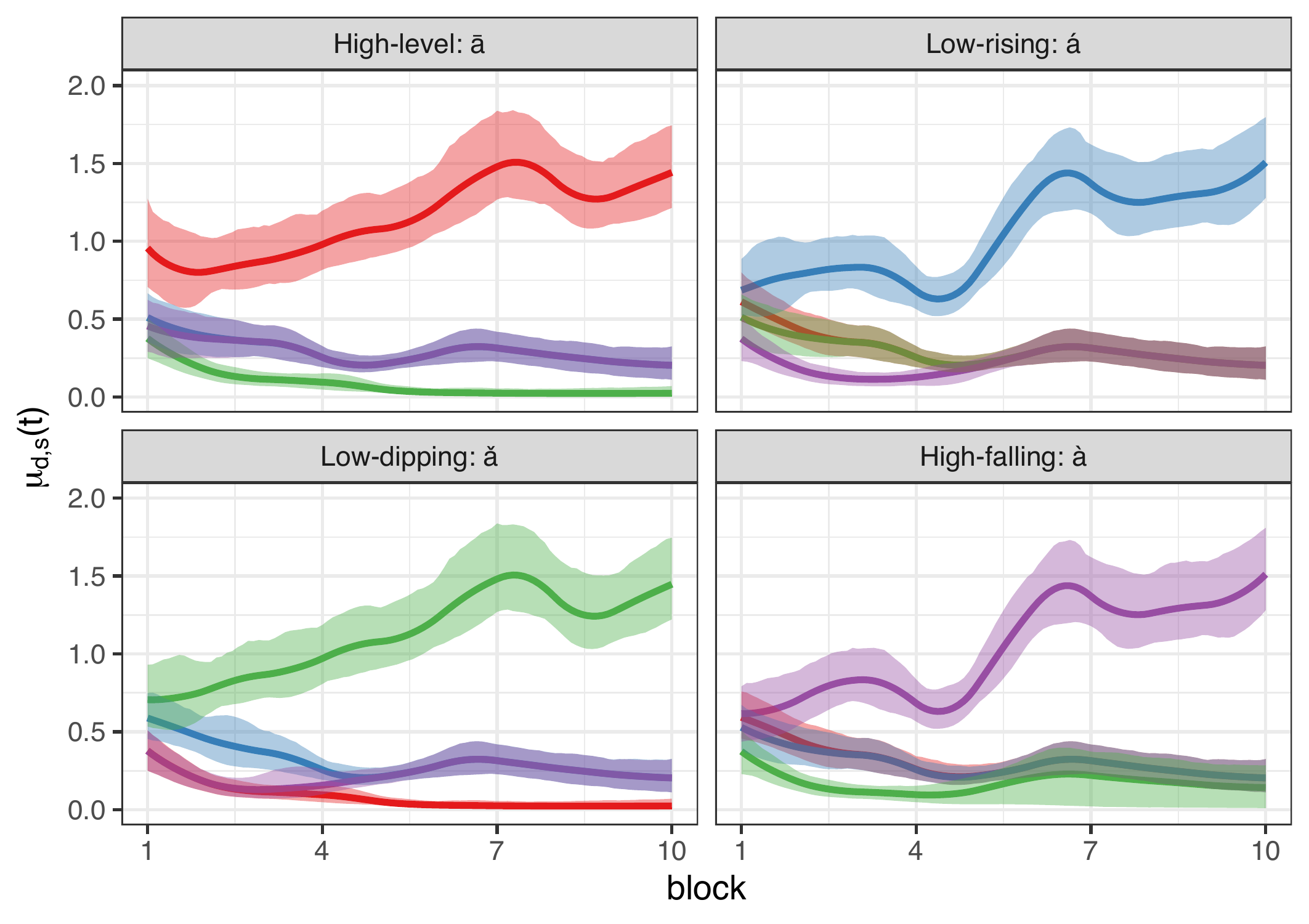} \hspace*{1cm}
	\includegraphics[width=6.5cm, trim=2cm 0.25cm 1cm 0.25cm]{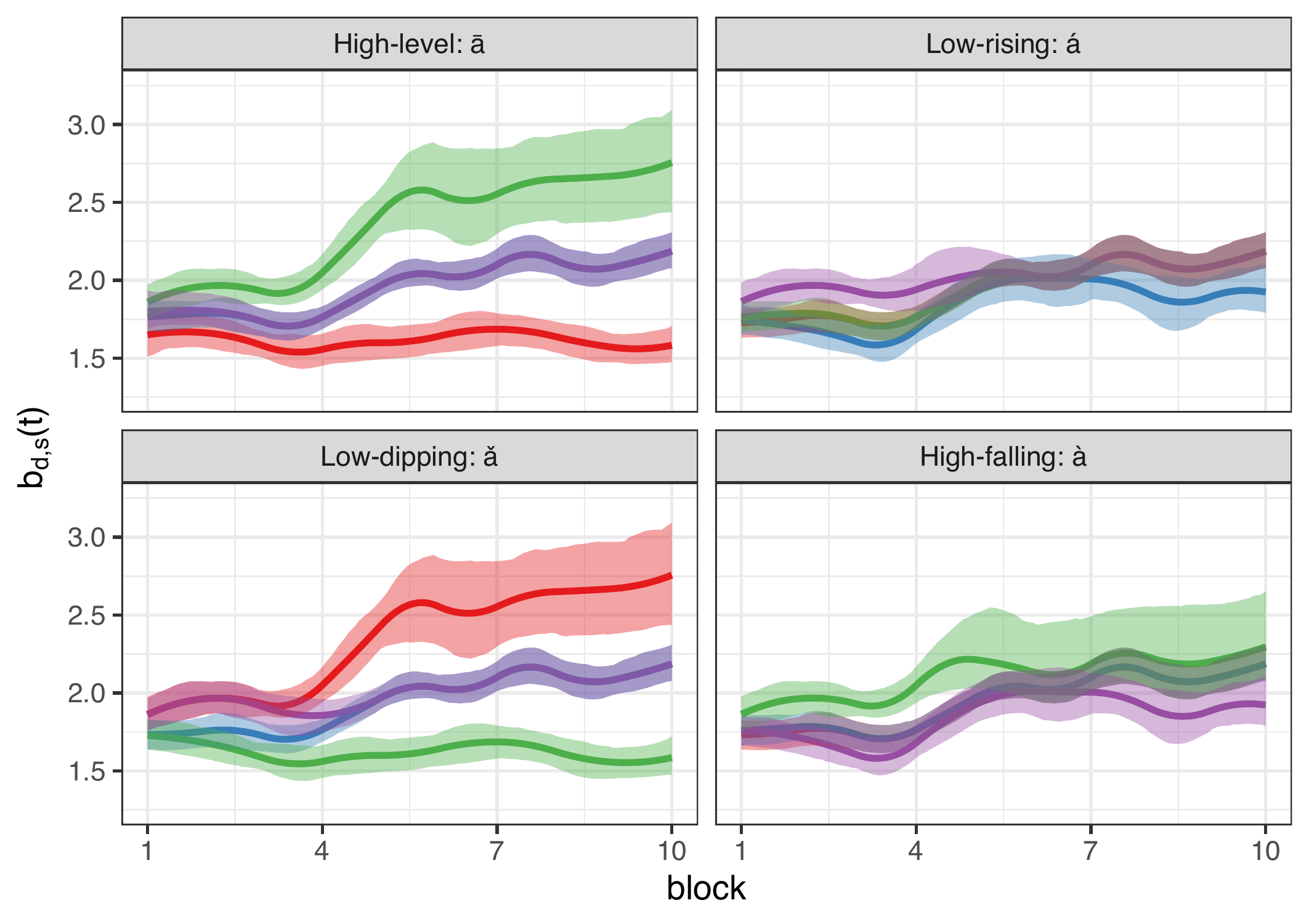}
	\caption{Results for tone learning data: Estimated posterior mean trajectories of the population level drifts $\mu_{d,s}(t)$ (left panel) 
	and boundaries $b_{d,s}(t)$ (right panel) for the proposed longitudinal inverse Gaussian drift-diffusion mixed model. 
	The shaded areas represent the corresponding $90\%$ point wise credible intervals.
	Parameters for the high-level tone response category T1 are shown in red; low-rising T2 in blue; low-dipping T3 in green; and high-falling T4 in purple.}
	\label{fig: population_effect}
\end{figure}

\begin{figure}[ht!]
	\centering
	\includegraphics[width=10cm, trim=0cm 1.25cm 2cm 0.25cm]{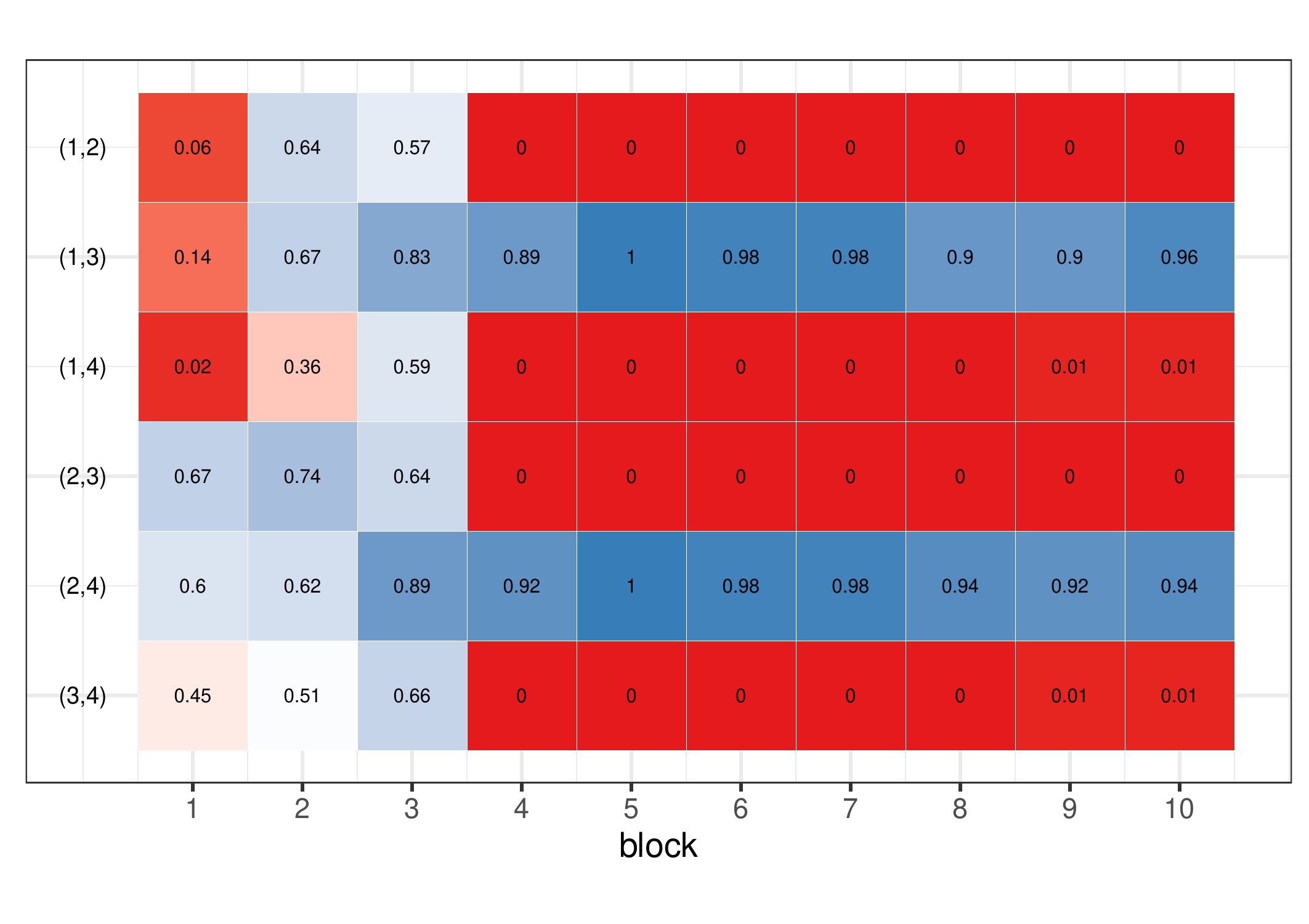} \hspace*{1cm}
	\caption{Results for tone learning data: Pairwise posterior co-clustering probabilities of the parameter trajectories for successful identification ($d=s$) of different input tones in different learning phases. 
	The estimated posterior probability of $(\mu_{2,2},b_{2,2})$ and $(\mu_{3,3},b_{3,3})$ being clustered together, and hence being equal, in the $3\th$ block is thus $0.74$, as shown in row $(2,3)$ and column $3$. 
	Equivalently, the estimated posterior probability of $(\mu_{2,2},b_{2,2})$ and $(\mu_{3,3},b_{3,3})$ being  different in the $3\th$ block is $0.26$.
	}
	\label{fig: coclust_prob}
\end{figure}

Figure \ref{fig: population_effect} shows the posterior mean trajectories and associated $90\%$ credible intervals for the boundaries $b_{d,s}(t)$ and the drift rates $\mu_{d,s}(t)$ estimated by our method 
for different combinations of $(d,s)$. 
Figure \ref{fig: coclust_prob} reports the estimated posterior probabilities of each of the ${4 \choose 2}=6$ pairs of success $(d=s)$ parameters to cluster together in different blocks. 
Figure \ref{fig: local_diff_drift_real} in the supplementary materials additionally presents the drift curves for successful identifications $(d=s)$ superimposed on each other.
These results suggest that after an initial learning phase, where the underlying processes are all similar across all input tones, 
there are two main learning groups. 
Two of the tones \{T1, T3\} seem to be easier to learn, 
as the corresponding drift parameters are larger, 
and tones \{T2, T4\} are more challenging. 
These findings are corroborated by empirical evidence and have significant biological relevance.
The similarity groups of the mandarin tones are in fact \{T1, T3\}, which are characterized by the height of the pitch, and tones \{T2, T4\}, which are characterized by the direction of the pitch and are more challenging to learn. 
Tone T3, in particular, 
has a unique 'dipping' pitch pattern that is rarely encountered in English \citep{song2008plasticity}, and therefore is easier to categorize.
Our proposed method allows similar inferential questions to be answered for the drift parameters corresponding to misclassifications, 
as well as for all the boundary parameters. 
The misclassification drift curves are mostly similar to each other, although some minor local differences can be found. 
Notable exceptions are $\mu_{1,3}(t)$ and $\mu_{3,1}(t)$ which are significantly smaller than all other drifts after the third block.
As the participants get trained and experienced, for input tone T1, evidence in favor of tone T3 is thus collected more slowly compared to evidence in favor of T2 and T4, and vice versa. 
Likewise, while the boundary curve estimates mostly remain constant over the training blocks and similar to each other, 
$b_{1,3}(t)$ and $b_{3,1}(t)$ again differ from the rest and actually increase over the blocks. 
As the participants get trained and experienced, more evidence in favor of tone T3 is thus needed to misclassify tone T1 as tone T3 and vice versa. 
These suggest that, as the participants get trained and experienced, tones T1 and T3 become harder to misclassify for one another.

\begin{figure}[ht!]
	\centering
	\includegraphics[width=6.5cm, trim=2cm 0.25cm 1cm 0.25cm]{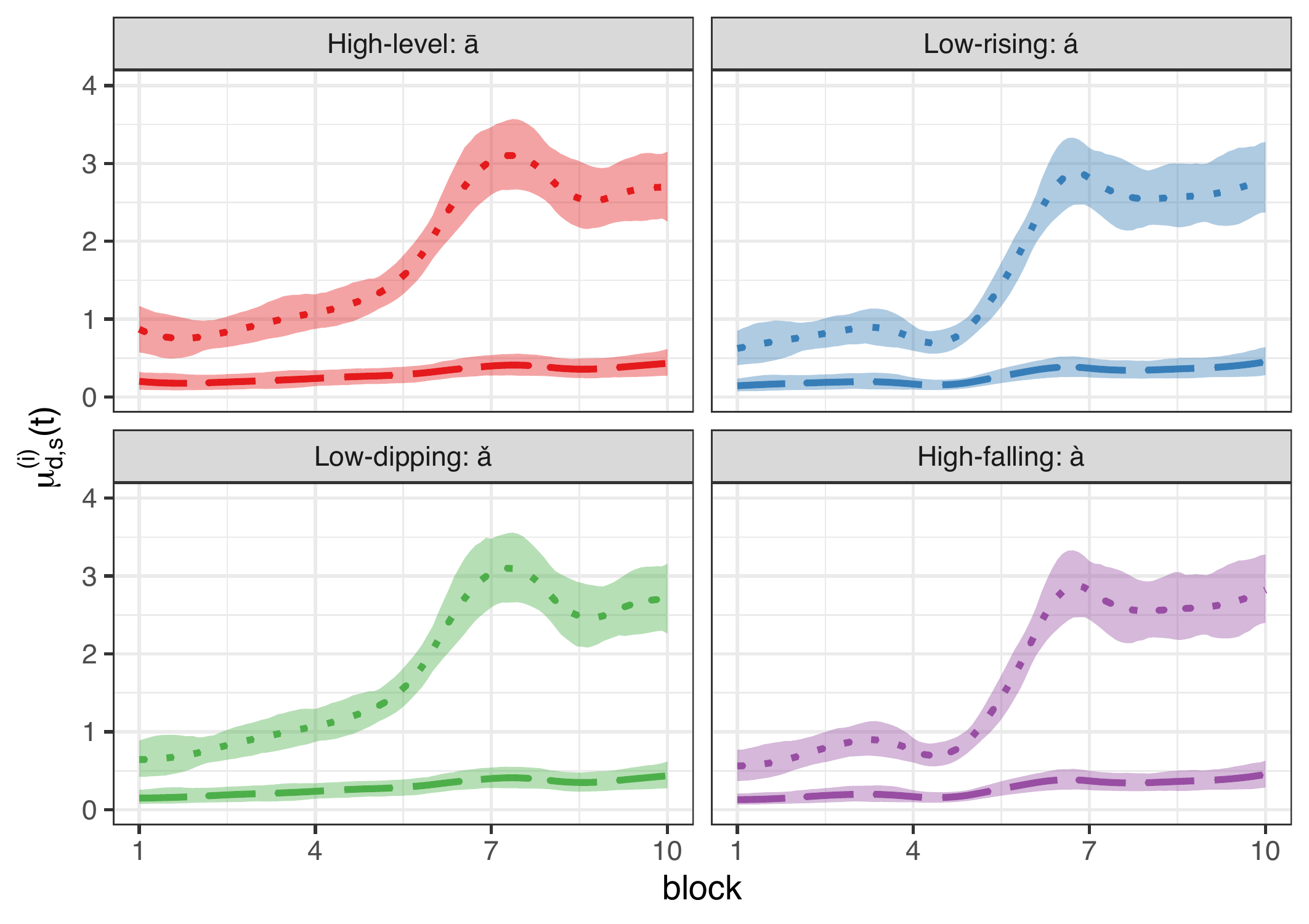} \hspace*{1cm}
	\includegraphics[width=6.5cm, trim=2cm 0.25cm 1cm 0.25cm]{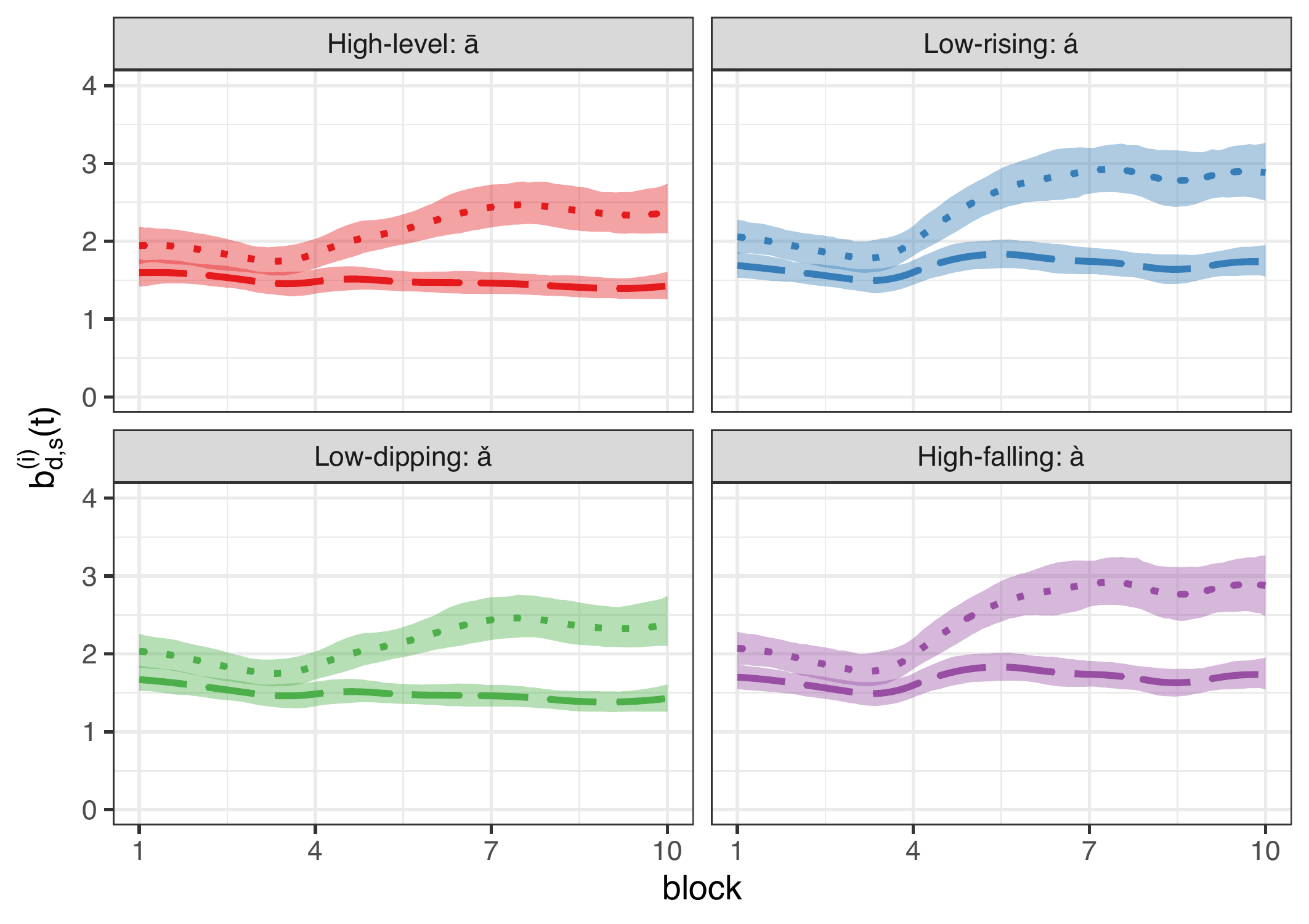}
	\caption{Results for tone learning data: Estimated posterior mean trajectories for individual specific drifts $\mu_{d,s}^{(i)}(t) = \exp \{ f_{\mu, d,s}(t) + u_{\mu,C}^{(i)}(t) \}$ (left panel)
	and boundaries $b_{d,s}^{(i)}(t) = \exp \{ f_{b, d,s}(t) + u_{b,C}^{(i)}(t) \}$ (right panel) for successful identification $(d=s)$ for two different participants 
	- one performing well (dotted line) and one performing poorly (dashed line). 
	The shaded areas represent the corresponding $90\%$ point wise credible intervals.
	Parameters for the high-level tone response category T1 are shown in red; low-rising T2 in blue; low-dipping T3 in green; and high-falling T4 in purple.}
	\label{fig: random_effect}
\end{figure}

Importantly, our proposed drift-diffusion mixed model not only allows population level inference about the underlying processes 
but also allows us to assess individual specific parameter trajectories.  
Figure \ref{fig: random_effect} shows the posterior mean trajectories and associated $90\%$ credible intervals 
for the drift rates $\mu_{s,d}^{(i)}$ and the boundaries $b_{s,d}^{(i)}$ estimated by our method for the different success combinations of $(d,s)$ for two participants 
- the one with the best accuracy averaged across all blocks, and the one with the worst accuracy averaged across all blocks.
These results suggest significant individual specific heterogeneity. 
Importantly, the differences in the performances can again be explained mostly by differences in the drift trajectories. 
For the well performing participant, the drift trajectories increase rapidly with the training blocks before plateauing down around block 6 at which stage the participant has already attained native-like proficiency. 
For the poorly performing candidate on the other hand, the drift trajectories remain approximately constant across all 10 blocks.

We compare the performance of our method with that of the linear ballistic accumulator (LBA) model \citep{brown2008simplest}. 
Similar to our model, the LBA uses independent evidence accumulators starting at $\delta$ that continue until a response threshold $b$ is reached. 
The accumulator that first reaches the boundary corresponds to the decision outcome, and the time taken to reach this decision boundary is the observed response time. 
The LBA model, however, assumes that 
the evidence accumulates linearly at the rate $\mu$, reaching the boundary $b$ precisely at time $\tau=b/\mu$. 
Unlike in drift-diffusion models, where trial-by-trial variability is explained by stochastically different diffusion paths, 
the LBA model explains trial-by-trial variability assuming the slopes $\mu$ for different trials to be drawn from a $\Normal(m_{d,s},v_{d,s})$ distribution. 
(Figure \ref{fig: comparison_drift_LBA} in the supplementary materials). 

The literature on LBA models has many serious limitations. 
The normality assumption on the slopes $\mu$ clearly does not satisfy any non-negativity constraints.  
Existing LBA models are also limited in their use of a common boundary $b_{s}$ for all decision categories $d$. 
There is also no principled way to incorporate systematic stimulus and decision category specific fixed or individual specific random effects into the LBA model. 
Existing literature is also limited to static settings, there is no mechanism to estimate smoothly varying longitudinal parameter trajectories as the participants get trained and experienced in their decision tasks. 
In our implementation, we thus fitted the LBA model separately for each block. 
Finally, the likelihood function of the LBA model is non-convex in the parameters. 
Parameter estimation based on optimization of the likelihood function is thus fraught with convergence issues. 
We used the \texttt{rtdists} package \citeplatex{rtdists} in \texttt{R}, using several random initializations and tracking the objective function to ensure convergence. 
A more detailed review of the LBA model can be found in Section \ref{sec: sm lba} of the supplementary materials.

\begin{figure}[ht!]
	\centering
	\includegraphics[width=6.5cm, trim=2cm 0.25cm 1cm 0.25cm]{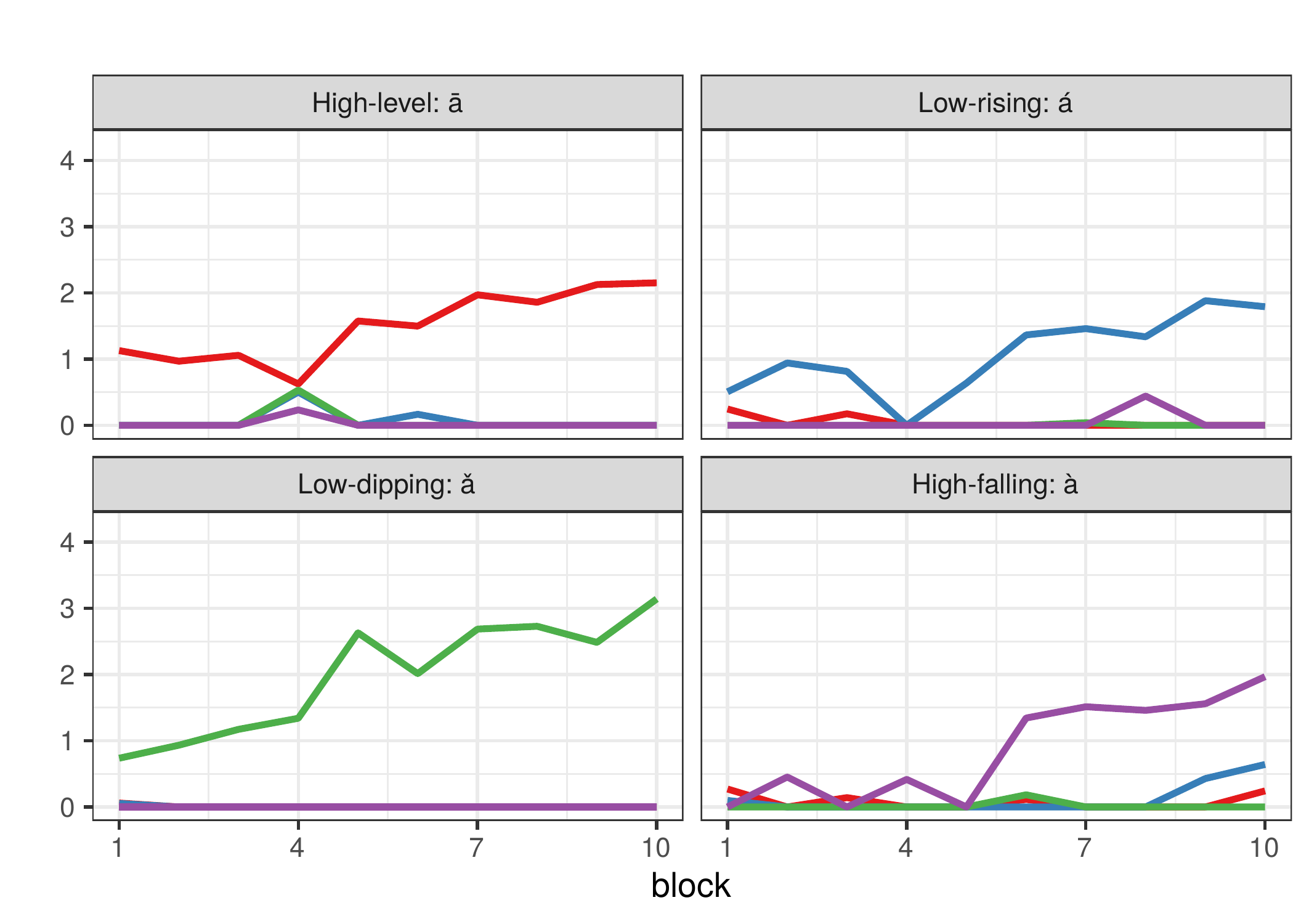} \hspace*{1cm}
	\includegraphics[width=6.5cm, trim=2cm 0.25cm 1cm 0.25cm]{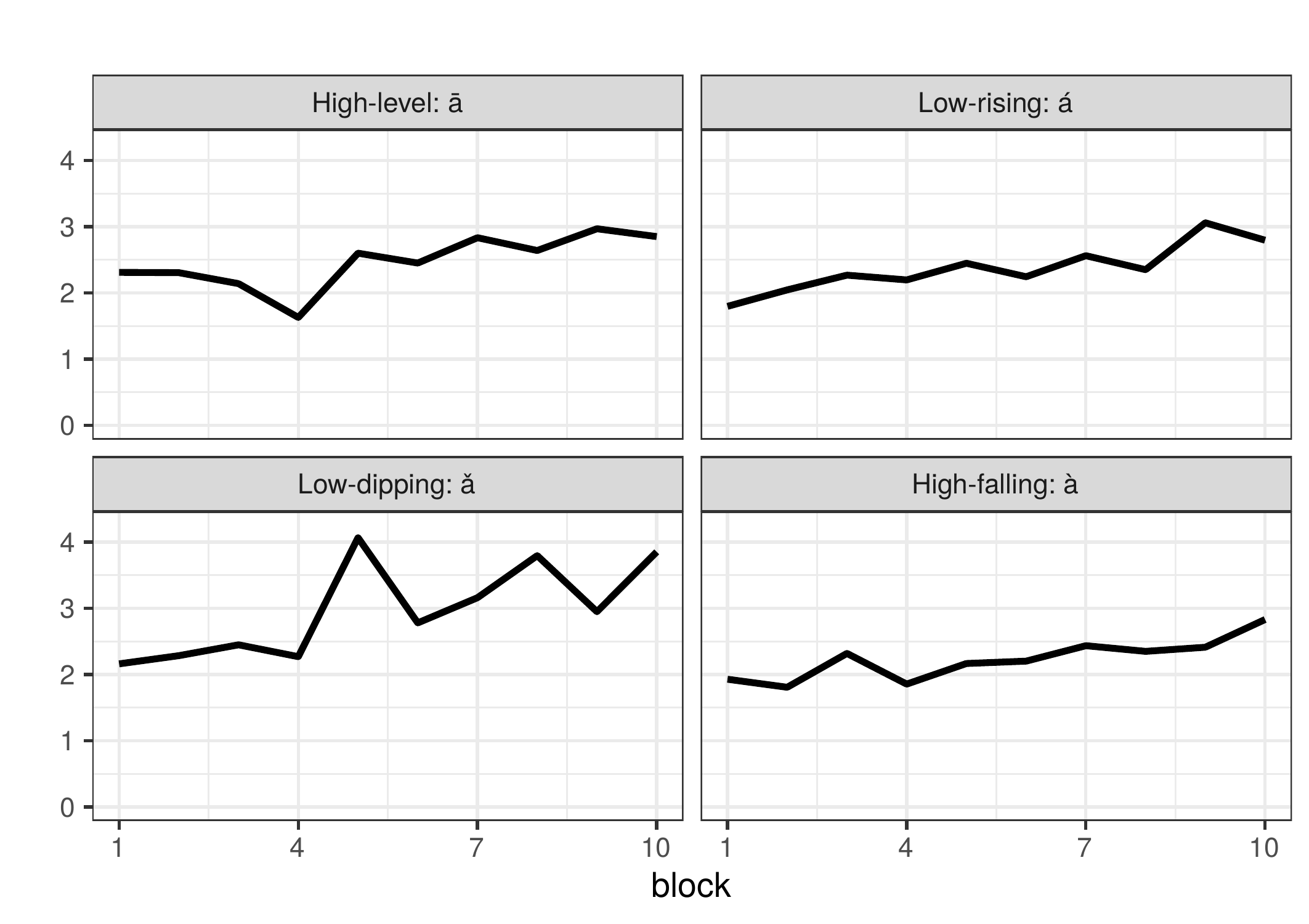}
	\caption{Results for tone learning data: 
	Left: Estimated mean slopes $m_{d,s,t}$ for the LBA model. 
	Right: Estimated boundaries $b_{s,t}$ for the LBA model. 
	In the left panel, $m_{d,s,t}$'s for the high-level tone response category T1 are shown in red; low-rising T2 in blue; low-dipping T3 in green; and high-falling T4 in purple.
}
	\label{fig: inference_LBA}
\end{figure}

Results produced by the LBA model applied to our motivating tone-learning data are reported in Figure \ref{fig: inference_LBA}. 
Owing to the limitations discussed above, 
the inference we make with such models is very limited. 
For instance, only non-smooth population level estimates are available, individual specific trajectories can not be assessed, etc. 
Some of our findings can, however, be confirmed by the LBA method. 
For example, looking at the drift parameter estimates, one can see that tone T3 is consistently associated with larger drifts.
As was also seen in the estimates returned by our method, tones \{T2, T4\} have similar values for the drift and the boundary parameters. 
Except such general overall findings, 
the LBA model, however, can not answer scientific questions related to the dynamics of category learning with fine detail. 

Our method, on the other hand, 
provides a biologically interpretable, statistically principled approach to accommodate fixed effects of input stimuli and decision categories as well as random subject specific heterogeneity, 
allows MCMC algorithm based efficient estimation of longitudinally smoothly evolving parameter trajectories, 
borrowing information across sample subgroups, participants as well as adjacent time stamps through many layers of hierarchy. 
Crucially, building on a novel local cluster inducing mechanism, 
our method also allows automated assessment of local similarities and differences in the parameter trajectories in very fine detail 
as the participants get trained and experienced in their decision tasks.

On the scientific side, the detailed insights obtained here point toward interesting and novel hypotheses about learning. 
For example, we demonstrate that a difference in drift rates, associated with the speed of sensory evidence accumulation, is critical in determining good vs poor learners. 
Evidence thresholds, on the other hand, remain relatively stable over training blocks as well as across participants. 
Recent studies have shown that the process of evidence accumulation can be selectively targeted by brain stimulation \citep{van2018stochastic}. 
Novel tone learning studies are currently being designed to test if such neurostimulation primarily improves the drift rates but not the evidence thresholds.

On the practical side, the insights obtained above can have important implications for developing advanced training regimens in language learning platforms used by millions of adults. 
Due to poor understanding of the temporal dynamics of learning, especially in multi-category learning problems, current training regimens are neither time adaptive nor individualized. 
Similar to personalized medicine, next-generation speech training paradigms seek to optimize and individualize training to reduce vast inter-individual differences in learning success \citep{wong2017personalized, birdsong2004second}. 
With our ability to assess detailed longitudinal confusion patterns, we can set up efficient training paradigms that can change the dynamics of learning in specific ways. 
For example, learners may generally benefit from introducing greater variability in pitch height that allows them to shift their focus on pitch direction and hence can reduce disparities in tone confusions like that between T2 and T4; 
poor learners may additionally benefit from `perceptual fading' - beginning with easy tones like \{T1,T3\} and making the training more challenging afterward with the introduction of tones like \{T2,T4\}; 
etc. 
As mentioned before, non-invasive and safe brain stimulation approaches like transcranial random noise stimulation and vagus nerve stimulation can be leveraged to selectively improve the process of sensory accumulation that could enhance the performance in poor learners. 

\section{Discussion} \label{sec: discussion}


\hspace*{7mm}{\bf Summary:}
In this article, we proposed a novel longitudinal drift-diffusion mixed model for perceptual decision making, 
allowing the underlying mechanisms to be similar or different at different longitudinal stages. 
Our research was motivated primarily by auditory neuroscience experiments where scientists are interested in understanding 
how the decision making mechanisms evolve 
as the participants get more training in the decision tasks. 
Our model was built on a novel statistical framework for longitudinal data 
that exploited local support properties of B-spline bases and (factorial) HMMs 
to allow automated assessment of local similarities and differences in the underlying parameter trajectories. 

Application to our motivating tone categorization experiments provided interesting novel insights into the underlying learning mechanisms. 
Notably, we discovered that the improvements and the local variations in tone categorization performance 
can be explained mostly by variations in the underlying drift parameters while the boundaries mostly remain constant. 
We also discovered local groupings among the underlying parameter curves in various phases of the learning experiments, 
how they differ between well and poorly performing participants etc. 
Such inferences were outside the scope of the previously existing literature.

{\bf Methodological extensions:}
Methodological extensions and topics of our ongoing research include 
adapting the proposed models to time constrained learning experiments, 
developing nested models to capture the dynamics within the blocks, 
accommodating sleep induced overnight `consolidation' effects, 
fully developing the inverse-probit model (\ref{eq: multinomial inv-probit}) for accuracies introduced in Section \ref{sec: lddmm}, 
etc. 

{\bf Broader scientific impact:} 
The proposed approach, we believe, takes the existing literature on drift-diffusion decision making models many significant steps forward, 
enabling neuroscientists to study the longitudinal behavior of biologically interpretable model parameters 
in much finer detail than what previous methods could achieve. 

As reported in Section \ref{sec: application}, the findings of our motivating speech learning experiment help formulate interesting novel scientific hypotheses about speech learning. 
The findings are also practically highly significant 
in providing 
exciting opportunities for developing time adaptive and individualized training regimens for language learning. 

Efficient estimation of group and individual level trajectories also open exciting avenues for potential adaptations in clinical settings, especially in conjunction with simultaneously performed imaging studies. 

Finally, 
the scope of proposed method is also not restricted to auditory neuroscience problems 
but the approach can be readily applied to study decision making mechanisms in other areas of neuroscience as well.

\baselineskip=14pt
\section*{Supplementary Materials}
Supplementary materials present substantive additional details. 
These include brief reviews of 
fHMMs, B-splines, locally informed Hamming ball samplers, the linear ballistic accumulator model, etc. to make the article relatively self-contained. 
The supplementary materials also discuss the choice of hyper-parameters for our model, the MCMC algorithm used to sample from the posterior of our model and its convergence diagnostics. 
The supplementary materials also present simulation studies and a comparison with a reduced model further illustrating the efficacy and the advantages of our proposed method.
In separate files, the supplementary materials additionally include the tone categorization data set described in Section \ref{sec: background} and analyzed in Section \ref{sec: application}, audio recordings of the four input Mandarin tones, 
and \texttt{R} programs implementing the longitudinal drift-diffusion mixed model developed in this article. 

\vspace{-3ex}
\section*{Acknowledgments}
We thank the editor, Dr. Heping Zhang, for comments leading to a significantly improved version of the initial manuscript. 
We also thank Dr. Peter Mueller, Dr. Mario Peruggia, Dr. Rachel Reetzke and Dr. Tobias Teichert for helpful discussions on the research presented here. 
This work was supported by 
the National Institute on Deafness and Other Communication Disorders grants R01DC013315 and R01DC015504 awarded to 
Chandrasekaran.

\baselineskip=14pt
\bibliographystyle{natbib}
\bibliography{Categorical,Diffusion,FDA_LDA,HMM,HOHMM,MCMC_Latent_Var_Models}

\begin{thebibliography}{}

\bibitem[Brown and Heathcote(2008)Brown and Heathcote]{brown2008simplest}
Brown, S.~D. and Heathcote, A. (2008).
\newblock The simplest complete model of choice response time: Linear ballistic
  accumulation.
\newblock {\em Cognitive Psychology\/}, {\bf 57}, 153--178.

\bibitem[de~Boor(1978)de~Boor]{de1978practical}
de~Boor, C. (1978).
\newblock {\em A practical guide to splines\/}.
\newblock Springer-Verlag.

\bibitem[Escobar and West(1995)Escobar and West]{escobar1995bayesian}
Escobar, M.~D. and West, M. (1995).
\newblock Bayesian density estimation and inference using mixtures.
\newblock {\em Journal of the American Statistical Association\/}, {\bf 90},
  577--588.

\bibitem[Fr{\"u}hwirth-Schnatter(2006)Fr{\"u}hwirth-Schnatter]{fruhwirth2006finite}
Fr{\"u}hwirth-Schnatter, S. (2006).
\newblock {\em Finite mixture and Markov switching models\/}.
\newblock Springer, New York.

\bibitem[Geweke(1991)Geweke]{geweke1991evaluating}
Geweke, J. (1991).
\newblock Evaluating the accuracy of sampling-based approaches to the
  calculation of posterior moments.
\newblock In {\em Proceedings of the Fourth Valencia International Conference
  on Bayesian Statistics\/}, pages 169--193.

\bibitem[Ghahramani and Jordan(1997)Ghahramani and
  Jordan]{ghahramani1996factorial}
Ghahramani, Z. and Jordan, M.~I. (1997).
\newblock Factorial hidden {M}arkov models.
\newblock {\em Machine Learning\/}, {\bf 29}, 245--273.

\bibitem[McDonald and Zucchini(1997)McDonald and
  Zucchini]{mcdonald_zuchhini:1997}
McDonald, S. and Zucchini, W. (1997).
\newblock {\em Hidden {M}arkov and other models for discrete-valued time
  series\/}.
\newblock Chapman \& Hall, London.

\bibitem[Neal(2003)Neal]{neal2003slice}
Neal, R.~M. (2003).
\newblock Slice sampling.
\newblock {\em The Annals of Statistics\/}, {\bf 31}, 705--767.

\bibitem[Rabiner(1989)Rabiner]{Rabiner:1989}
Rabiner, L. (1989).
\newblock A tutorial on hidden {M}arkov models and selected applications in
  speech recognition.
\newblock {\em \IEEE\/}, {\bf 77}, 257--286.

\bibitem[Roberts and Rosenthal(2009)Roberts and Rosenthal]{roberts2009examples}
Roberts, G.~O. and Rosenthal, J.~S. (2009).
\newblock Examples of adaptive {MCMC}.
\newblock {\em Journal of Computational and Graphical Statistics\/}, {\bf 18},
  349--367.

\bibitem[Roberts {\em et~al.}(2001)Roberts, Rosenthal, {\em
  et~al.}]{roberts2001optimal}
Roberts, G.~O., Rosenthal, J.~S., {\em et~al.} (2001).
\newblock Optimal scaling for various {M}etropolis-{H}astings algorithms.
\newblock {\em Statistical Science\/}, {\bf 16}, 351--367.

\bibitem[Scott(2002)Scott]{Scott:2002}
Scott, S.~L. (2002).
\newblock Bayesian methods for hidden {M}arkov models recursive computing in
  the 21st century.
\newblock {\em \JASA\/}, {\bf 97}, 337--351.

\bibitem[Singmann {\em et~al.}(2019)Singmann, Brown, Gretton, and
  Heathcote]{rtdists}
Singmann, H., Brown, S., Gretton, M., and Heathcote, A. (2019).
\newblock rtdists: Response time distributions.
\newblock R package version 0.10-0.

\bibitem[Titsias and Yau(2014)Titsias and Yau]{titsias2016hamming}
Titsias, M.~K. and Yau, C. (2014).
\newblock Hamming ball auxiliary sampling for factorial hidden {M}arkov models.
\newblock In {\em Advances in Neural Information Processing Systems\/}, pages
  2960--2968.

\bibitem[Zanella(2019)Zanella]{zanella2019informed}
Zanella, G. (2019).
\newblock Informed proposals for local {MCMC} in discrete spaces.
\newblock {\em Journal of the American Statistical Association\/}, pages 1--14.

\end{thebibliography}


\begin{thebibliography}{}

\bibitem[Agresti(2018)Agresti]{agresti2018introduction}
Agresti, A. (2018).
\newblock {\em An introduction to categorical data analysis\/}.
\newblock Wiley.

\bibitem[Birdsong(2004)Birdsong]{birdsong2004second}
Birdsong, D. (2004).
\newblock Second language acquisition and ultimate attainment.
\newblock {\em Handbook of Applied Linguistics\/}, pages 82--105.

\bibitem[Bogacz {\em et~al.}(2010)Bogacz, Wagenmakers, Forstmann, and
  Nieuwenhuis]{bogacz2010neural}
Bogacz, R., Wagenmakers, E.-J., Forstmann, B.~U., and Nieuwenhuis, S. (2010).
\newblock The neural basis of the speed-accuracy tradeoff.
\newblock {\em Trends in Neurosciences\/}, {\bf 33}, 10--16.

\bibitem[Borooah(2002)Borooah]{borooah2002logit}
Borooah, V.~K. (2002).
\newblock {\em Logit and probit: ordered and multinomial models\/}.
\newblock Sage.

\bibitem[Brody and Hanks(2016)Brody and Hanks]{brody2016neural}
Brody, C.~D. and Hanks, T.~D. (2016).
\newblock Neural underpinnings of the evidence accumulator.
\newblock {\em Current Opinion in Neurobiology\/}, {\bf 37}, 149--157.

\bibitem[Brown and Heathcote(2008)Brown and Heathcote]{brown2008simplest}
Brown, S.~D. and Heathcote, A. (2008).
\newblock The simplest complete model of choice response time: Linear ballistic
  accumulation.
\newblock {\em Cognitive Psychology\/}, {\bf 57}, 153--178.

\bibitem[Capp{\'e} {\em et~al.}(2005)Capp{\'e}, Moulines, and
  Ryd{\'e}n]{cappe2009inference}
Capp{\'e}, O., Moulines, E., and Ryd{\'e}n, T. (2005).
\newblock {\em Inference in hidden {M}arkov models\/}.
\newblock Springer Verlag, Berlin.

\bibitem[Cavanagh {\em et~al.}(2011)Cavanagh, Wiecki, Cohen, Figueroa, Samanta,
  Sherman, and Frank]{cavanagh2011subthalamic}
Cavanagh, J.~F., Wiecki, T.~V., Cohen, M.~X., Figueroa, C.~M., Samanta, J.,
  Sherman, S.~J., and Frank, M.~J. (2011).
\newblock Subthalamic nucleus stimulation reverses mediofrontal influence over
  decision threshold.
\newblock {\em Nature Neuroscience\/}, {\bf 14}, 1462--1467.

\bibitem[Chandrasekaran {\em et~al.}(2010)Chandrasekaran, Sampath, and
  Wong]{chandrasekaran2010individual}
Chandrasekaran, B., Sampath, P.~D., and Wong, P.~C. (2010).
\newblock Individual variability in cue-weighting and lexical tone learning.
\newblock {\em The Journal of the Acoustical Society of America\/}, {\bf 128},
  456--465.

\bibitem[Chandrasekaran {\em et~al.}(2014)Chandrasekaran, Yi, and
  Maddox]{chandrasekaran2014dual}
Chandrasekaran, B., Yi, H.-G., and Maddox, W.~T. (2014).
\newblock Dual-learning systems during speech category learning.
\newblock {\em Psychonomic Bulletin \& Review\/}, {\bf 21}, 488--495.

\bibitem[Chhikara(1988)Chhikara]{chhikara1988inverse}
Chhikara, R. (1988).
\newblock {\em The inverse Gaussian distribution: Theory, methodology, and
  applications\/}.
\newblock CRC Press.

\bibitem[Chib and Hamilton(2002)Chib and Hamilton]{chib2002semiparametric}
Chib, S. and Hamilton, B.~H. (2002).
\newblock Semiparametric {B}ayes analysis of longitudinal data treatment
  models.
\newblock {\em Journal of Econometrics\/}, {\bf 110}, 67--89.

\bibitem[Cox and Miller(1965)Cox and Miller]{cox1965theory}
Cox, D.~R. and Miller, H.~D. (1965).
\newblock {\em The theory of stochastic processes\/}.
\newblock CRC Press.

\bibitem[Craigmile {\em et~al.}(2010)Craigmile, Peruggia, and
  Van~Zandt]{craigmile2010hierarchical}
Craigmile, P.~F., Peruggia, M., and Van~Zandt, T. (2010).
\newblock Hierarchical {B}ayes models for response time data.
\newblock {\em Psychometrika\/}, {\bf 75}, 613--632.

\bibitem[Daniels and Pourahmadi(2002)Daniels and
  Pourahmadi]{daniels2002bayesian}
Daniels, M.~J. and Pourahmadi, M. (2002).
\newblock Bayesian analysis of covariance matrices and dynamic models for
  longitudinal data.
\newblock {\em Biometrika\/}, {\bf 89}, 553--566.

\bibitem[de~Boor(1978)de~Boor]{de1978practical}
de~Boor, C. (1978).
\newblock {\em A practical guide to splines\/}.
\newblock Springer-Verlag.

\bibitem[Diggle {\em et~al.}(2002)Diggle, Diggle, Heagerty, Heagerty, Liang,
  Zeger, {\em et~al.}]{diggle2002analysis}
Diggle, P., Diggle, P.~J., Heagerty, P., Heagerty, P.~J., Liang, K.-Y., Zeger,
  S., {\em et~al.} (2002).
\newblock {\em Analysis of longitudinal data\/}.
\newblock Oxford University Press.

\bibitem[Ding and Gold(2013)Ding and Gold]{ding2013basal}
Ding, L. and Gold, J.~I. (2013).
\newblock The basal ganglia's contributions to perceptual decision making.
\newblock {\em Neuron\/}, {\bf 79}, 640--649.

\bibitem[Dufau {\em et~al.}(2012)Dufau, Grainger, and Ziegler]{dufau2012say}
Dufau, S., Grainger, J., and Ziegler, J.~C. (2012).
\newblock How to say ``no'' to a nonword: A leaky competing accumulator model
  of lexical decision.
\newblock {\em Journal of Experimental Psychology: Learning, Memory, and
  Cognition\/}, {\bf 38}, 1117--1128.

\bibitem[Eilers and Marx(1996)Eilers and Marx]{eilers1996flexible}
Eilers, P.~H. and Marx, B.~D. (1996).
\newblock Flexible smoothing with {B}-splines and penalties.
\newblock {\em Statistical Science\/}, {\bf 11}, 89--102.

\bibitem[Feng {\em et~al.}(2019)Feng, Yi, and Chandrasekaran]{feng2018role}
Feng, G., Yi, H.~G., and Chandrasekaran, B. (2019).
\newblock The role of the human auditory corticostriatal network in speech
  learning.
\newblock {\em Cerebral Cortex\/}, {\bf 29}, 4077--4089.

\bibitem[Fitzmaurice {\em et~al.}(2008)Fitzmaurice, Davidian, Verbeke, and
  Molenberghs]{fitzmaurice2008longitudinal}
Fitzmaurice, G., Davidian, M., Verbeke, G., and Molenberghs, G. (2008).
\newblock {\em Longitudinal data analysis\/}.
\newblock CRC Press.

\bibitem[Fontanesi {\em et~al.}(2019)Fontanesi, Gluth, Spektor, and
  Rieskamp]{fontanesi2019reinforcement}
Fontanesi, L., Gluth, S., Spektor, M.~S., and Rieskamp, J. (2019).
\newblock A reinforcement learning diffusion decision model for value-based
  decisions.
\newblock {\em Psychonomic Bulletin \& Review\/}, {\bf 26}, 1099--1121.

\bibitem[Fr{\"u}hwirth-Schnatter(2006)Fr{\"u}hwirth-Schnatter]{fruhwirth2006finite}
Fr{\"u}hwirth-Schnatter, S. (2006).
\newblock {\em Finite mixture and Markov switching models\/}.
\newblock Springer, New York.

\bibitem[Gelman(2006)Gelman]{gelman2006prior}
Gelman, A. (2006).
\newblock Prior distributions for variance parameters in hierarchical models.
\newblock {\em Bayesian Analysis\/}, {\bf 1}, 515--534.

\bibitem[Ghahramani and Jordan(1997)Ghahramani and
  Jordan]{ghahramani1996factorial}
Ghahramani, Z. and Jordan, M.~I. (1997).
\newblock Factorial hidden {M}arkov models.
\newblock {\em Machine Learning\/}, {\bf 29}, 245--273.

\bibitem[Glimcher and Fehr(2013)Glimcher and Fehr]{glimcher2013neuroeconomics}
Glimcher, P.~W. and Fehr, E. (2013).
\newblock {\em Neuroeconomics: Decision making and the brain\/}.
\newblock Academic Press.

\bibitem[Gold and Shadlen(2007)Gold and Shadlen]{gold2007neural}
Gold, J.~I. and Shadlen, M.~N. (2007).
\newblock The neural basis of decision making.
\newblock {\em Annual Review of Neuroscience\/}, {\bf 30}, 535--574.

\bibitem[Guo(2002)Guo]{guo2002functional}
Guo, W. (2002).
\newblock Functional mixed effects models.
\newblock {\em Biometrics\/}, {\bf 58}, 121--128.

\bibitem[Heekeren {\em et~al.}(2004)Heekeren, Marrett, Bandettini, and
  Ungerleider]{heekeren2004general}
Heekeren, H.~R., Marrett, S., Bandettini, P.~A., and Ungerleider, L.~G. (2004).
\newblock A general mechanism for perceptual decision-making in the human
  brain.
\newblock {\em Nature\/}, {\bf 431}, 859--862.

\bibitem[Iverson {\em et~al.}(2003)Iverson, Kuhl, Akahane-Yamada, Diesch,
  Tohkura, Kettermann, and Siebert]{iverson2003perceptual}
Iverson, P., Kuhl, P.~K., Akahane-Yamada, R., Diesch, E., Tohkura, Y.,
  Kettermann, A., and Siebert, C. (2003).
\newblock A perceptual interference account of acquisition difficulties for
  non-native phonemes.
\newblock {\em Cognition\/}, {\bf 87}, 47--57.

\bibitem[Johnson and Newport(1989)Johnson and Newport]{johnson1989critical}
Johnson, J.~S. and Newport, E.~L. (1989).
\newblock Critical period effects in second language learning: The influence of
  maturational state on the acquisition of {E}nglish as a second language.
\newblock {\em Cognitive Psychology\/}, {\bf 21}, 60--99.

\bibitem[Kim {\em et~al.}(2017)Kim, Potter, Craigmile, Peruggia, and
  Van~Zandt]{kim2017bayesian}
Kim, S., Potter, K., Craigmile, P.~F., Peruggia, M., and Van~Zandt, T. (2017).
\newblock A {B}ayesian race model for recognition memory.
\newblock {\em Journal of the American Statistical Association\/}, {\bf 112},
  77--91.

\bibitem[Kunkel {\em et~al.}(2019)Kunkel, Potter, Craigmile, Peruggia, and
  Van~Zandt]{kunkel2019bayesian}
Kunkel, D., Potter, K., Craigmile, P.~F., Peruggia, M., and Van~Zandt, T.
  (2019).
\newblock A {B}ayesian race model for response times under cyclic stimulus
  discriminability.
\newblock {\em The Annals of Applied Statistics\/}, {\bf 13}, 271--296.

\bibitem[Leite and Ratcliff(2010)Leite and Ratcliff]{leite2010modeling}
Leite, F.~P. and Ratcliff, R. (2010).
\newblock Modeling reaction time and accuracy of multiple-alternative
  decisions.
\newblock {\em Attention, Perception, \& Psychophysics\/}, {\bf 72}, 246--273.

\bibitem[Li {\em et~al.}(2010)Li, Lin, and M{\"u}ller]{li2010bayesian}
Li, Y., Lin, X., and M{\"u}ller, P. (2010).
\newblock Bayesian inference in semiparametric mixed models for longitudinal
  data.
\newblock {\em Biometrics\/}, {\bf 66}, 70--78.

\bibitem[Lu(1995)Lu]{lu1995degradation}
Lu, J. (1995).
\newblock {\em Degradation processes and related reliability models\/}.
\newblock Ph.D. thesis, McGill University, Montreal, Canada.

\bibitem[Maddox and Chandrasekaran(2014)Maddox and
  Chandrasekaran]{maddox2014tests}
Maddox, W.~T. and Chandrasekaran, B. (2014).
\newblock Tests of a dual-system model of speech category learning.
\newblock {\em Bilingualism: Language and Cognition\/}, {\bf 17}, 709--728.

\bibitem[McDonald and Zucchini(1997)McDonald and
  Zucchini]{mcdonald_zuchhini:1997}
McDonald, S. and Zucchini, W. (1997).
\newblock {\em Hidden {M}arkov and other models for discrete-valued time
  series\/}.
\newblock Chapman \& Hall, London.

\bibitem[Milosavljevic {\em et~al.}(2010)Milosavljevic, Malmaud, Huth, Koch,
  and Rangel]{milosavljevic2010drift}
Milosavljevic, M., Malmaud, J., Huth, A., Koch, C., and Rangel, A. (2010).
\newblock The drift diffusion model can account for the accuracy and reaction
  time of value-based choices under high and low time pressure.
\newblock {\em Judgment and Decision Making\/}, {\bf 5}, 437--449.

\bibitem[Morris(2015)Morris]{morris2015functional}
Morris, J.~S. (2015).
\newblock Functional regression.
\newblock {\em Annual Review of Statistics and Its Application\/}, {\bf 2},
  321--359.

\bibitem[M{\"u}ller {\em et~al.}(2013)M{\"u}ller, Quintana, Rosner, and
  Maitland]{muller2013bayesian}
M{\"u}ller, P., Quintana, F.~A., Rosner, G.~L., and Maitland, M.~L. (2013).
\newblock Bayesian inference for longitudinal data with non-parametric
  treatment effects.
\newblock {\em Biostatistics\/}, {\bf 15}, 341--352.

\bibitem[Navarro and Fuss(2009)Navarro and Fuss]{navarro2009fast}
Navarro, D.~J. and Fuss, I.~G. (2009).
\newblock Fast and accurate calculations for first-passage times in {W}iener
  diffusion models.
\newblock {\em Journal of Mathematical Psychology\/}, {\bf 53}, 222--230.

\bibitem[Neal(2003)Neal]{neal2003slice}
Neal, R.~M. (2003).
\newblock Slice sampling.
\newblock {\em The Annals of Statistics\/}, {\bf 31}, 705--767.

\bibitem[Nguyen and Gelfand(2011)Nguyen and Gelfand]{nguyen2011dirichlet}
Nguyen, X. and Gelfand, A.~E. (2011).
\newblock The {D}irichlet labeling process for clustering functional data.
\newblock {\em Statistica Sinica\/}, {\bf 21}, 1249--1289.

\bibitem[Nguyen and Gelfand(2014)Nguyen and Gelfand]{nguyen2014bayesian}
Nguyen, X. and Gelfand, A.~E. (2014).
\newblock Bayesian nonparametric modeling for functional analysis of variance.
\newblock {\em Annals of the Institute of Statistical Mathematics\/}, {\bf 66},
  495--526.

\bibitem[Paulon {\em et~al.}(2019)Paulon, Reetzke, Chandrasekaran, and
  Sarkar]{paulon2019functional}
Paulon, G., Reetzke, R., Chandrasekaran, B., and Sarkar, A. (2019).
\newblock Functional logistic mixed-effects models for learning curves from
  longitudinal binary data.
\newblock {\em Journal of Speech, Language, and Hearing Research\/}, {\bf 62},
  543--553.

\bibitem[Pedersen {\em et~al.}(2017)Pedersen, Frank, and
  Biele]{pedersen2017drift}
Pedersen, M.~L., Frank, M.~J., and Biele, G. (2017).
\newblock The drift diffusion model as the choice rule in reinforcement
  learning.
\newblock {\em Psychonomic Bulletin \& Review\/}, {\bf 24}, 1234--1251.

\bibitem[Peters and D'Esposito(2020)Peters and D'Esposito]{peters2020drift}
Peters, J. and D'Esposito, M. (2020).
\newblock The drift diffusion model as the choice rule in inter-temporal and
  risky choice: {A} case study in medial orbitofrontal cortex lesion patients
  and controls.
\newblock {\em PLOS Computational Biology\/}, {\bf 16}.

\bibitem[Petrone {\em et~al.}(2009)Petrone, Guindani, and
  Gelfand]{petrone2009hybrid}
Petrone, S., Guindani, M., and Gelfand, A.~E. (2009).
\newblock Hybrid {D}irichlet mixture models for functional data.
\newblock {\em Journal of the Royal Statistical Society: Series B\/}, {\bf 71},
  755--782.

\bibitem[Polson and Scott(2012)Polson and Scott]{polson2012half}
Polson, N.~G. and Scott, J.~G. (2012).
\newblock On the half-{C}auchy prior for a global scale parameter.
\newblock {\em Bayesian Analysis\/}, {\bf 7}, 887--902.

\bibitem[Purcell(2013)Purcell]{purcell2013neural}
Purcell, B.~A. (2013).
\newblock {\em Neural mechanisms of perceptual decision making\/}.
\newblock Vanderbilt University.

\bibitem[Quintana {\em et~al.}(2016)Quintana, Johnson, Waetjen, and
  B.~Gold]{quintana2016bayesian}
Quintana, F.~A., Johnson, W.~O., Waetjen, L.~E., and B.~Gold, E. (2016).
\newblock Bayesian nonparametric longitudinal data analysis.
\newblock {\em Journal of the American Statistical Association\/}, {\bf 111},
  1168--1181.

\bibitem[Rabiner(1989)Rabiner]{Rabiner:1989}
Rabiner, L. (1989).
\newblock A tutorial on hidden {M}arkov models and selected applications in
  speech recognition.
\newblock {\em \IEEE\/}, {\bf 77}, 257--286.

\bibitem[Ramsay and Silverman(2007)Ramsay and Silverman]{ramsay2007applied}
Ramsay, J.~O. and Silverman, B.~W. (2007).
\newblock {\em Applied functional data analysis: methods and case studies\/}.
\newblock Springer.

\bibitem[Ratcliff(1978)Ratcliff]{ratcliff1978theory}
Ratcliff, R. (1978).
\newblock A theory of memory retrieval.
\newblock {\em Psychological Review\/}, {\bf 85}, 59--108.

\bibitem[Ratcliff and McKoon(2008)Ratcliff and McKoon]{ratcliff2008diffusion}
Ratcliff, R. and McKoon, G. (2008).
\newblock The diffusion decision model: Theory and data for two-choice decision
  tasks.
\newblock {\em Neural Computation\/}, {\bf 20}, 873--922.

\bibitem[Ratcliff and Rouder(1998)Ratcliff and Rouder]{ratcliff1998modeling}
Ratcliff, R. and Rouder, J.~N. (1998).
\newblock Modeling response times for two-choice decisions.
\newblock {\em Psychological Science\/}, {\bf 9}, 347--356.

\bibitem[Ratcliff {\em et~al.}(2016)Ratcliff, Smith, Brown, and
  McKoon]{ratcliff2016diffusion}
Ratcliff, R., Smith, P.~L., Brown, S.~D., and McKoon, G. (2016).
\newblock Diffusion decision model: Current issues and history.
\newblock {\em Trends in Cognitive Sciences\/}, {\bf 20}, 260--281.

\bibitem[Reetzke {\em et~al.}(2018)Reetzke, Xie, Llanos, and
  Chandrasekaran]{reetzke2018tracing}
Reetzke, R., Xie, Z., Llanos, F., and Chandrasekaran, B. (2018).
\newblock Tracing the trajectory of sensory plasticity across different stages
  of speech learning in adulthood.
\newblock {\em Current Biology\/}, {\bf 28}, 1419--1427.

\bibitem[Ross {\em et~al.}(1996)Ross, Kelly, Sullivan, Perry, Mercer, Davis,
  Washburn, Sager, Boyce, and Bristow]{ross1996stochastic}
Ross, S.~M., Kelly, J.~J., Sullivan, R.~J., Perry, W.~J., Mercer, D., Davis,
  R.~M., Washburn, T.~D., Sager, E.~V., Boyce, J.~B., and Bristow, V.~L.
  (1996).
\newblock {\em Stochastic processes\/}.
\newblock Wiley New York.

\bibitem[Schall(2001)Schall]{schall2001neural}
Schall, J.~D. (2001).
\newblock Neural basis of deciding, choosing and acting.
\newblock {\em Nature Reviews Neuroscience\/}, {\bf 2}, 33--42.

\bibitem[Scott(2002)Scott]{Scott:2002}
Scott, S.~L. (2002).
\newblock Bayesian methods for hidden {M}arkov models recursive computing in
  the 21st century.
\newblock {\em \JASA\/}, {\bf 97}, 337--351.

\bibitem[Singer {\em et~al.}(2003)Singer, Willett, Willett, {\em
  et~al.}]{singer2003applied}
Singer, J.~D., Willett, J.~B., Willett, J.~B., {\em et~al.} (2003).
\newblock {\em Applied longitudinal data analysis: Modeling change and event
  occurrence\/}.
\newblock Oxford university press.

\bibitem[Smith and Ratcliff(2004)Smith and Ratcliff]{smith2004psychology}
Smith, P.~L. and Ratcliff, R. (2004).
\newblock Psychology and neurobiology of simple decisions.
\newblock {\em Trends in Neurosciences\/}, {\bf 27}, 161--168.

\bibitem[Smith and Vickers(1988)Smith and Vickers]{smith1988accumulator}
Smith, P.~L. and Vickers, D. (1988).
\newblock The accumulator model of two-choice discrimination.
\newblock {\em Journal of Mathematical Psychology\/}, {\bf 32}, 135--168.

\bibitem[Song {\em et~al.}(2008)Song, Skoe, Wong, and
  Kraus]{song2008plasticity}
Song, J.~H., Skoe, E., Wong, P.~C., and Kraus, N. (2008).
\newblock Plasticity in the adult human auditory brainstem following short-term
  linguistic training.
\newblock {\em Journal of Cognitive Neuroscience\/}, {\bf 20}, 1892--1902.

\bibitem[Teichert {\em et~al.}(2016)Teichert, Grinband, and
  Ferrera]{teichert2016importance}
Teichert, T., Grinband, J., and Ferrera, V. (2016).
\newblock The importance of decision onset.
\newblock {\em Journal of Neurophysiology\/}, {\bf 115}, 643--661.

\bibitem[Titsias and Yau(2014)Titsias and Yau]{titsias2016hamming}
Titsias, M.~K. and Yau, C. (2014).
\newblock Hamming ball auxiliary sampling for factorial hidden {M}arkov models.
\newblock In {\em Advances in Neural Information Processing Systems\/}, pages
  2960--2968.

\bibitem[Tuerlinckx(2004)Tuerlinckx]{tuerlinckx2004efficient}
Tuerlinckx, F. (2004).
\newblock The efficient computation of the cumulative distribution and
  probability density functions in the diffusion model.
\newblock {\em Behavior Research Methods, Instruments, \& Computers\/}, {\bf
  36}, 702--716.

\bibitem[Tuerlinckx {\em et~al.}(2001)Tuerlinckx, Maris, Ratcliff, and
  De~Boeck]{tuerlinckx2001comparison}
Tuerlinckx, F., Maris, E., Ratcliff, R., and De~Boeck, P. (2001).
\newblock A comparison of four methods for simulating the diffusion process.
\newblock {\em Behavior Research Methods, Instruments, \& Computers\/}, {\bf
  33}, 443--456.

\bibitem[Usher and McClelland(2001)Usher and McClelland]{usher2001time}
Usher, M. and McClelland, J.~L. (2001).
\newblock The time course of perceptual choice: The leaky, competing
  accumulator model.
\newblock {\em Psychological Review\/}, {\bf 108}, 550--592.

\bibitem[Van~der Groen {\em et~al.}(2018)Van~der Groen, Tang, Wenderoth, and
  Mattingley]{van2018stochastic}
Van~der Groen, O., Tang, M.~F., Wenderoth, N., and Mattingley, J.~B. (2018).
\newblock Stochastic resonance enhances the rate of evidence accumulation
  during combined brain stimulation and perceptual decision-making.
\newblock {\em PLOS Computational Biology\/}, {\bf 14}, 1--17.

\bibitem[Van~Gael {\em et~al.}(2008)Van~Gael, Saatci, Teh, and
  Ghahramani]{van2008beam}
Van~Gael, J., Saatci, Y., Teh, Y.~W., and Ghahramani, Z. (2008).
\newblock Beam sampling for the infinite hidden {M}arkov model.
\newblock In {\em Proceedings of the 25th International Conference on Machine
  Learning\/}, pages 1088--1095. ACM.

\bibitem[Vandekerckhove and Tuerlinckx(2007)Vandekerckhove and
  Tuerlinckx]{vandekerckhove2007fitting}
Vandekerckhove, J. and Tuerlinckx, F. (2007).
\newblock Fitting the {R}atcliff diffusion model to experimental data.
\newblock {\em Psychonomic Bulletin \& Review\/}, {\bf 14}, 1011--1026.

\bibitem[Vandekerckhove {\em et~al.}(2008)Vandekerckhove, Tuerlinckx, and
  Lee]{vandekerckhove2008bayesian}
Vandekerckhove, J., Tuerlinckx, F., and Lee, M.~D. (2008).
\newblock A {B}ayesian approach to diffusion process models of decision-making.
\newblock In {\em Proceedings of the 30th Annual Conference of the Cognitive
  Science Society\/}, pages 1429--1434. Washington, DC.

\bibitem[Wang {\em et~al.}(2016)Wang, Chiou, and
  M{\"u}ller]{wang2016functional}
Wang, J.-L., Chiou, J.-M., and M{\"u}ller, H.-G. (2016).
\newblock Functional data analysis.
\newblock {\em Annual Review of Statistics and Its Application\/}, {\bf 3},
  257--295.

\bibitem[Wang {\em et~al.}(1999)Wang, Spence, Jongman, and
  Sereno]{wang1999training}
Wang, Y., Spence, M.~M., Jongman, A., and Sereno, J.~A. (1999).
\newblock Training {A}merican listeners to perceive {M}andarin tones.
\newblock {\em The Journal of the Acoustical Society of America\/}, {\bf 106},
  3649--3658.

\bibitem[Whitmore and Seshadri(1987)Whitmore and
  Seshadri]{whitmore1987heuristic}
Whitmore, G. and Seshadri, V. (1987).
\newblock A heuristic derivation of the inverse gaussian distribution.
\newblock {\em The American Statistician\/}, {\bf 41}, 280--281.

\bibitem[Wong {\em et~al.}(2017)Wong, Vuong, and Liu]{wong2017personalized}
Wong, P.~C., Vuong, L.~C., and Liu, K. (2017).
\newblock Personalized learning: {F}rom neurogenetics of behaviors to designing
  optimal language training.
\newblock {\em Neuropsychologia\/}, {\bf 98}, 192--200.

\bibitem[Xie {\em et~al.}(2017)Xie, Reetzke, and
  Chandrasekaran]{xie2017stability}
Xie, Z., Reetzke, R., and Chandrasekaran, B. (2017).
\newblock Stability and plasticity in neural encoding of linguistically
  relevant pitch patterns.
\newblock {\em Journal of Neurophysiology\/}, {\bf 117}, 1409--1424.

\bibitem[Zanella(2019)Zanella]{zanella2019informed}
Zanella, G. (2019).
\newblock Informed proposals for local {MCMC} in discrete spaces.
\newblock {\em Journal of the American Statistical Association\/}, pages 1--14.

\end{thebibliography}


\clearpage\pagebreak\newpage
\newgeometry{textheight=9in, textwidth=6.5in,}
\pagestyle{fancy}
\fancyhf{}
\rhead{\bfseries\thepage}
\lhead{\bfseries SUPPLEMENTARY MATERIALS}

\baselineskip 20pt
\begin{center}
{\LARGE{Supplementary Materials for\\} 
\bf Bayesian Semiparametric Longitudinal\\ 
\vskip -16pt Drift-Diffusion Mixed Models\\ 
for Tone Learning in Adults
}
\end{center}

\setcounter{equation}{0}
\setcounter{page}{1}
\setcounter{table}{1}
\setcounter{figure}{0}
\setcounter{section}{0}
\numberwithin{table}{section}
\renewcommand{\theequation}{S.\arabic{equation}}
\renewcommand{\thesubsection}{S.\arabic{section}.\arabic{subsection}}
\renewcommand{\thesection}{S.\arabic{section}}
\renewcommand{\thepage}{S.\arabic{page}}
\renewcommand{\thetable}{S.\arabic{table}}
\renewcommand{\thefigure}{S.\arabic{figure}}
\baselineskip=15pt

\vspace{0cm}

\begin{center}
Giorgio Paulon$^{1}$ (giorgio.paulon@utexas.edu)\\
Fernando Llanos$^{2,3}$ (f.llanos@pitt.edu)\\
Bharath Chandrasekaran$^{3}$(b.chandra@pitt.edu)\\
Abhra Sarkar$^{1}$ (abhra.sarkar@utexas.edu)\\

\vskip 7mm
$^{1}$Department of Statistics and Data Sciences, \\
University of Texas at Austin,\\ 2317 Speedway D9800, Austin, TX 78712-1823, USA\\
\vskip 8pt 
$^{2}$Department of Linguistics, \\
University of Texas at Austin,\\ 305 East 23rd Street  B5100, Austin, TX 78712, USA\\
\vskip 8pt 
$^{3}$Department of Communication Science and Disorders,\\ 
University of Pittsburgh,\\
4028 Forbes Tower, Pittsburgh, PA 15260, USA\end{center}

\vskip 10mm
Supplementary materials present 
brief reviews of B-splines, 
additional illustrations of our proposed smoothness inducing priors, 
brief reviews of fHMMs and associated computational machinery, 
details of the MCMC algorithm we designed to sample from the posterior, 
MCMC performance diagnostics, 
a review of linear ballistic accumulator models, 
comparisons with a simpler sub-model, 
results of simulation experiments, 
and some additional figures. 
Separate files additionally include the tone categorization data set described in Section \ref{sec: background} and analyzed in Section \ref{sec: application}, audio recordings of the four input Mandarin tones, 
and \texttt{R} programs implementing the longitudinal drift-diffusion mixed model developed in this article.

\newpage
\section{B-splines} \label{sec: b-splines}
In the main article, we employed quadratic B-spline bases in the construction of functional factorial HMMs.  
The construction of quadratic B-spline bases is detailed below \citeplatex{de1978practical}. 
Consider knot-points $t_{1} = t_{2} = t_{3} = A < t_{4} < \dots < B = t_{K+3} = t_{K+4} = t_{K+5}$,
where $t_{3:(K+3)}$ are equidistant with $\delta = (t_{4} - t_{3})$.
For $j=3,4,\dots,(K+2)$, quadratic B-splines $B_{j}$ are then defined as 
\vspace{-4ex}\\
\bse
 B_{j}(X) &= \left\{\begin{array}{ll}
        \{(X-t_{J-1})/\delta\}^{2}/2  				& ~~~~\text{if } t_{J-1} \leq X < t_{J},  \\
        -\{(X-t_{J})/\delta\}^{2} + (X-t_{J})/\delta + 1/2	     	& ~~~~\text{if } t_{J} \leq X < t_{j+2},  \\
        \{1-(X-t_{j+2})/\delta\}^{2}		       	    	& ~~~~\text{if } t_{j+2} \leq X < t_{j+3},  \\
        0  							& ~~~~ \text{otherwise}.
        \end{array}\right.
\ese
The components at the ends are likewise defined as 
\bse
B_{1}(X) &=&  \left\{\begin{array}{ll}
        \{1-(X-t_{1})/\delta\}^{2} /2           	& ~~~~~~~~~~~~~~~~~~~~~~~\text{if } t_{3} \leq X < t_{4},  \\
        0  						& ~~~~~~~~~~~~~~~~~~~~~~~ \text{otherwise}.
        \end{array}\right.\\
B_{2}(X) &=&  \left\{\begin{array}{ll}
        -\{(X-t_{3})/\delta\}^2 + (X-t_{4})/\delta + 1/2     	& ~~~~\text{if } t_{3} \leq X < t_{4},  \\
        \{1-(X-t_{4})/\delta\}^{2} /2            			& ~~~~\text{if } t_{4} \leq X < t_{5},  \\
        0  							& ~~~~ \text{otherwise}.
        \end{array}\right.\\
B_{K+1}(X) &=&  \left\{\begin{array}{ll}
        \{(X-t_{K+1})/\delta\}^{2} /2  					& ~~~~\text{if } t_{K+1} \leq X < t_{K+2},  \\
        -\{(X-t_{K+2})/\delta\}^{2} + (X-t_{K+2})/\delta + 1/2     	& ~~~~\text{if } t_{K+2} \leq X < t_{K+3},  \\
        0  								& ~~~~ \text{otherwise}.
        \end{array}\right.\\
B_{K+2}(X) &=&  \left\{\begin{array}{ll}
        \{(X-t_{K+2})/\delta\}^{2} /2  		& ~~~~~~~~~~~~~~~~~~~~~~~~~\text{if } t_{K+2} \leq X < t_{K+3},  \\
        0  					& ~~~~~~~~~~~~~~~~~~~~~~~~~ \text{otherwise}.
        \end{array}\right.
\ese
Figure \ref{fig: b-splines} in the main paper provides a graphical illustration of these functions.

\section{Illustration of the Smoothness Inducing Prior} \label{sec: prior_details}

\begin{figure}[ht!]
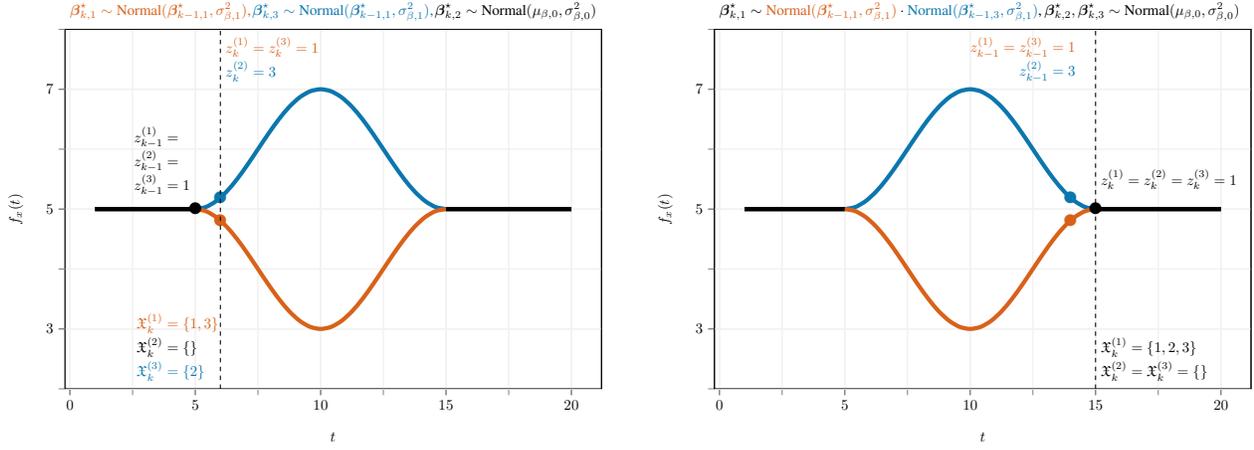

	\centering
	\hspace*{-0.00cm}\includegraphics[width=0.47\linewidth, trim=2cm 1.25cm 1cm 1.25cm]{./Figures/plot_prior_HMM} \hspace*{1cm}
	\hspace*{-0.40cm}\includegraphics[width=0.47\linewidth, trim=2cm 1.25cm 1cm 1.25cm]{./Figures/plot_prior_2_HMM}
	\caption{\baselineskip=10pt 
	An illustration of the prior on the spline core coefficients $\beta_{k,z_{k}}^{\star}$ at location $k$ (marked by the dashed vertical lines) in 
	the fixed effects model developed in Section \ref{sec: fixed effects} 
	for a scenario with $x \in \{1,2,3\}$, 
	where the curves corresponding to the three levels of $x$ are initially equal, the curves for $x=1,3$ (in red) and $x=2$ (in blue) then diverge at $t=6$, merging back again at $t=15$. 
	}
	\label{fig: prior core beta 2}
\end{figure}

Our proposed model penalizes the difference between the pairs of core coefficients in the same latent state.
Figure \ref{fig: prior core beta} in the main paper, reproduced here as Figure \ref{fig: prior core beta 2} for easy access, shows the effect of the smoothing prior on the core coefficients in a synthetic scenario with $x \in \{1,2,3\}$.

In the example in the left panel, at location $k-1 = 5$, all of the levels for the covariate $x$ are assigned to the first latent state, yielding the same curve for the three levels of $x$.
At location $k = 6$, levels $1$ and $3$ are assigned to the first latent state, whereas level $2$ is assigned to the third latent state. 
This corresponds to the case in which the curves for $x=1,3$ and $x=2$ diverge.
Therefore, using \eqref{eq: prior on beta star},
\begin{itemize}
	\item $\mathfrak{X}_{k}^{(1)} = \{x : z_{k}^{(x)} = 1\} = \{1, 3\}$ and the conditional prior for the core coefficient of the first latent state is 
$\beta_{k,1}^{\star} \sim \prod_{j \in \{z_{k-1}^{(1)}, z_{k-1}^{(3)}\}} \Normal(\beta_{k-1,j}^{\star}, \sigma_{\beta,1}^2) = \Normal(\beta_{k-1,1}^{\star}, \sigma_{\beta,1}^2)$,
	\item $\mathfrak{X}_{k}^{(2)} = \{x : z_{k}^{(x)} = 2\} =\emptyset$ and the conditional prior for the core coefficient of the second latent state is 
$\beta_{k,2}^{\star} \sim \Normal(\mu_{\beta,0}, \sigma_{\beta,0}^{2})$,
	\item $\mathfrak{X}_{k}^{(3)} = \{x : z_{k}^{(x)} = 3\} = \{2\}$ and the conditional prior for the core coefficient of the third latent state is 
$\beta_{k,3}^{\star} \sim \prod_{j \in \{z_{k-1}^{(2)}\}} \Normal(\beta_{k-1,j}^{\star}, \sigma_{\beta,1}^2) = \Normal(\beta_{k-1,1}^{\star}, \sigma_{\beta,1}^2)$.
\end{itemize}

In the example in the right panel, at location $k-1 = 14$, levels $1$ and $3$ are assigned to the first latent state, whereas level $2$ is assigned to the third latent state. 
At location $k = 15$, all of the levels for the covariate $x$ are assigned to the first latent state. This corresponds to the case in which the curves for $x=1,3$ and $x=2$ merge back. 
Therefore,
\begin{itemize}
	\item $\mathfrak{X}_{k}^{(1)} = \{x : z_{k}^{(x)} = 1\} = \{1, 2, 3\}$ and the conditional prior for the core coefficient of the first latent state is 
$\beta_{k,1}^{\star} \sim \prod_{j \in \{z_{k-1}^{(1)}, z_{k-1}^{(2)}, z_{k-1}^{(3)}\}} \Normal(\beta_{k-1,j}^{\star}, \sigma_{\beta,1}^2) = \prod_{j \in \{1, 3\}} \Normal(\beta_{k-1,j}^{\star}, \sigma_{\beta,1}^2)$,
	\item $\mathfrak{X}_{k}^{(2)} = \{x : z_{k}^{(x)} = 2\} =\emptyset$ and the conditional prior for the core coefficient of the second latent state is 
$\beta_{k,2}^{\star} \sim \Normal(\mu_{\beta,0}, \sigma_{\beta,0}^{2})$,
	\item $\mathfrak{X}_{k}^{(3)} = \{x : z_{k}^{(x)} = 3\} =\emptyset$ and the conditional prior for the core coefficient of the third latent state is 
$\beta_{k,3}^{\star} \sim \Normal(\mu_{\beta,0}, \sigma_{\beta,0}^{2})$.
\end{itemize}

\newpage
\section{Factorial HMM (fHMM)} \label{sec: fHMM}

The basic HMM \citeplatex[][etc.]{fruhwirth2006finite,mcdonald_zuchhini:1997}
consists of two processes: an \emph{observed} process $\{\by_{t}\}$ recorded sequentially over a set of discrete time points $t=1,2,\dots,T$
and an associated \emph{hidden} process $\{z_{t}\}$ which evolves according to a first order Markov chain with discrete state space.
Specifically, an HMM makes the following set of conditional independence assumptions to model the hidden and the observed processes
\vskip-5ex
\bse
& p(z_{t} \mid \bz_{1:(t-1)})  = p(z_{t}\mid z_{t-1}),	\\
& p(\by_{t} \mid \by_{1:(t-1)},\bz_{1:t}) = p(y_{t} \mid z_{t}).
\ese
\vskip-1.5ex
The distributions $p(z_{t} \mid z_{t-1})$ and $p(y_{t} \mid z_{t})$ are often referred to as the \emph{transition distribution} and the \emph{emission distribution}, respectively. 

In factorial HMMs \citeplatex{ghahramani1996factorial}, the latent states are represented by a collection of variables $\{\bz_{t}\} = \{(z_{t}^{(1)},\dots,z_{t}^{(L)})\}$ where each component $\{z_{t}^{(\ell)}\}$ now evolves according to a first order Markov chain with discrete state spaces, 
and the \emph{observed} process $\{y_{t}\}$ is observed sequentially as before over a set of discrete time points $t=1,2,\dots,T$.
An fHMM thus makes the following set of conditional independence assumptions to model the hidden and the observed processes
\vskip-5ex
\bse
& \textstyle p(\bz_{t} \mid \bz_{1:(t-1)})  = \prod_{\ell=1}^{L} p(z_{t}^{(\ell)} \mid z_{t-1}^{(\ell)}),	\\
& p(y_{t} \mid \by_{1:(t-1)},\bz_{1:t}) = p(\by_{t} \mid \bz_{t}) = p(\by_{t} \mid z_{t}^{(1)},\dots,z_{t}^{(L)}).
\ese

\vspace*{-0.5cm}
\begin{figure}[ht!]
	\centering
	\hspace*{-0.00cm}\includegraphics[width=6.75cm, trim=2cm 1.25cm 1cm 1.25cm]{./Figures/Graphical_Model_HMM_1} 
	\hspace*{-0.40cm}\includegraphics[width=8.75cm, trim=1cm 1.25cm 1cm 1.25cm]{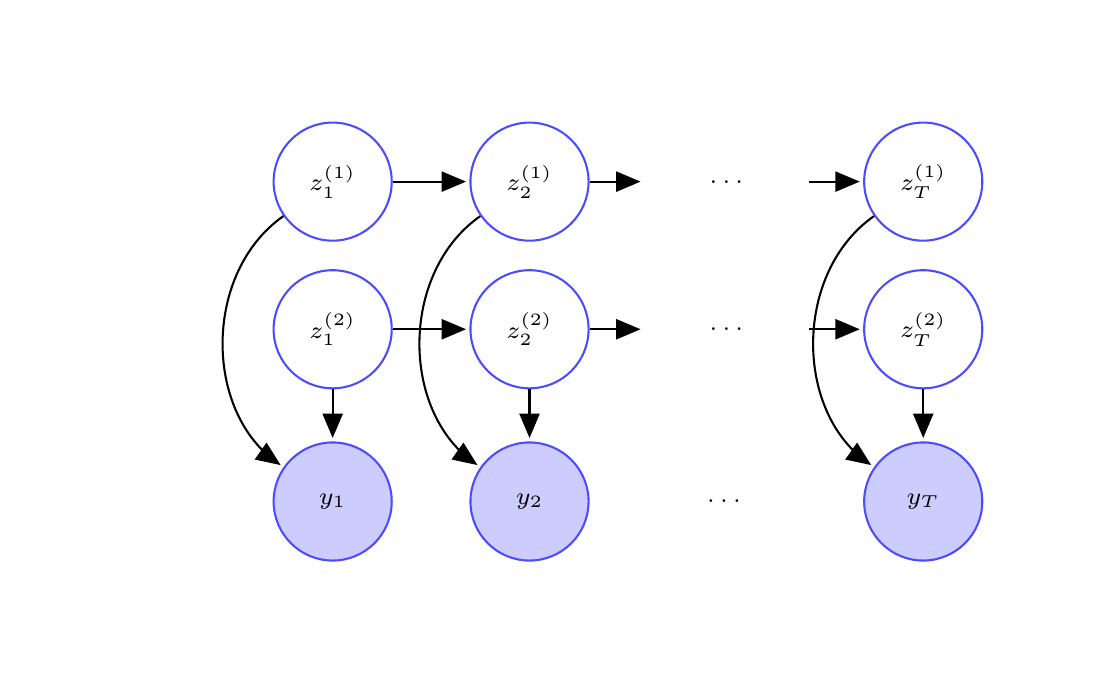}
	\caption{\baselineskip=10pt 
	Left panel: Graph of an HMM. 
	Right panel: Graph of an fHMM with two layers. 
	}
	\label{fig: dag HMM and fHMM}
\end{figure}


In our work, we adapted the basic fHMM to characterize local influences of the categorical predictor in longitudinal functional models.  
In the drift-diffusion model of Section \ref{sec: lddmm}, for each input tone $s \in \{1,\dots,d_{1}\}$, 
we introduced an fHMM $\{\bz_{k}^{(s)}=(z_{k}^{(1,s)},\dots,z_{k}^{(d_{0},s)})\}$ with $d_{0}$ layers, one for each level of the response $d$. 
Conditional on  $z_{k}^{(d,s)} = z_{k}$, 
we then associated the coefficients $\beta_{k,d,s}$ of a predictor dependent B-spline mixture model with atoms $\beta_{k,z_{k}}^{\star}$. 
Specifically, we let
\vskip-5ex
\bse
& \textstyle p(\bz_{k}^{(s)} \mid \bz_{1:(k-1)}^{(s)}) = \prod_{d=1}^{d_{0}} p(z_{k}^{(d,s)} \mid z_{k-1}^{(d,s)}),	\\
& \{\beta_{k,d,s} \mid z_{k}^{(d,s)}=z_{k}\} = \beta_{k,z_{k}}^{\star}.
\ese

\section{Locally Informed Hamming Ball Sampler} \label{sec: hamming ball}
Forward-backward (or backward-forward) algorithms for HMMs rely on passing messages forward (or backward) and then sampling backward (or forward) \citeplatex{Rabiner:1989,Scott:2002}. 
While adapting such algorithms to fHMMs, the requirement to sum over all possible configurations in computing the messages becomes a challenge. 
Hamming ball samplers for fHMMs \citeplatex{titsias2016hamming} avoid this computationally expensive step by introducing and conditioning on an auxiliary variable 
that restricts the sampling to only a slice \citeplatex{neal2003slice} of the entire high-dimensional space. 
In doing so, the sampler also allows localized joint updating of all constituent chains, making it less prone to get trapped in local modes.   

Let $h(\bz_{t},\bv_{t}) = \sum_{\ell=1}^{L}1\{z_{t}^{\ell} \neq v_{t}^{\ell}\}$ denote the Hamming distance between the vectors $\bz_{t}=(z_{t}^{(1)},\dots,z_{t}^{(L)})\trans$ and $\bv_{t}=(v_{t}^{(1)},\dots,v_{t}^{(L)})\trans$
and $\H_{m}(\bz_{t}) = \left\{\bv_{t}: h(\bz_{t},\bv_{t}) \leq m \right\}$ denote a Hamming ball of radius $m$ around $\bz_{t}$. 

Consider an fHHM, as shown in Figure \ref{fig: dag HMM and fHMM} but with $L$ component chains each with state space $\{1,\dots,d\}$. 
Introducing an auxiliary variable $\bv$ following a conditional probability distribution $p(\bv \mid \bz) = \prod_{t=1}^{T} p(\bv_{t} \mid \bz_{t})$, 
the augmented joint model becomes $p(\by,\bz,\bv) = p(\bv \mid \bz) p(\by \mid \bz) p(\bz) = \big\{\prod_{t=1}^{T} p(\bv_{t} \mid \bz_{t}) p(\by_{t} \mid \bz_{t}) \big\} p(\bz_{1}) \prod_{t=2}^{T}p(\bz_{t} \mid \bz_{t-1})$. 
Sampling $\bv$ from the posterior can then be done by sampling independently from the full conditionals 
$p(\bv_{t} \mid \bz_{t})$. 
Sampling $\bz$ from the posterior can still be carried out using forward-backward (or backward-forward) message passing algorithms but with the augmented full conditional 
$p(\bz \mid \by,\bv) \propto \big\{\prod_{t=1}^{T}p(\bv_{t} \mid \bz_{t}) p(\by_{t} \mid \bz_{t}) \big\} \big\{\prod_{t=2}^{T}p(\bz_{t} \mid \bz_{t-1}) \big\} p(\bz_{1})$.  
The set of possible configurations needed to compute the messages at time $t$ is now restricted to the support of $p(\bv_{t} \mid \bz_{t})$. 
If this can be made much smaller compared to the original size of the state space, computational burden can be greatly reduced.

The Hamming ball algorithm does this by setting $p(\bv_{t} \mid \bz_{t}) \propto 1\{\bv_{t} \in \H_{m}(\bz_{t})\}$, that is, by sampling the $\bv_{t}$'s uniformly from $\H_{m}(\bz_{t})$. 
By symmetry, since $\bv_{t} \in \H_{m}(\bz_{t})$ if and only if $\bz_{t} \in \H_{m}(\bv_{t})$, the support of each $\bz_{t}$ in the full conditional $p(\bz \mid \by,\bv)$ is then restricted only to $\H_{m}(\bv_{t})$. 

\begin{figure}[ht!]
	\centering
	\includegraphics[scale=0.85, trim=2cm 1cm 1cm 1cm]{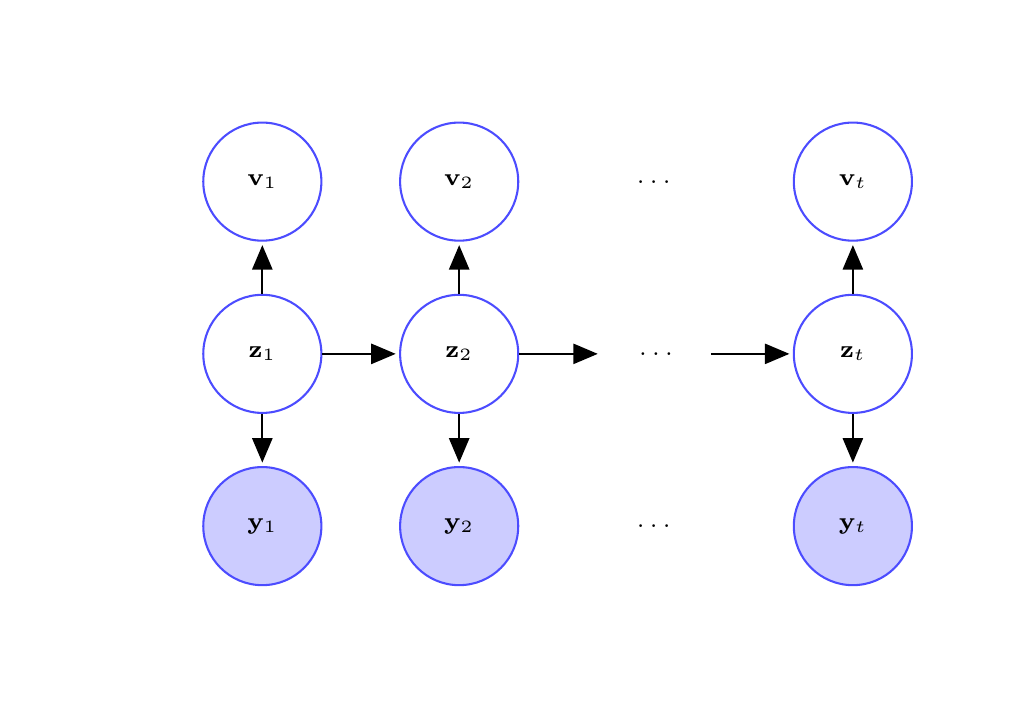} \quad\quad\quad\quad
	\includegraphics[scale=0.85, trim=2cm 1cm 1cm 1cm]{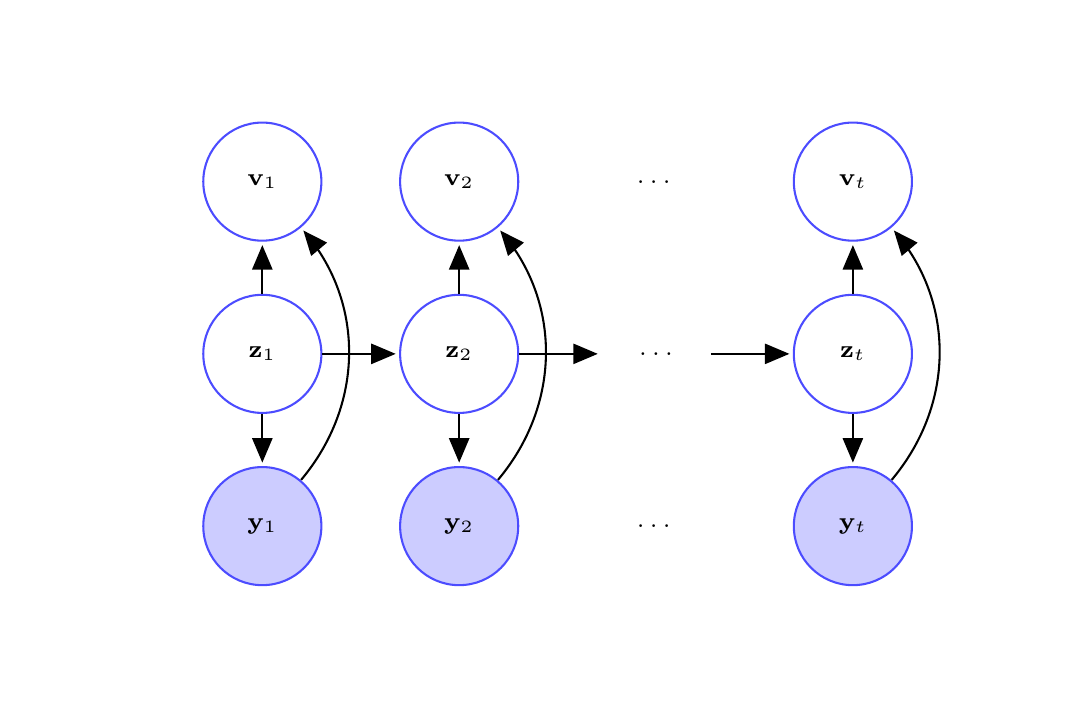}
	\caption{\baselineskip=10pt 
	Graph of a Hamming ball sampler (left panel) and a locally informed Hamming ball sampler (right panel) for fHMM. 
	}
	\label{fig: dag HMM and loc fHMM}
\end{figure}

The Hamming ball sampler is still limited in its ability to efficiently explore the neighborhood of $\bz_{t}$ as it blindly proposes new values along arbitrarily chosen directions within the ball. 
More informed moves can be proposed utilizing the information contained in the likelihood function \citeplatex{zanella2019informed}. 
For instance, $p(\bv_{t} \mid \bz_{t}, \by_{t})  \propto  g\{p(\by_{t} \mid \bv_{t})\} 1\{\bv_{t} \in \H_{m}(\bz_{t})\}$, for proper choices of $g(\cdot)$, favors moves along directions that increase the conditional likelihood $p(\by_{t} \mid \bv_{t})$ (Figure \ref{fig: dag HMM and loc fHMM}). 
The augmented joint model now becomes $p(\by,\bz,\bv) = p(\bv \mid \by, \bz) p(\by \mid \bz) p(\bz) = \big\{\prod_{t=1}^{T} p(\bv_{t} \mid \by_{t},\bz_{t}) p(\by_{t} \mid \bz_{t}) \big\} \big\{\prod_{t=2}^{T}p(\bz_{t} \mid \bz_{t-1}) \big\} p(\bz_{1})$. 
Sampling $\bz$ from the posterior can be carried out using message passing algorithms as before with each $\bz_{t}$ restricted to $\H_{m}(\bv_{t})$ 
but with the updated full conditionals  
$p(\bz \mid \by,\bv) \propto \big\{\prod_{t=1}^{T} p(\bv_{t} \mid \bz_{t}, \by_{t})  p(\by_{t} \mid \bz_{t})\big\} \big\{\prod_{t=2}^{T}p(\bz_{t} \mid \bz_{t-1}) \big\} p(\bz_{1})$.

\begin{figure}[ht!]
	\centering
	\hspace*{-0.40cm}\includegraphics[width=0.75\linewidth, trim=2cm 1.25cm 1cm 1.25cm]{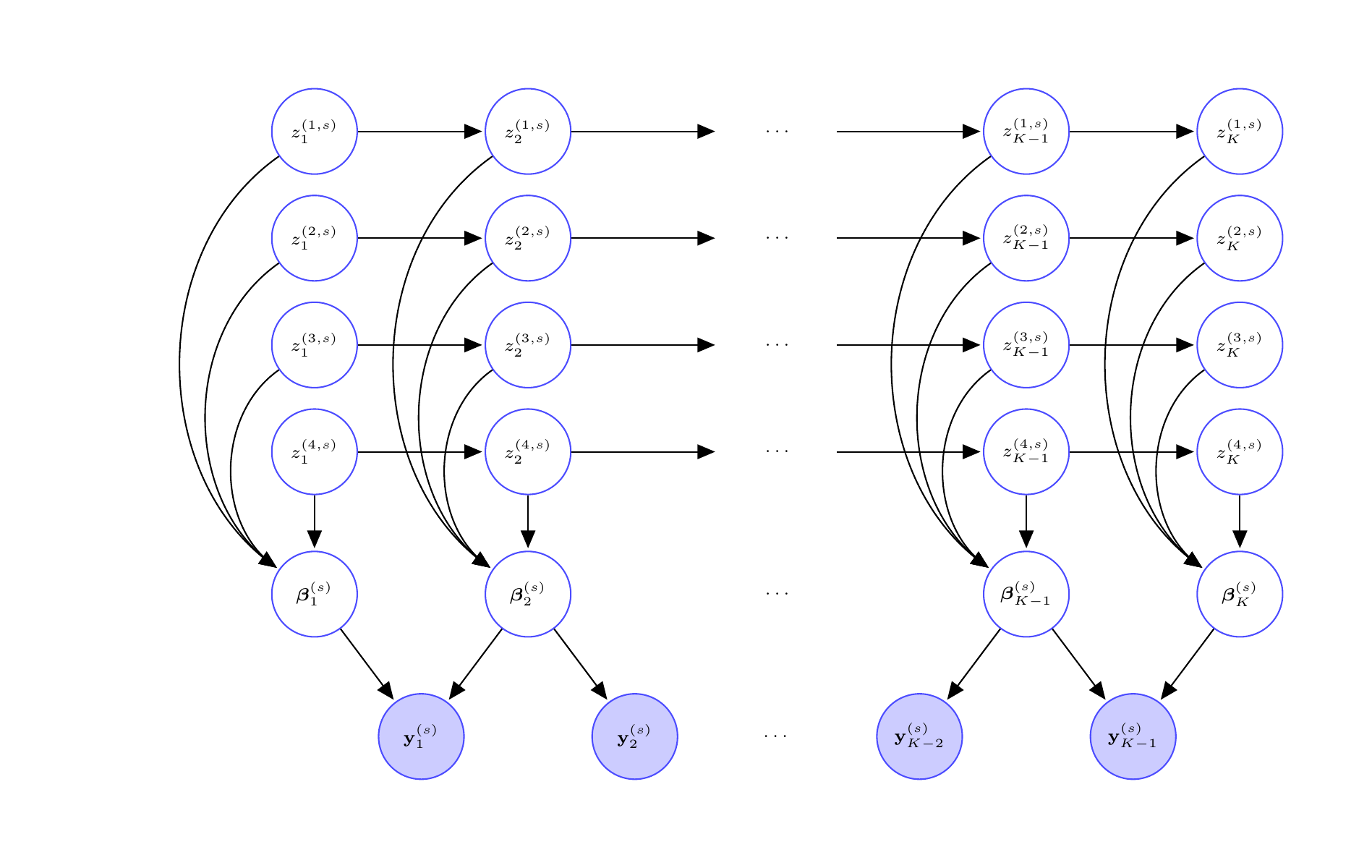}
	\caption{\baselineskip=10pt 
	Graph of the proposed longitudinal drift-diffusion mixed model for tone learning with $\bbeta_{k}^{(1,s)},\dots,\bbeta_{k}^{(4,s)}$ collected in single nodes $\bbeta_{k}^{(s)}$ for each $k$. 
	}
	\label{fig: dag fHMM 2}
\end{figure}

\section{Posterior Inference} \label{sec: post inference sm}

\subsection{Prior Hyper-parameters and MCMC Initializations} \label{sec: prior hyper-parameters}
The fixed effects parameters of the drift-diffusion mixed effects model (\ref{eq: function 1})
are initialized with an empirical Bayes type approach. 
As discussed in Section \ref{sec: lddmm}, the boundary and the drift parameters are related to the first two moments of the response times. 
Thus, we can use the empirical distribution of the response times to choose the initial 
guess for both drift and boundary parameters for each combination of input stimulus and response.  
The random effects are instead initialized at zero. 
The clustering configuration is initialized with all the success curves in different clusters, and all the failure curves in the same cluster.

Other crucial hyper-parameters are the mean and the standard deviation for the prior term of the unassigned components of $\bbeta_{\mu}^{(x)}$ and $\bbeta_{b}^{(x)}$, 
that is, the second term in the prior \eqref{eq: prior on beta star} in the main paper. 
We use the empirical distributions of the response times at every time point to set $\mu_{\beta,0}, \sigma_{\beta,0}^2$. 

The hyper-parameters in the $\text{Gamma}(a_{\alpha},b_{\alpha})$ prior for the concentration parameters $\alpha^{(C)}$ and $\alpha^{(I)}$ of the Dirichlet distributions characterizing the latent variable dynamics are set at $a_{\alpha}=b_{\alpha}=1$, as recommended in \citetlatex{escobar1995bayesian}.

\begin{figure}[ht!]
	\centering
	\hspace*{-0.00cm}\includegraphics[width=0.47\linewidth]{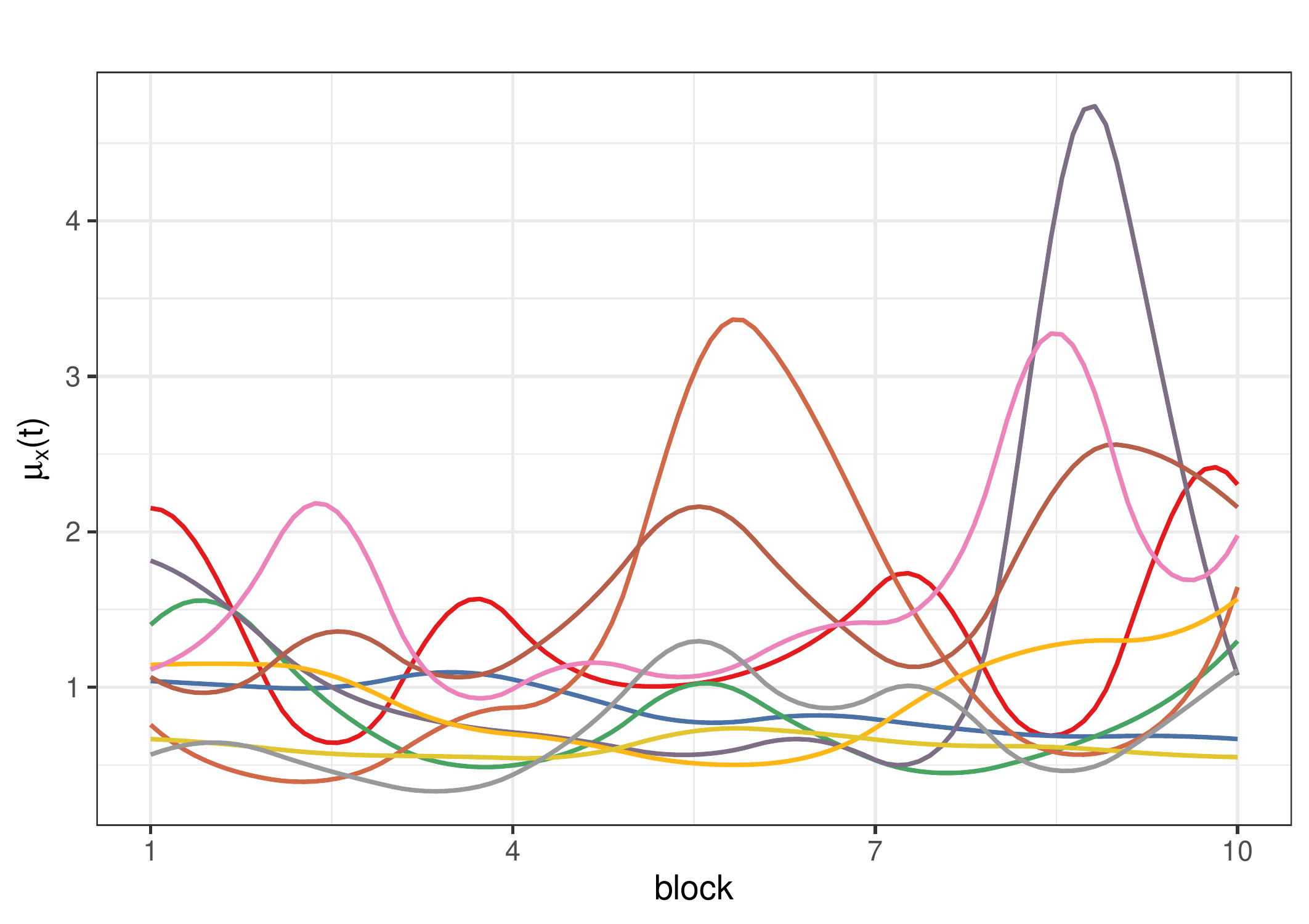} \hspace*{1cm}
	\hspace*{-0.40cm}\includegraphics[width=0.47\linewidth,trim=0cm 0.3cm 0cm 0cm]{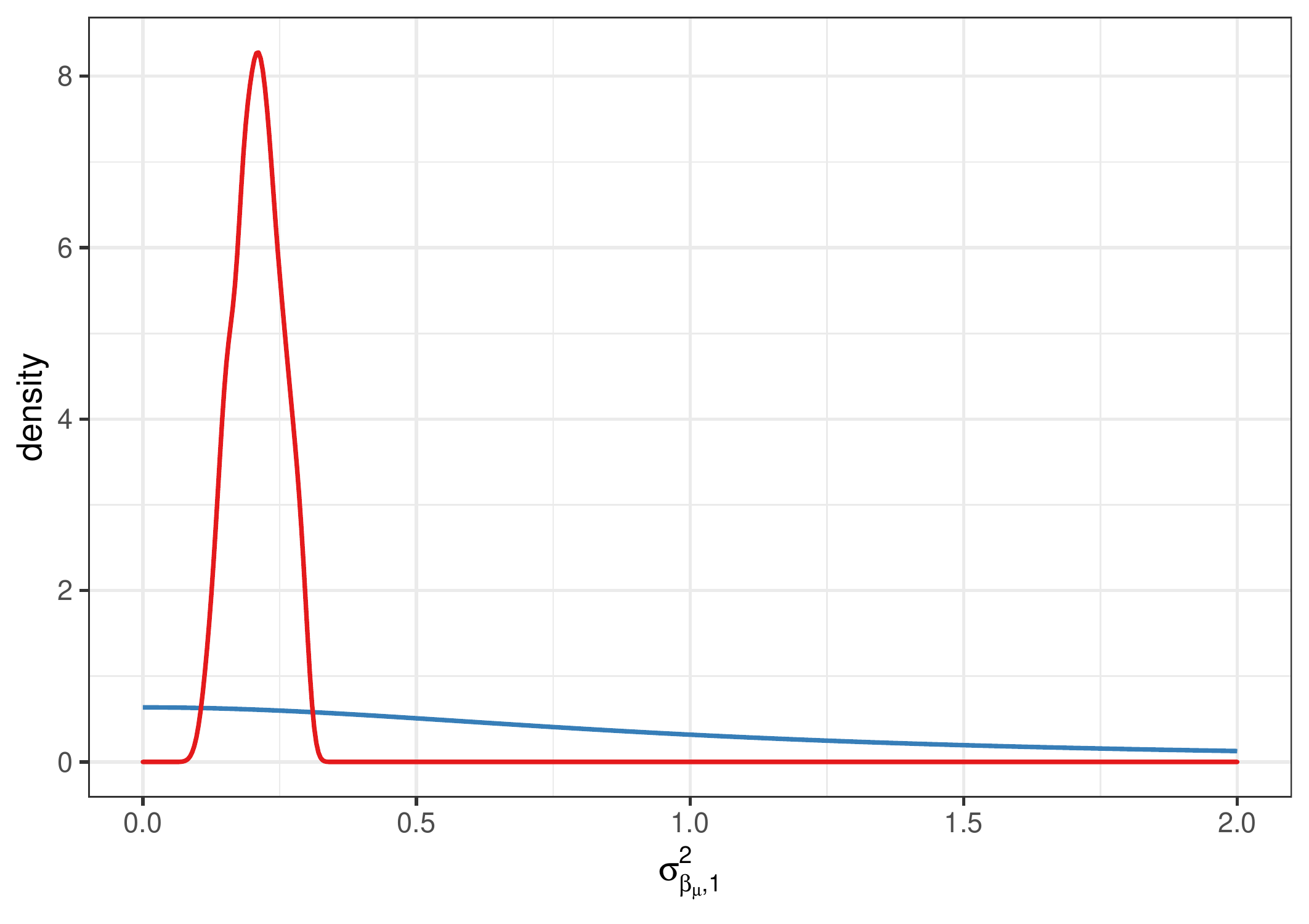}
	\caption{\baselineskip=10pt 
	Left: $10$ conditionally independent draws from $\mu_{x}(t) \mid \sigma_{\beta_{\mu},1}^{2}$ with independent draws of $\sigma_{\beta_{\mu},1}^{2}$ from a $\HC (0, 1)$ prior.
	Right: The $\HC (0, 1)$ prior distribution (in blue) and the corresponding posterior distribution (in red) for the smoothness parameter $\sigma_{\beta_{\mu},1}^{2}$.}
	\label{fig:prior_sample}
\end{figure}

The half-Cauchy priors $\HC (0, 1)$ on the smoothness parameters are non-informative for the smoothness of the corresponding longitudinal curves. 
The $\HC (0, 1)$ distribution attains its mode at zero and hence is capable of capturing strong smoothness but also has heavy tails and is thus also capable of capturing wiggly functions.
The left panel of Figure \ref{fig:prior_sample} shows some  draws from $\mu_{x}(t) \mid \sigma_{\beta_{\mu},1}^{2}$ with independent draws of the corresponding smoothness controlling parameter $\sigma_{\beta_{\mu},1}^{2}$ from a $\HC (0, 1)$ prior. 
A wide variety of curves are clearly sampled - some very smooth, some very wiggly, and many in between.
Also, as the right panel of Figure \ref{fig:prior_sample} illustrates, the posterior distributions of the smoothness parameters in our model all concentrate well within a region of flat $\HC (0, 1)$ prior probability density.
This is additional evidence that our prior is not producing any consistent bias in the posterior estimates.

\vspace*{-1cm}
\subsection{Posterior Computation} \label{sec: pc}
Posterior inference for the longitudinal drift-diffusion mixed  model, described in Section \ref{sec: lddmm} in the main paper, 
is based on samples drawn from the posterior using a message passing MCMC algorithm.

In what follows, $\bzeta$ denotes a generic variable that collects all other variables not explicitly mentioned, including the data points.
Also, $p_{0}$ will sometimes be used as a generic for a prior distribution without explicitly mentioning its hyper-parameters.  
The sampler for the drift diffusion model of Section \ref{sec: lddmm} comprises the following steps. 

\begin{enumerate}[leftmargin=0cm,itemindent=.5cm,labelwidth=\itemindent,labelsep=0cm,align=left]

	\item Update the offset parameters $\delta_{s}^{(i)}, s = 1, \dots, d_0$. 
	The full conditionals $p(\delta_{s}^{(i)} \mid \bzeta) \propto p_{0}(\delta_{s}^{(i)}) L(\by \mid \bs, \btheta)$ do not have closed forms. 
	Metropolis-Hastings (MH) steps with log-normal proposals centered on the previous sampled values are used to update these parameters.

	\item Jointly update the drift and boundary spline coefficients $(\beta^{\star}_{\mu, k, z_{k}}, \beta^{\star}_{b,k, z_{k}}), k = 1, \dots, K$. 
	
	\begin{enumerate}
		\item ~ If the parameters are assigned to one of the clusters, the full conditionals do not have closed forms. 
	MH steps are therefore used with the smoothness inducing priors (\ref{eq: prior on beta star}) on $(\beta^{\star}_{\mu, k, z_{k}}, \beta^{\star}_{b,k, z_{k}})$ as the proposal distributions.
		
		\item ~ If the parameters are not assigned to any of the clusters, the full conditional distribution is the second term of the prior in \eqref{eq: prior on beta star}.
	\end{enumerate}

	\item Update the latent cluster assignments $\bz^{(s)}_{k}= (z_{k}^{(1,s)},\dots,z_{k}^{(4,s)})\trans$: 
	\begin{enumerate}
		\item ~ Sample the auxiliary variables $\bv_{k}^{(s)} = (v_{k}^{(1,s)},\dots,v_{k}^{(4,s)})\trans$ as 
		\bse
		&& p(\bv_{k}^{(s)} \mid \bz_{k}^{(s)},  \bz_{k+1}^{(s)}, \by_{k}^{(s)}, \bzeta) \propto g\{p(\by_{k}^{(s)} \mid \bv_{k}^{(s)}, \bz_{k+1}^{(s)}, \bzeta)\}  1\{\bv_{k}^{(s)} \in \H_{m}(\bz_{k}^{(s)})\},~~~~~k=1,\dots,K-1,\\
		&& p(\bv_{K}^{(s)} \mid \bz_{K}^{(s)}, \bzeta) \propto 1\{\bv_{K}^{(s)} \in \H_{m}(\bz_{K}^{(s)})\}.
		\ese 
				
		\item ~ Back-propagate the messages $m_{k}(\bz_{k}^{(s)}) = p( \by_{k:(K-1)}^{(s)}, \bv_{k:K}^{(s)} \mid \bz_{k}^{(s)}, \bzeta)$ using the recursion 
		\bse			
		&& \hspace{-1.5cm}  m_{k}(\bz_{k}^{(s)}) = p( \by_{k:(K-1)}^{(s)}, \bv_{k:K}^{(s)} \mid \bz_{k}^{(s)}, \bzeta) \\
		&& \hspace{-1.5cm}  = \sum_{\bz_{k+1}^{(s)}} p( \by_{k:(K-1)}^{(s)}, \bv_{k:K}^{(s)} \mid \bz_{k}^{(s)}, \bz_{k+1}^{(s)}, \bzeta) p( \bz_{k+1}^{(s)} \mid \bz_{k}^{(s)}, \bzeta)\\
		&& \hspace{-1.5cm}  = \sum_{\bz_{k+1}^{(s)}}  p(\by_{k}^{(s)}, \bv_{k}^{(s)} \mid \bz_{k}^{(s)}, \bz_{k+1}^{(s)}, \bzeta)   p(\by_{(k+1):(K-1)}^{(s)}, \bv_{(k+1):K}^{(s)} \mid \bz_{k}^{(s)}, \bz_{k+1}^{(s)}, \bzeta) p( \bz_{k+1}^{(s)} \mid \bz_{k}^{(s)}, \bzeta)\\
		&& \hspace{-1.5cm}  = \sum_{\bz_{k+1}^{(s)}}  p(\by_{k}^{(s)}, \bv_{k}^{(s)} \mid \bz_{k}^{(s)}, \bz_{k+1}^{(s)}, \bzeta)   p(\by_{(k+1):(K-1)}^{(s)}, \bv_{(k+1):K}^{(s)} \mid \bz_{k+1}^{(s)}, \bzeta) p( \bz_{k+1}^{(s)} \mid \bz_{k}^{(s)}, \bzeta)\\
		&& \hspace{-1.5cm}  = \sum_{\bz_{k+1}^{(s)}}  p(\bv_{k}^{(s)} \mid  \bz_{k}^{(s)}, \bz_{k+1}^{(s)}, \by_{k}^{(s)}, \bzeta)  p( \by_{k}^{(s)} \mid \bz_{k}^{(s)}, \bz_{k+1}^{(s)}, \bzeta)  p(\bz_{k+1}^{(s)} \mid \bz_{k}^{(s)}, \bzeta) m_{k+1}(\bz_{k+1}^{(s)}), \\
		&& \hspace{-1.5cm}  \propto \sum_{\bz_{k+1}^{(s)} \in \H_{m}(\bv_{k+1}^{(s)}) }  g\{p(\by_{k}^{(s)} \mid \bv_{k}^{(s)}, \bz_{k+1}^{(s)}, \bzeta)\} 1\{\bv_{k}^{(s)} \in \H_{m}(\bz_{k}^{(s)})\}  p( \by_{k}^{(s)} \mid \bz_{k}^{(s)}, \bz_{k+1}^{(s)}, \bzeta)  p( \bz_{k+1}^{(s)} \mid \bz_{k}^{(s)}, \bzeta) m_{k+1}(\bz_{k+1}^{(s)}),
		\ese
		starting with the final condition $m_{K}(\bz_{K}^{(s)}) = 1\{\bz_{K}^{(s)} \in \H_{m}(\bv_{K}^{(s)})\}$.
		
		\item ~ Sample the latent cluster assignments forward one step at a time from 
		\bse
		p(\bz_{1:K}^{(s)} \mid \by_{1:(K-1)}^{(s)}, \bv_{1:K}^{(s)}, \bzeta) = p(\bz_{K}^{(s)} \mid \bz_{1:(K-1)}^{(s)}, \by_{1:(K-1)}^{(s)}, \bv_{1:K}^{(s)}, \bzeta) \dots p(\bz_{1}^{(s)} \mid \by_{1:(K-1)}^{(s)}, \bv_{1:K}^{(s)}, \bzeta),
		\ese
		where 
		\bse
		&& \hspace{-1.5cm} p(\bz_{k}^{(s)} \mid \bz_{1:(k-1)}^{(s)}, \by_{1:(K-1)}^{(s)}, \bv_{1:K}^{(s)}, \bzeta) \propto p(\by_{k:(K-1)}^{(s)}, \bv_{k:K}^{(s)} \mid \bz_{1:k}^{(s)}, \bzeta) p(\bz_{k}^{(s)} \mid \bz_{1:(k-1)}^{(s)}, \bzeta)  \\
		&& \hspace{-1.5cm}  = p( \by_{1:(k-2)}^{(s)}, \bv_{1:(k-2)}^{(s)} \mid \bz_{1:k}^{(s)}, \bzeta) p(\by_{k-1}^{(s)}, \bv_{k-1}^{(s)} \mid \bz_{k-1}^{(s)}, \bz_{k}^{(s)}, \bzeta)  p(\by_{k:(K-1)}^{(s)}, \bv_{k:K}^{(s)} \mid \bz_{k}^{(s)}, \bzeta) p(\bz_{k}^{(s)} \mid \bz_{1:(k-1)}^{(s)}, \bzeta)  \\
		&& \hspace{-1.5cm}  \propto p( \by_{k-1}^{(s)}, \bv_{k-1}^{(s)} \mid \bz_{k-1}^{(s)}, \bz_{k}^{(s)}, \bzeta)  p(\bz_{k}^{(s)} \mid \bz_{1:(k-1)}^{(s)}, \bzeta)  m_{k}(\bz_{k}^{(s)}) \\
		&& \hspace{-1.5cm}  = p(\bv_{k-1}^{(s)} \mid \bz_{k-1}^{(s)},  \bz_{k}^{(s)}, \by_{k-1}^{(s)}, \bzeta)  p(\by_{k-1}^{(s)} \mid \bz_{k-1}^{(s)}, \bz_{k}^{(s)}, \bzeta) p(\bz_{k}^{(s)} \mid \bz_{k-1}^{(s)}, \bzeta) m_{k}(\bz_{k}^{(s)})  \\
		&& \hspace{-1.5cm}  \propto  g\{p(\by_{k-1}^{(s)} \mid \bv_{k-1}^{(s)}, \bz_{k}^{(s)}, \bzeta)\}  p(\by_{k-1}^{(s)} \mid \bz_{k-1}^{(s)}, \bz_{k}^{(s)}, \bzeta) p(\bz_{k}^{(s)} \mid \bz_{k-1}^{(s)}, \bzeta) m_{k}(\bz_{k}^{(s)}).
		\ese
	\end{enumerate}

	\item Update the cluster specific fixed effects spline coefficients:
		\bse
		&(\beta_{\mu,k}^{(x)} \mid z_{k}^{(x)} = z_{k}, \bzeta) \sim 1\{\beta_{\mu,k}^{(x)}  =  \beta_{\mu,k,z_{k}}^{\star}\}, \quad k = 1, \dots, K.
		\\
		&(\beta_{b,k}^{(x)} \mid z_{k}^{(x)} = z_{k}, \bzeta) \sim 1\{\beta_{b,k}^{(x)}  =  \beta_{b,k,z_{k}}^{\star}\}, \quad k = 1, \dots, K.
		\ese
		
	\begin{figure}[ht!]
		\centering
		\hspace*{-0.40cm}\includegraphics[trim=2cm 1.25cm 1cm 1.25cm]{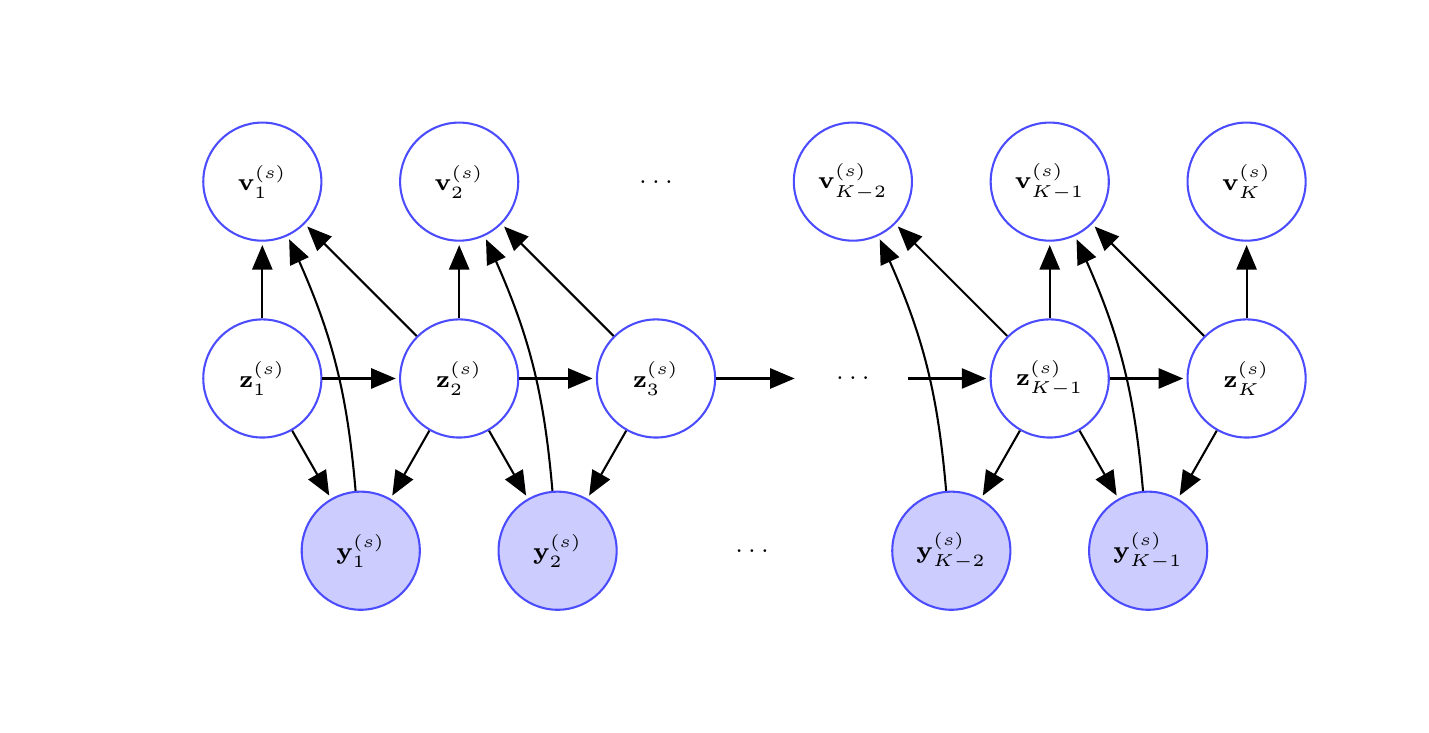}
		\caption{\baselineskip=10pt 
		Locally informed Hamming ball sampling of the latent states in our tone-learning longitudinal drift-diffusion mixed model. 
		See also Figure \ref{fig: dag fHMM} in the main paper. 
		}
		\label{fig: dag fHMM supp}
	\end{figure}

	\item Update the transition probability matrices:
	    \vspace{-4ex}\\
		\bse
		\left( \bpi_{z}^{(C)} \mid \bzeta \right) \sim \Dir(\alpha^{(C)}/z_{\max} + n_{z,1}^{(C)}, \dots, \alpha^{(C)}/z_{\max} + n_{z,z_{\max}}^{(C)}) 
		\\
		\left( \bpi_{z}^{(I)} \mid \bzeta \right) \sim \Dir(\alpha^{(I)}/z_{\max} + n_{z,1}^{(I)}, \dots, \alpha^{(I)}/z_{\max} + n_{z,z_{\max}}^{(I)}), 
		\ese
    	\vspace{-4ex}\\
	where $n_{z,z^{\prime}}^{(C)} = \sum_{k}1\{z_{k}^{(x)} = z,z_{k+1}^{(x)}=z'\}$ is the number of transitions from $z$ to $z^{\prime}$ for the HMMs associated with the correct identification of the tones, that is, with $x$ s.t. $d = s$. A similar definition holds for $n_{z,z^{\prime}}^{(I)}$.

	\item Update the cluster specific smoothness parameter
	\vspace{-4ex}\\
	\bse
	p(\sigma_{\beta_{\mu},1}^{2} \mid \bzeta) \propto \left(\sigma_{\beta_{\mu},1}\right)^{-K x_{\max}} \exp \left( - \frac{1}{2 \sigma_{\beta_{\mu},1}^{2}} \sum_{x} \bbeta_{\mu}^{(x)\rm{T}} \bP_{u} \bbeta_{\mu}^{(x)} \right) p_{0}(\sigma_{\mu,u,a}^2).
	\ese
	\vspace{-4ex}\\
	MH steps with log-normal proposals centered on the previous sampled values are used to update these parameters.
	
	\item Update the random effects spline coefficients $\beta_{\mu,k,u}^{(i)}$ and $\beta_{b,k,u}^{(i)}$: 
	The full conditional does not have a closed form. An MH step with a normal proposal centered on the previous value was used. 
	

	\item Update the random effects variance parameters $\sigma_{\mu,u,a}^{2}$, $\sigma_{\mu,u,s}^{2}$, $\sigma_{b,u,a}^{2}$ and $\sigma_{b,u,s}^{2}$: \\
	The full conditional for $\sigma_{\mu,u,a}^{2}$ is given by 
	    \vspace{-4ex}\\
		\bse
		p(\sigma_{\mu,u,a}^{2} \mid \bzeta) \propto \det(\sigma_{\mu,u,s}^{-2} \bP_{u} + \sigma_{\mu,u,a}^{-2} \bI_{K})^{n/2} \exp \left( - \frac{1}{2\sigma_{\mu,u,a}^{2}} \sum_{i = 1}^{n} \bbeta_{\mu,u}^{(i) \rm{T}} \bP_{u} \bbeta_{\mu,u}^{(i)} \right) p_{0}(\sigma_{\mu,u,a}^2).
		\ese
    	\vspace{-4ex}\\
	Analogous expressions can be found for the full conditionals of $\sigma_{\mu,u,s}^{2}$, $\sigma_{b,u,a}^{2}$ and $\sigma_{b,u,s}^{2}$.
	MH steps with log-normal proposals centered on the previous sampled values are used to update these parameters.


\end{enumerate}

The main challenge here arises from the nonconjugacy of the inverse Gaussian distribution based likelihood function, requiring MH steps for updating $\delta_{s}^{(i)},\beta_{b,k,z_{k}}^{\star},\beta_{\mu,k,z_{k}}^{\star}$. 
We employed the adaptive MH algorithm \citeplatex{roberts2009examples} for updating $\delta_{s}^{(i)}$ and the variance parameters, 
avoiding the difficult task of choosing the parameters of their proposal distributions while also improving mixing. 
Specifically, for every batch of $50$ iterations, we inflate or deflate the standard deviation of the proposal distribution such that the optimal acceptance rate of $44\%$ is achieved \citeplatex{roberts2001optimal}.  
The adaptive MH could not be employed for the cluster specific parameters $(\beta_{b,k,z_{k}}^{\star},\beta_{\mu,k,z_{k}}^{\star})$ due to label switching, so we used tempered MH steps instead. 
For the proposal distributions for $(\beta_{b,k,z_{k}}^{\star},\beta_{\mu,k,z_{k}}^{\star})$, 
we used the smoothness inducing conditional prior distributions $p_{0}(\beta_{\mu,k,z_{k}}^{\star} \mid \bbeta_{\mu,k-1}^{\star}) \times p_{0}(\beta_{b,k,z_{k}}^{\star} \mid \bbeta_{b,k-1}^{\star})$.
Since the conditioning variables $\bbeta_{\mu,k-1}^{\star}$ and $\bbeta_{b,k-1}^{\star}$ are also updated at every iteration, 
the values sampled from the smoothness inducing priors are frequently accepted. 

Based on $M$ thinned samples $\{\btheta^{(m)}\}_{m=1}^{M}$ drawn from the posterior after the burn-in, 
the individual level drift parameters in the drift-diffusion mixed model are estimated as 
\bse
& \textstyle \mu_{x}^{(i)}(t) = \exp\{f_{\mu,x}(t) + u_{\mu}^{(i)}(t)\} 
= \frac{1}{M}\sum_{m=1}^{M}\exp\{\wh{f}_{\mu,x}^{(m)}(t) + \wh{u}_{\mu}^{(i,m)}(t)\},
\ese
where $\wh{f}_{\mu,x}^{(m)}(t) = \sum_{k=1}^{K} \beta_{\mu,k,z_{k}^{(x,m)}}^{\star (m)} B_{k}(t)$, 
$\wh{u}_{\mu}^{(i,m)}(t) = \sum_{k=1}^{K} \beta_{k,u,\mu}^{(i,m)} B_{k}(t)$ etc. 
The population level drift parameters are likewise estimated as 
\bse
& \textstyle \mu_{x}(t) = \int \exp\{f_{\mu,x}(t) + u_{\mu}^{(i)}(t)\} f\{u_{\mu}^{(i)}(t)\} du_{\mu}^{(i)}(t) = \exp\left[f_{\mu,x}(t) + \frac{\var\{u_{\mu}^{(i)}(t)\}}{2}\right] \\
& = \frac{1}{M} \sum_{m=1}^{M} \exp\left\{\wh{f}_{\mu,x}^{(m)}(t) + \frac{\var\{\wh{u}_{\mu}^{(i,m)}(t)\}}{2} \right\},
 \ese


\subsection{Software, Runtime, etc.}
The results reported in this article are all based on $5,000$ MCMC iterations with the initial $2,000$ iterations discarded as burn-in.
The remaining samples were further thinned by an interval of $5$.
We programmed in \texttt{R} and \texttt{C++}. 
The codes are available as part of the supplementary materials.
The MCMC algorithm takes 10 hours on a Dell machine with 16 Gb RAM. 
A `readme' file, providing additional details for a practitioner, is also included in the supplementary materials.

\newpage
\section{MCMC Diagnostics}\label{sec: mcmc_diagnostics}

This section presents some convergence diagnostics for the MCMC sampler described in the main manuscript. 
The results presented here are for the tone learning data set.
Diagnostics for the simulation experiments were similar and hence omitted.

\begin{figure}[ht!]
	\centering
	\includegraphics[width=12cm]{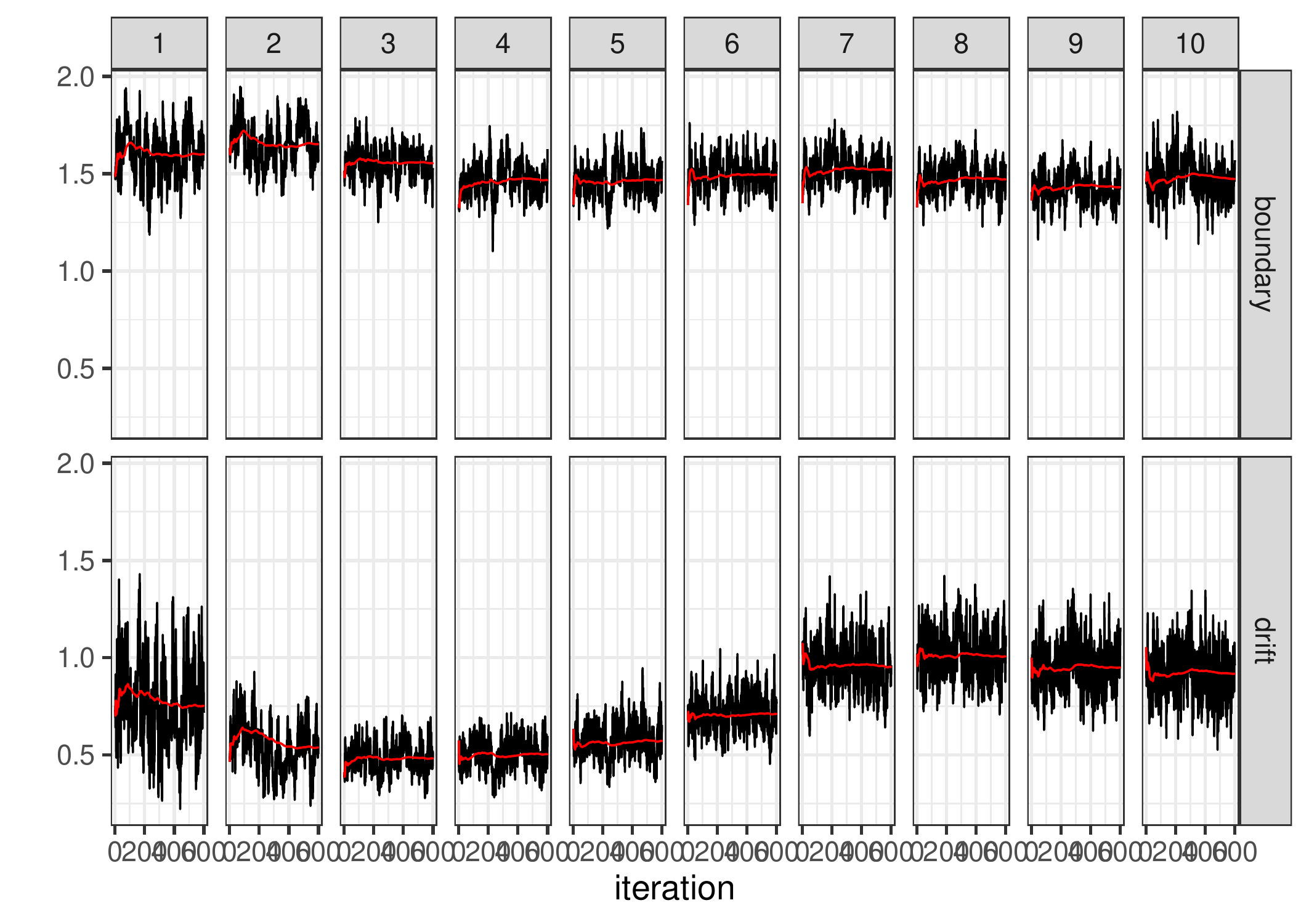} \\[5pt]
	\caption{Trace plots of the individual drift rates $\mu_{1,1}^{(i)}(t)$ and boundary parameters $b_{1,1}^{(i)}(t)$ corresponding to the success categorization of tone T1 evaluated at each of the training blocks. 
	The two rows correspond to the two different classes of parameters, and the ten columns to the training blocks.
	In each panel, the solid red line shows the running mean.
	Results for other drift and boundary parameters were very similar.}
	\label{fig: param_traceplots}
\end{figure}

\begin{figure}[ht!]
	\centering
	\includegraphics[width=12cm]{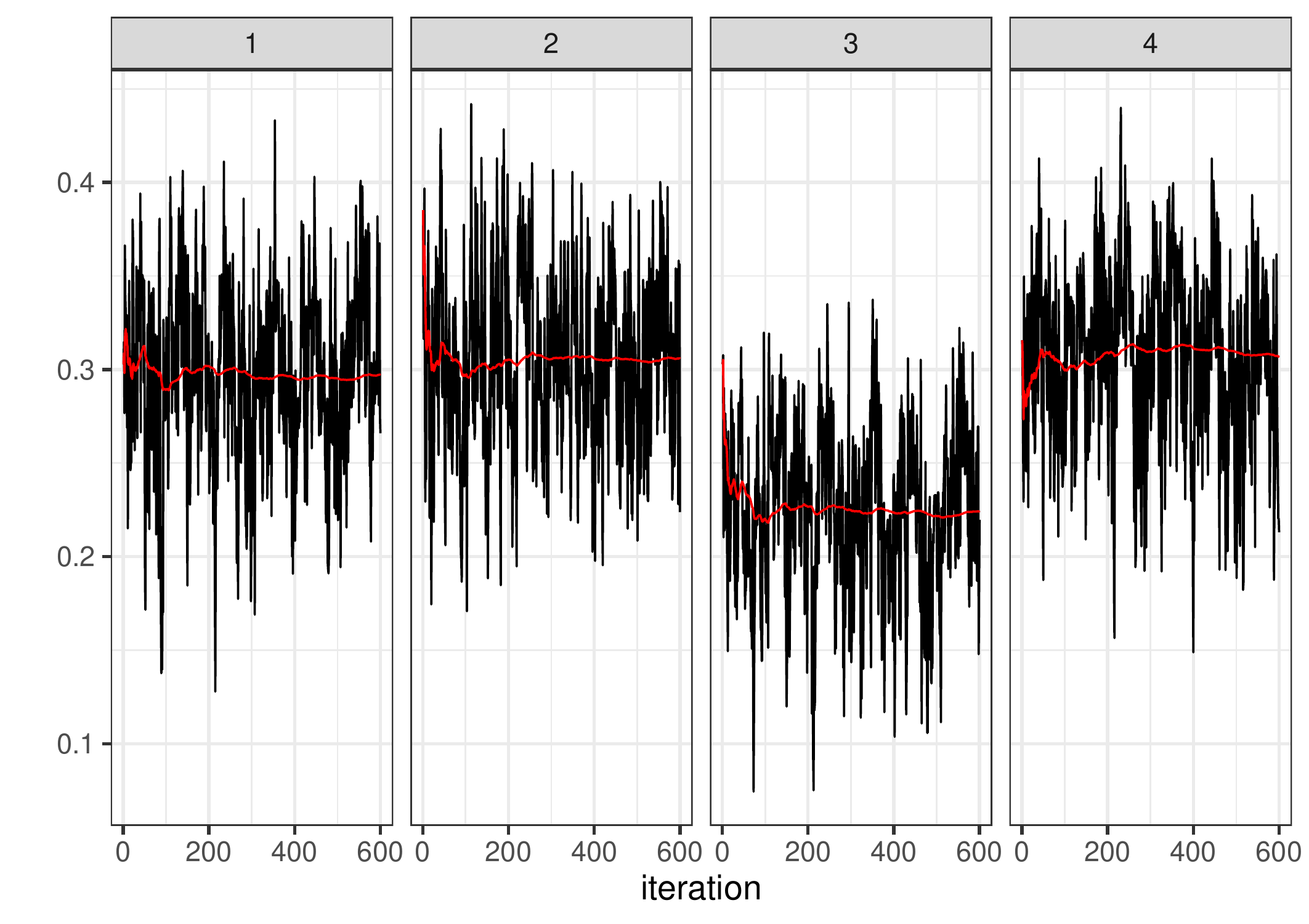} \\[5pt]
	\caption{Trace plots of the individual level offset parameters $\delta_{s}^{(i)}$ for the four possible input tones.
	The four columns correspond to the input stimuli $s$.
	In each panel, the solid red line shows the running mean. 
	Results for other offset parameters were very similar.}
	\label{fig: delta_traceplots}
\end{figure}

Figure \ref{fig: param_traceplots} shows the trace plots of some individual level parameters at different training blocks.
Figure \ref{fig: delta_traceplots} shows the trace plots of some individual level offset parameters.
These results are based on the MCMC thinned samples.
As these figures show, the running means are very stable and there seems to be no convergence issues.
Additionally, the Geweke test \citeplatex{geweke1991evaluating} for stationarity of the chains, which formally compares the means of the first and last part of a Markov chain, was also performed. 
If the samples are drawn from the stationary distribution of the chain, the two means are equal and Geweke's statistic has an asymptotically standard normal distribution. 
The results of the test, reported in Table \ref{tab: param_geweke} and Table \ref{tab: delta_geweke}, indicate that convergence was satisfactory for the parameters considered. 
Only one parameter, $\mu_{1,1}^{(i)}(2)$ in the second row of Table \ref{tab: param_geweke}, had a significant p-value. 
Some chance rejections are expected in multiple hypothesis testing scenarios. 
A visual inspection of the corresponding trace plot, however, does not indicate any serious issue.

\begin{table}[!ht]
	\centering
	\resizebox{\columnwidth}{!}{%
	\begin{tabular}{|l|cccccccccc|} \hline
		& $t=1$  & $t=2$  & $t=3$  & $t=4$  & $t=5$  & $t=6$  & $t=7$  & $t=8$  & $t=9$  & $t=10$ \\ \hline
		\multicolumn{1}{|c|}{\multirow{2}{*}{boundary}}	& 1.161 & 0.973  & 1.162  & -1.287  & -1.080 & -0.554 & 0.164 & -0.285 & 0.481 & 0.894 \\
		\multicolumn{1}{|c|}{}	& (0.25) & (0.33) & (0.25) & (0.20) & (0.28) & (0.58) & (0.87) & (0.78) & (0.63) & (0.37) \\ \hline
		\multirow{2}{*}{drift}	& 1.884 & 3.467 & -0.102 & -0.863  & -1.171  & -0.845 & 0.445 & 0.821  & 0.362  & 0.607  \\
		& (0.06) & (0.00) & (0.92) & (0.39) & (0.24) & (0.40) & (0.66) & (0.41) & (0.72) & (0.54) \\ \hline
	\end{tabular}
	}
	\caption{Geweke statistics and associated p-values assessing convergence of the individual drift rates $\mu_{1,1}^{(i)}(t)$ and boundary parameters $b_{1,1}^{(i)}(t)$ corresponding to the success categorization of tone T1 evaluated at each of the training blocks.
	Results for other drift and boundary parameters were very similar.
	}
	\label{tab: param_geweke}
\end{table}

\begin{table}[!ht]
	\centering
	\begin{tabular}{|cccc|} \hline
		$s=1$  & $s=2$  & $s=3$  & $s=4$  \\ \hline
		-0.395 & -0.848  & -0.019 & -0.217  \\
		(0.69) & (0.40) & (0.98) & (0.83)\\ \hline
	\end{tabular}
	\caption{Geweke statistics and associated p-values assessing convergence of the of the individual level offset parameters $\delta_{s}^{(i)}$ for the four possible input tones.
	Results for other offset parameters were very similar.}
	\label{tab: delta_geweke}
\end{table}

\clearpage
\newpage
\section{Linear Ballistic Accumulator Model}\label{sec: sm lba}

We present here a review of the LBA model \citeplatex{brown2008simplest} for easy reference with some repetition from the main paper to make this section relatively self-contained.

The LBA model is a popular framework for studying neural mechanisms underlying choice between multiple alternatives. 
Similar to our model, it uses independent evidence accumulators starting at $\delta_{s}$ that continue until a response boundary $b_{s}$ is reached. 
The accumulator that first reaches the boundary corresponds to the decision outcome, and the time at which the boundary is reached is the response time. 
The evidence, however, accumulates linearly at the rate $\mu_{d,s}$, reaching the boundary $b_{s}$ precisely at time $\tau_{d}=b_{s}/\mu_{d,s}$. 
To explain trial-by-trial variability, the LBA model assumes that the slopes $\mu$ for different trials are random draws from a $\Normal(m_{d,s},v_{d,s})$ distribution. 
The cumulative distribution function for the boundary crossing time $\tau_{d}$ for the $d^{th}$ category is thus given by 
\bse
F_{LBA}(\tau_{d} \mid \btheta_{d,s}) = 1 - \Phi\left(b_{s} / \tau_{d} \mid m_{d, s}, v_{d,s}\right), 
\ese
where $\btheta_{d,s} = (m_{d,s}, v_{d,s}, b_{s})\trans$. 
The likelihood of the LBA model at the $t\th$ time point is thus 
\bse
L_{t}(\by_{t} \mid \bs, \btheta) = \prod_{d=1}^{d_{0}} \prod_{i=1}^{n} \prod_{\ell=1}^{L} \left[f_{LBA}(\tau_{i,\ell,t} \mid \btheta_{d,s,t}) \prod_{d^{\prime} \neq d} \{1 - F_{LBA}(\tau_{i,\ell,t} \mid \btheta_{d^{\prime},s,t})\}\right]^{1\{d_{i,\ell,t} = d\}},
\ese
where $\btheta_{d,s,t} = (m_{d,s,t}, v_{d,s,t}, b_{s,t})\trans$, and $f_{LBA}(\tau) = \frac{dF_{LBA}(\tau)}{d\tau}$ is the pdf of $\tau$. 


\vskip 10 pt
\begin{figure}[ht!]
	\centering
	\includegraphics[width=6.95cm, trim=2cm 0.25cm 1cm 0.25cm]{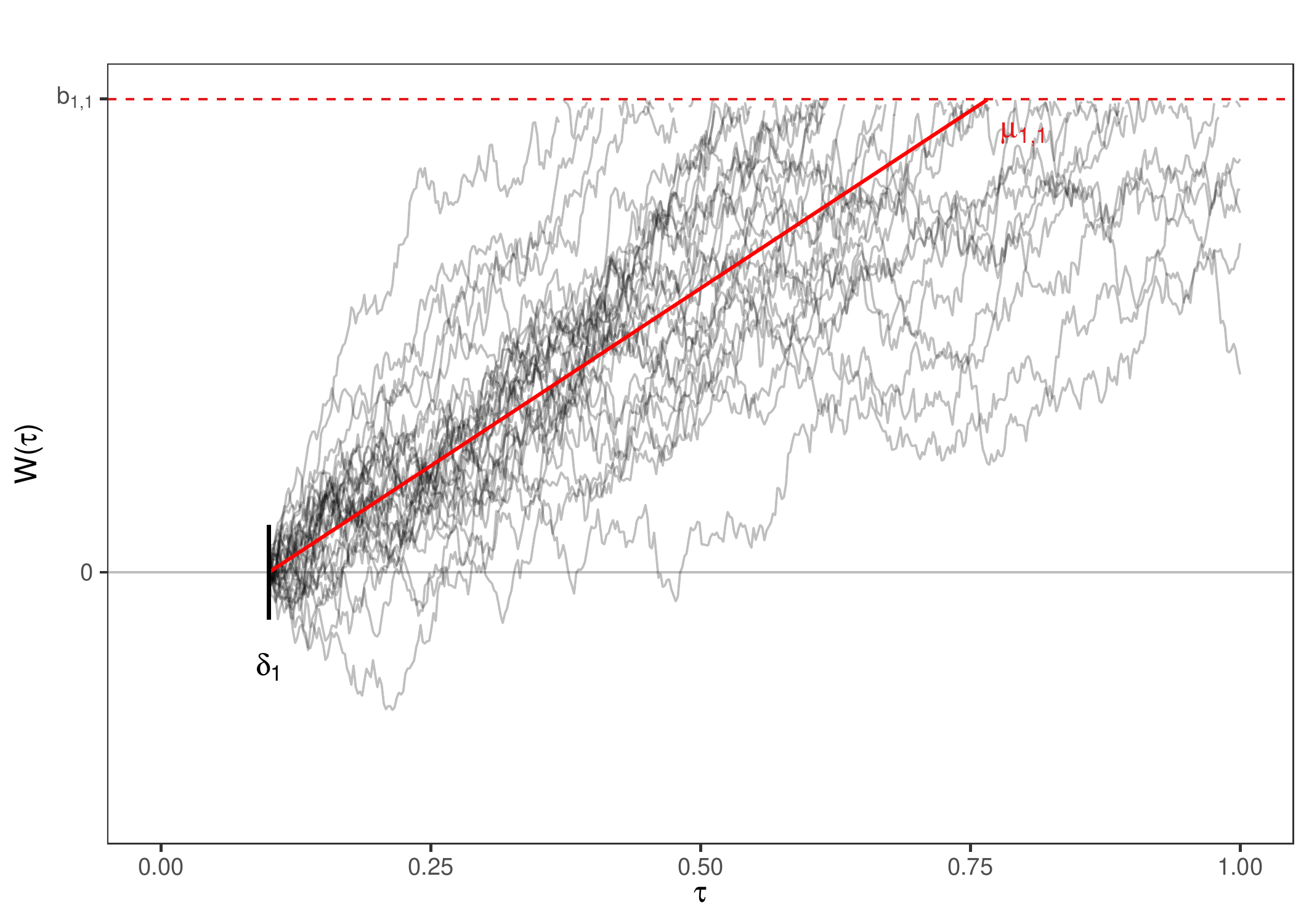} \hspace*{1cm}
	\includegraphics[width=6.95cm, trim=2cm 0.25cm 1cm 0.25cm]{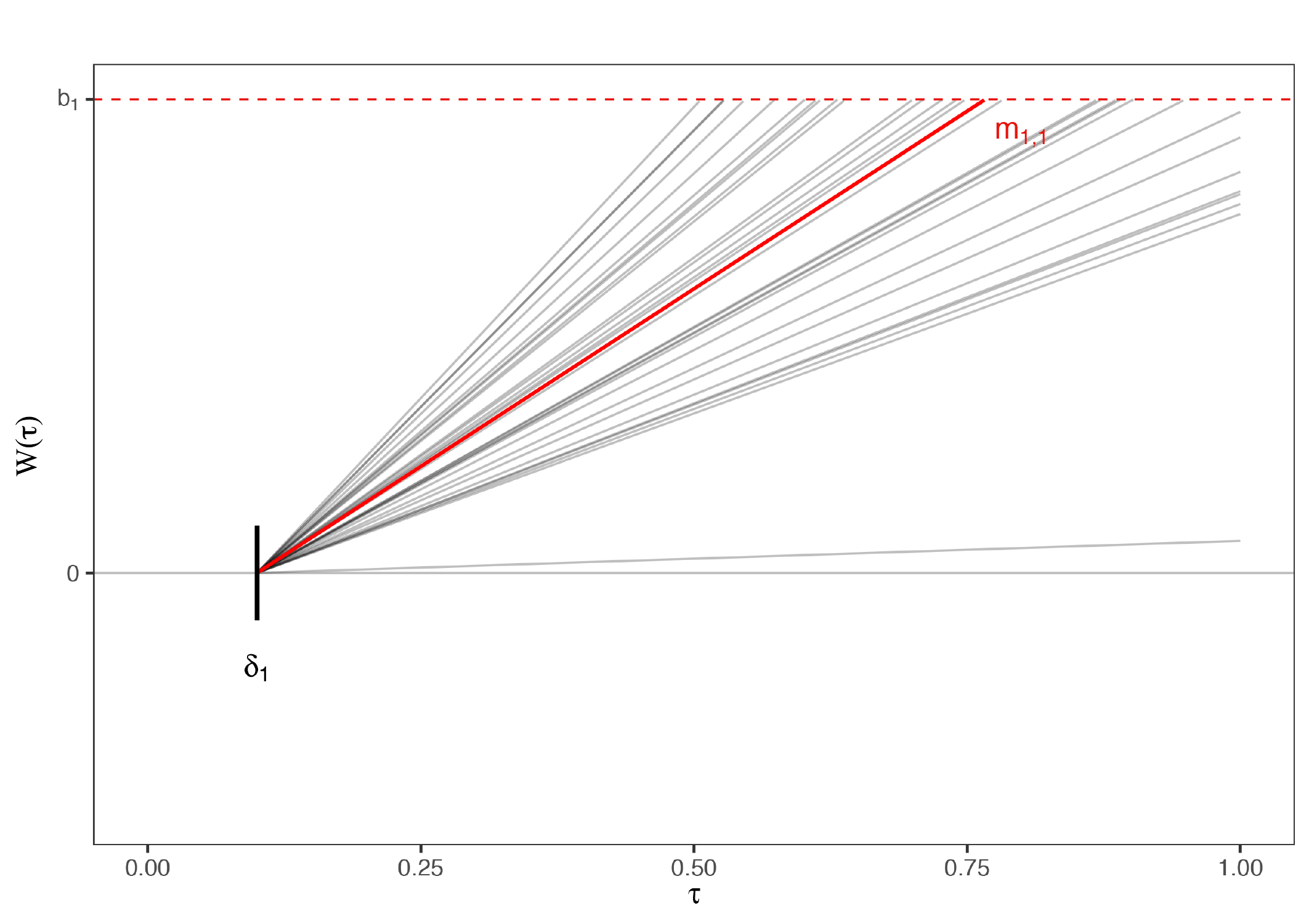}
	\caption{Representation of the underlying evidence accumulation processes for our drift-diffusion model (left) and the LBA model (right) 
	for 30 independent trials with fixed stimulus and decision categories $d = s = 1$. 
	The red line represents the drift parameter $\mu_{1,1}$ for the drift-diffusion model (left) and the mean of the drift parameters $m_{1,1}$ for the LBA (right).
	In drift-diffusion models, trial-by-trial variability is explained by stochastically different diffusion paths for different trials. 
	In the LBA model, trial-by-trial variability is explained by stochastically varying slopes drawn from a Normal distribution. 
	}
	\label{fig: comparison_drift_LBA}
\end{figure}

The existing literature on LBA models has many serious limitations. 
The normality assumption on the slopes $\mu$ in the LBA model does not satisfy a non-negativity constraint.  
A common boundary $b_{s}$ for all decision categories $d$ is also inflexible. 
Importantly, there is no principled method to incorporate systematic stimulus and decision category specific fixed or individual specific random effects into the LBA model. 
Existing literature is also limited to static settings, there is no mechanism to estimate smoothly varying longitudinal trajectories as the participants get trained and experienced in their decision tasks. 
In our implementation, we thus fitted these models separately for each time stamp. 
Finally, the likelihood function of the LBA model described above is non-convex in the parameters. 
Parameter estimation based on optimization of the likelihood function is thus fraught with convergence issues. 
We used the \texttt{rtdists} package \citeplatex{rtdists} in \texttt{R}, using several random initializations and tracking the objective function to ensure convergence.

\newpage
\section{Comparison with a Simpler Sub-Model} \label{sec: comparisons}

In this section, we summarize the results produced by a simpler alternative model, specifically, a reduced static version of our proposed longitudinal drift-diffusion mixed model fitted separately to data from each block as in the case of the LBA model.
Using notation similar to those in our proposed longitudinal mixed model, we now let $\mu_{x,t}^{(i)} = \exp \{ f_{\mu,x,t} + u_{\mu,x,t}^{(i)} \}$ be the drift rates and $b_{x,t}^{(i)} = \exp \{ f_{b,x,t} + u_{b,x,t}^{(i)} \}$ be the boundary parameters.
The time index $t$ now appears in subscript, as opposed to as an argument within parenthesis in our original longitudinal functional model.
Other relevant parts of the model, including the priors, remain unchanged. 

\begin{figure}[ht!]
	\centering
	\includegraphics[width=6.5cm, trim=2cm 0.25cm 1cm 0.25cm]{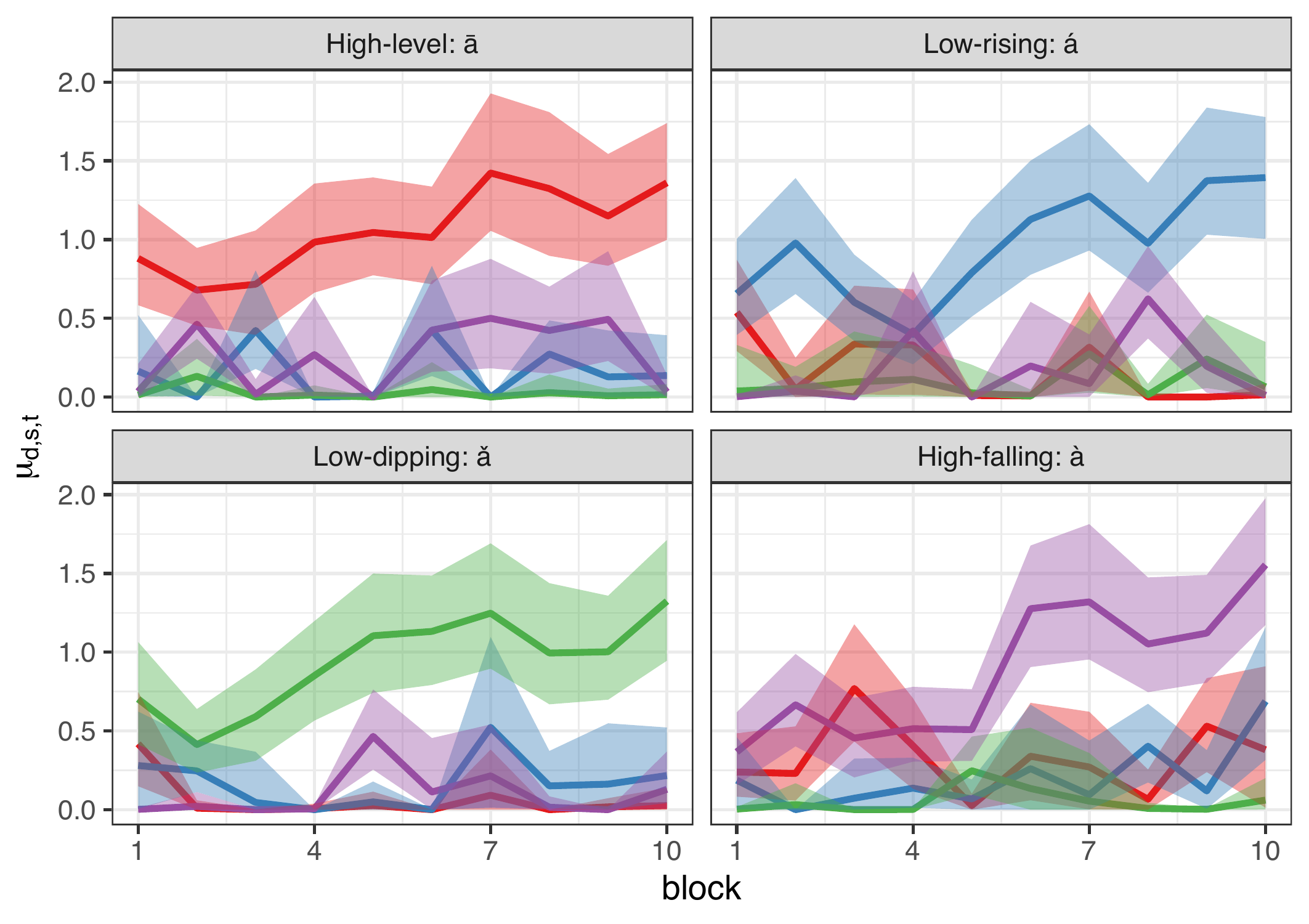} \hspace*{1cm}
	\includegraphics[width=6.5cm, trim=2cm 0.25cm 1cm 0.25cm]{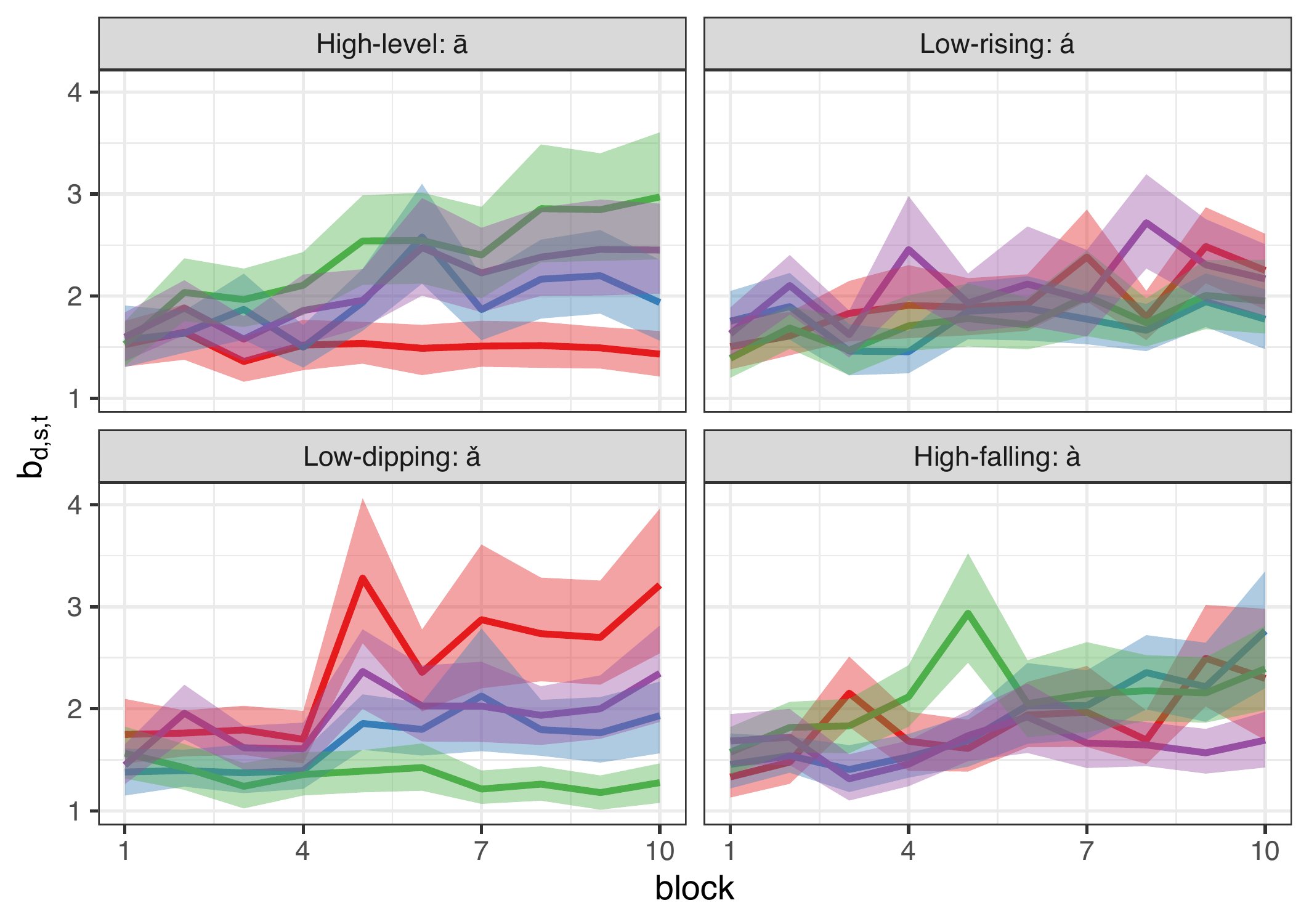}
	\caption{Results for tone learning data: Estimated posterior mean trajectories of the population level drifts $\mu_{d,s,t}$ (left panel) 
	and boundaries $b_{d,s,t}$ (right panel) for the inverse Gaussian drift-diffusion mixed model applied independently for each block. 
	The shaded areas represent the corresponding $90\%$ point wise credible intervals.
	Parameters for the high-level tone response category T1 are shown in red; low-rising T2 in blue; low-dipping T3 in green; and high-falling T4 in purple.}
	\label{fig: comparison_DDM}
\end{figure}

Figure \ref{fig: comparison_DDM} shows the posterior means and associated $90\%$ credible intervals for the population level boundaries $b_{d,s,t}$ and drift rates $\mu_{d,s,t}$ estimated by fitting the above described static drift-diffusion model fitted separately to data from each block.
These results are generally consistent with the ones illustrated in Figure \ref{fig: population_effect} in the main paper.
However, this reduced model yields less interpretable results for at least three reasons. 
First, the absence of functional dependence makes it harder to pinpoint a general trend  because the estimates are not smooth but very wiggly across the training blocks. 
Second, the fixed effects parameters are not allowed to cluster across input-response combinations, which results in many redundant configurations.
Third, the parameter estimates under our proposed model seem to have smaller uncertainty due borrowing of information across adjacent blocks as well as across input-output tone combinations via local clustering.

\newpage
\section{Simulation Studies} \label{sec: sim studies}

In this section, we discuss the results of some synthetic numerical experiments. 
We are not aware of any other method from the existing literature that can be readily applied or at least be easily adapted to our data settings and inferential challenges. 
We thus restrict our focus mostly on evaluating the performances of the proposed longitudinal inverse Gaussian drift-diffusion mixed model. 
We do present a comparison with the LBA model though, applying it separately for each block as in Section \ref{sec: application} in the main paper. 

In designing the simulation scenarios, we have tried to closely mimic our motivating tone learning data set. 
We thus chose $n=20$ participants being trained over $T=10$ blocks to identify $d_{0}=4$ tones.  
We set $\mu_{d,s}(t), b_{d,s}(t)$ to values that are very similar to the corresponding estimated values for the real data set. 
The local differences were all set to be in the drift curves; 
additionally, some boundary trajectories were globally different from each other.  
We slightly simplified the local clustering structure, however, to be able to better illustrate the workings of our proposed method.   
Moreover, we choose $u_{\mu}^{(C, i)}(t), u_{b}^{(C, i)}(t), u_{\mu}^{(I, i)}(t), u_{b}^{(I, i)}(t), \delta_{s}$ etc. to be the estimated posterior means obtained for the real data set. 

\begin{figure}[ht!]
	\centering
	\includegraphics[width=6.5cm, trim=2cm 0.25cm 1cm 0.25cm]{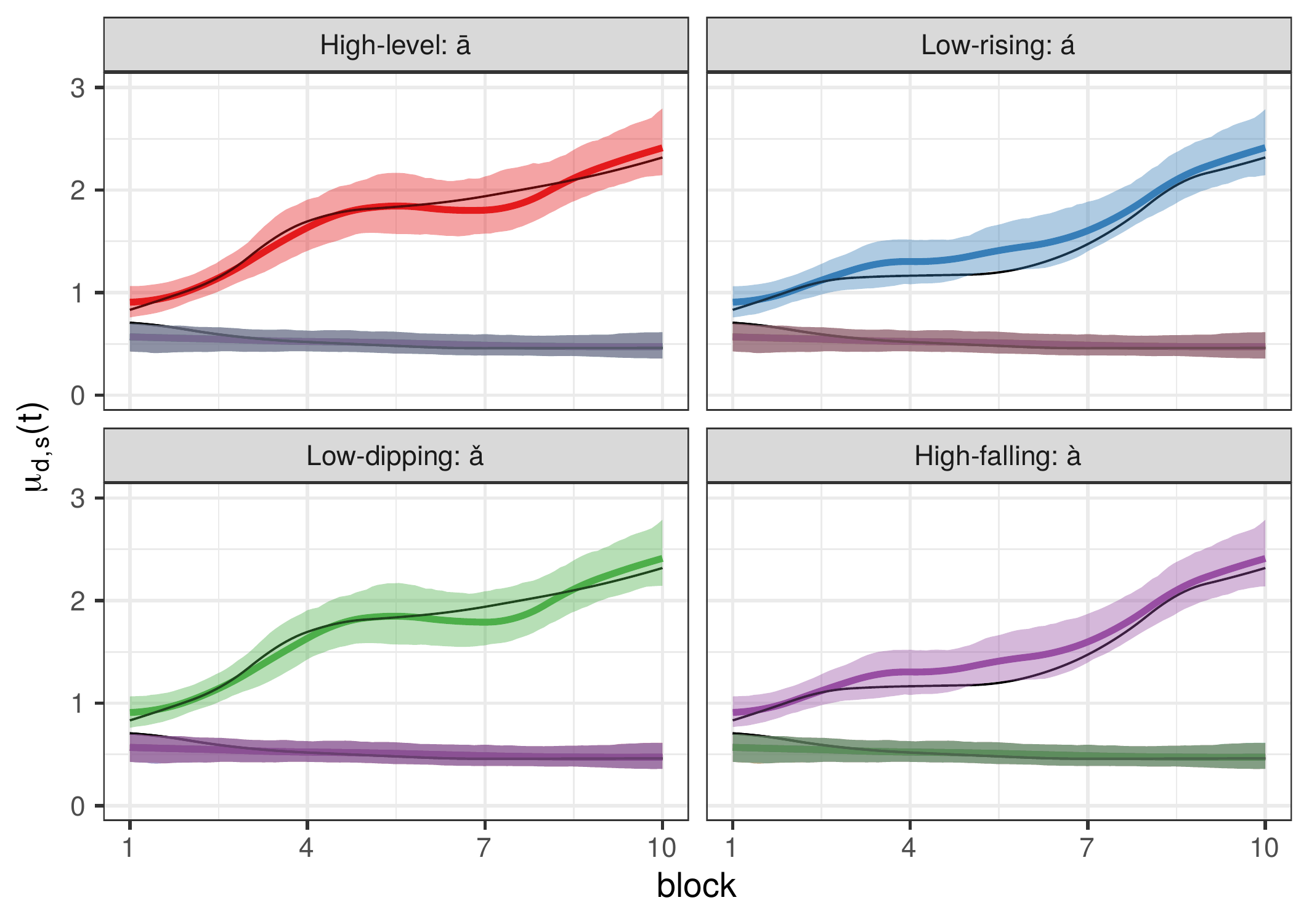} \hspace*{1cm}
	\includegraphics[width=6.5cm, trim=2cm 0.25cm 1cm 0.25cm]{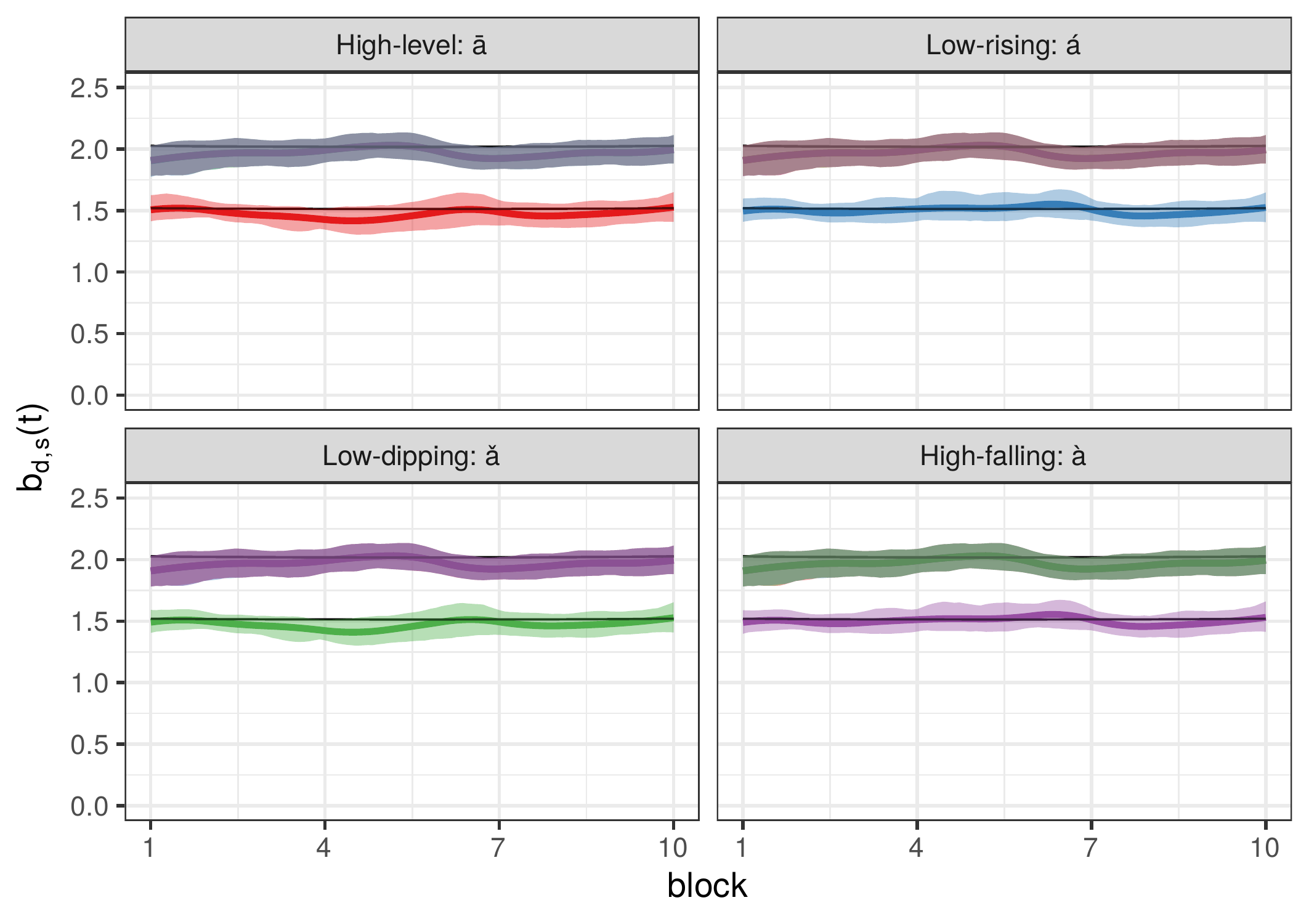}
	\caption{Results for synthetic data: Estimated posterior mean trajectories of the population level drifts $\mu_{d,s}(t)$ (left panel) 
	and boundaries $b_{d,s}(t)$ (right panel) for the proposed longitudinal inverse Gaussian drift-diffusion mixed model. 
	The shaded areas represent the corresponding $90\%$ point wise credible intervals. 
	The solid black lines represent underlying true curves.
 	Parameters for the high-level tone response category T1 are shown in red; low-rising T2 in blue; low-dipping T3 in green; and high-falling T4 in purple.}
	\label{fig: population_effect_sim}
\end{figure}

We experimented with $50$ synthetic data sets generated according to the design described above. 
The results produced by our method were highly stable and consistent across all data sets. 
The results summarized below represent a typical scenario. 

\begin{figure}[ht!]
	\centering
	\includegraphics[width=6.5cm, trim=2cm 0.25cm 1cm 0.25cm]{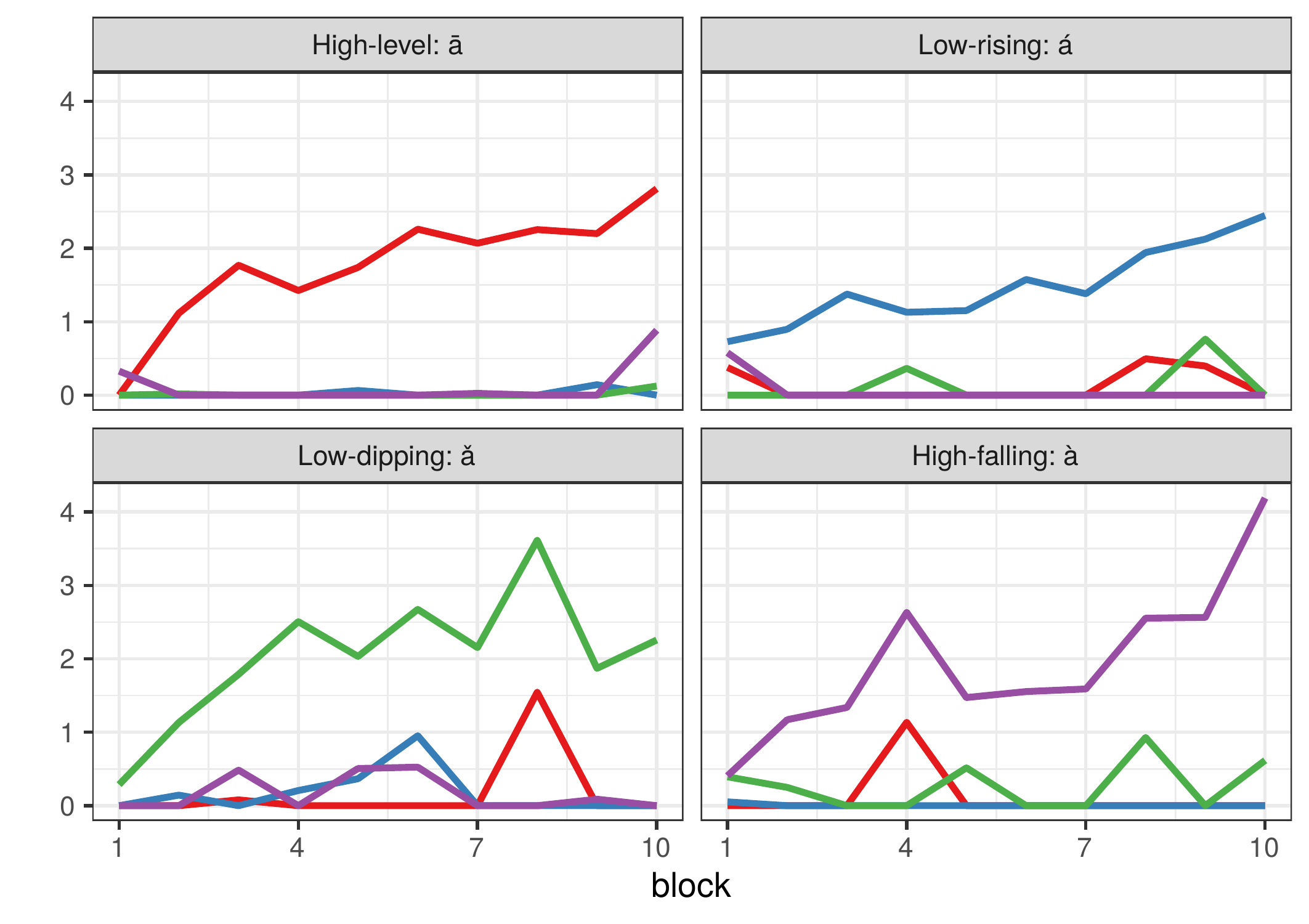} \hspace*{1cm}
	\includegraphics[width=6.5cm, trim=2cm 0.25cm 1cm 0.25cm]{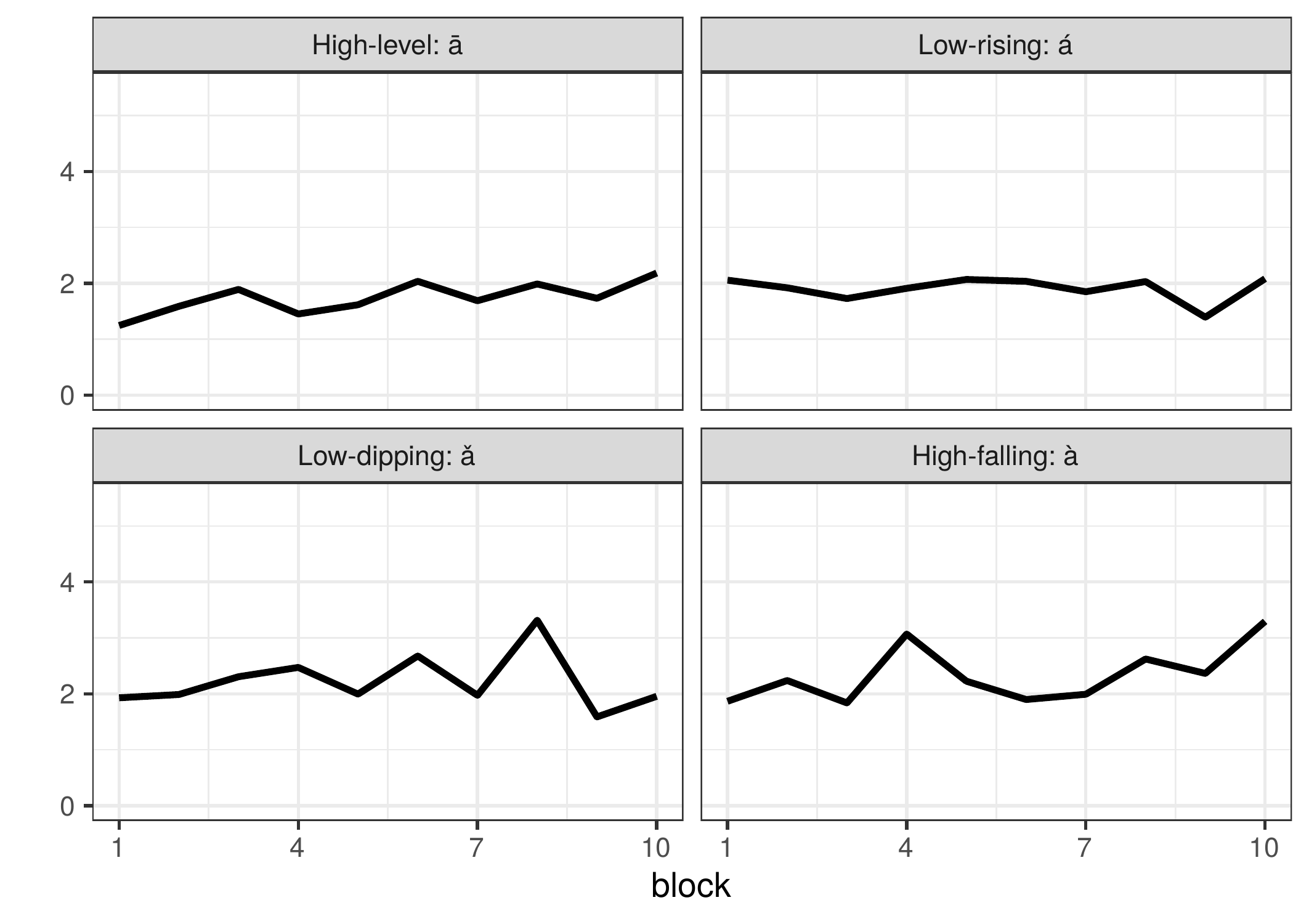}
	\caption{Results for synthetic data: 
	Left: Estimated mean slopes $m_{d,s,t}$ for the LBA model. 
	Right: Estimated boundaries $b_{s,t}$ for the LBA model. 
	In the left panel, $m_{d,s,t}$'s for the high-level tone response category T1 are shown in red; low-rising T2 in blue; low-dipping T3 in green; and high-falling T4 in purple.}
	\label{fig: LBA_sim}
\end{figure}

\begin{figure}[ht!]
	\centering
	\includegraphics[width=6.5cm, trim=2cm 1.25cm 1cm 1.25cm]{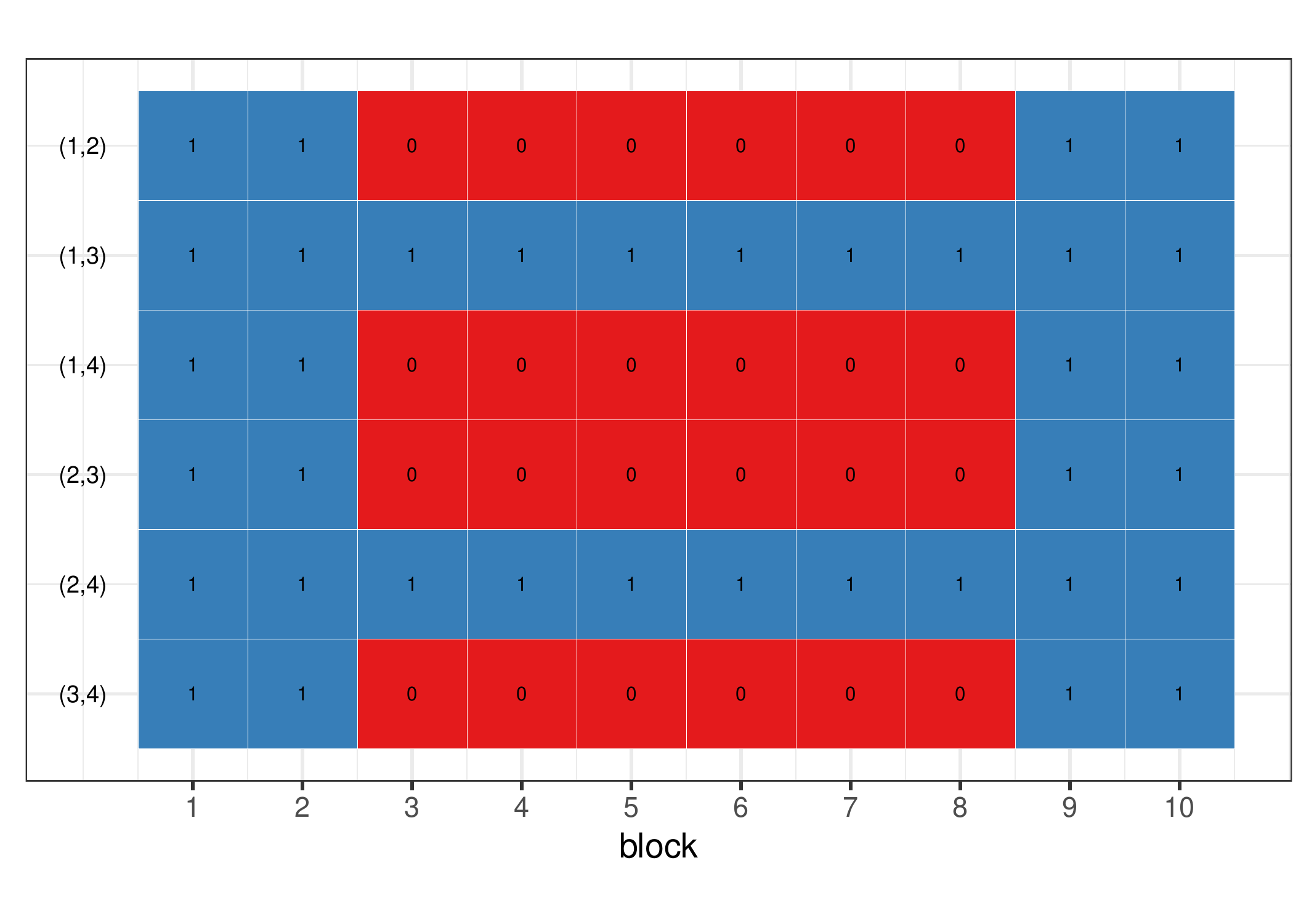} \hspace*{1cm}
	\includegraphics[width=6.5cm, trim=2cm 1.25cm 1cm 1.25cm]{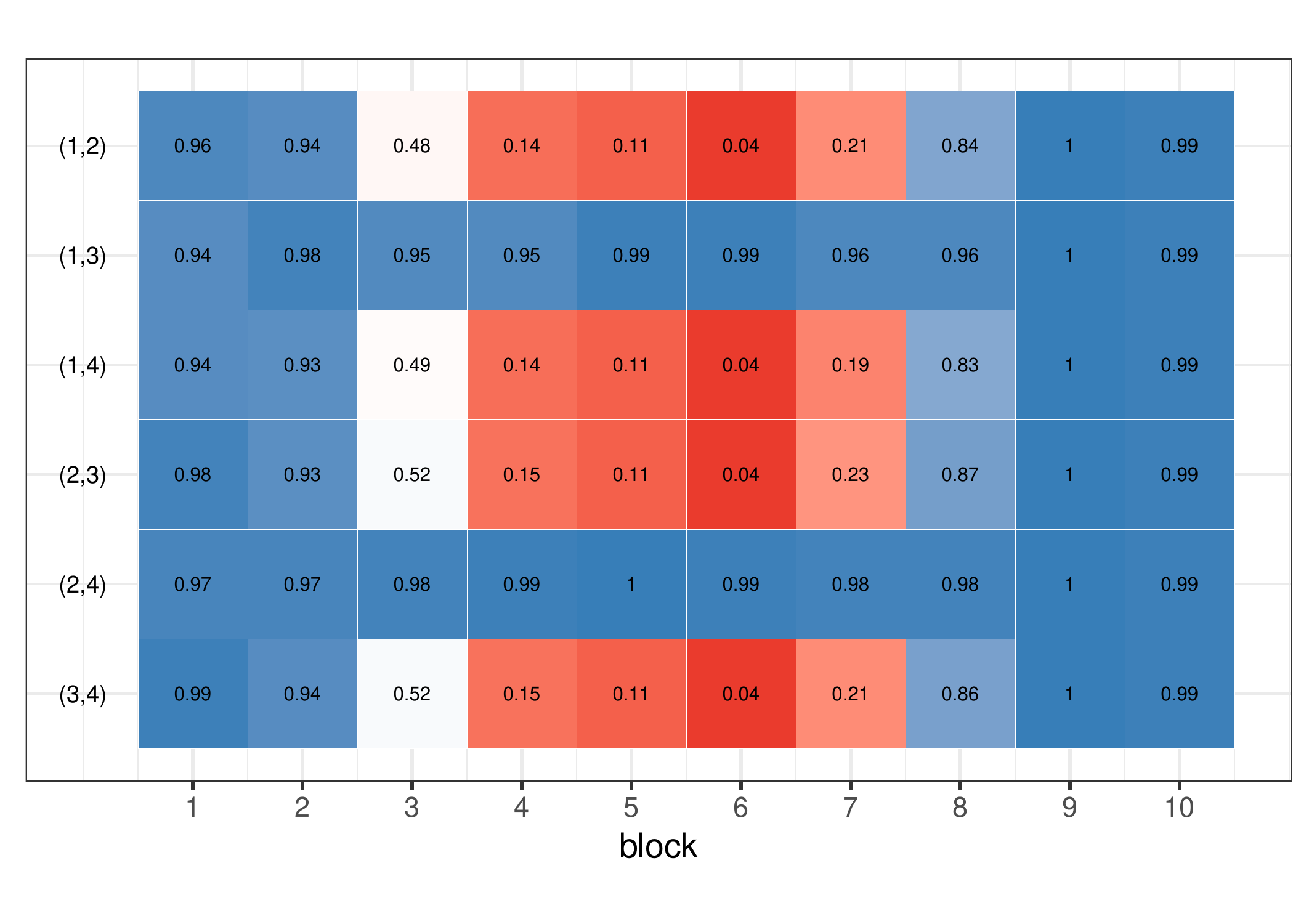}
	\caption{Results for synthetic data: The left panel shows the true clustering structure of the underlying parameter trajectories for successful identification ($d=s$) of different input tones in different learning phases. The right panel shows the corresponding posterior co-clustering probabilities estimated by our proposed method. 
	}
	\label{fig: coclust_sim}
\end{figure}

\begin{figure}[ht!]
	\centering
	\includegraphics[width=6.5cm, trim=2cm 1.25cm 1cm 1.25cm]{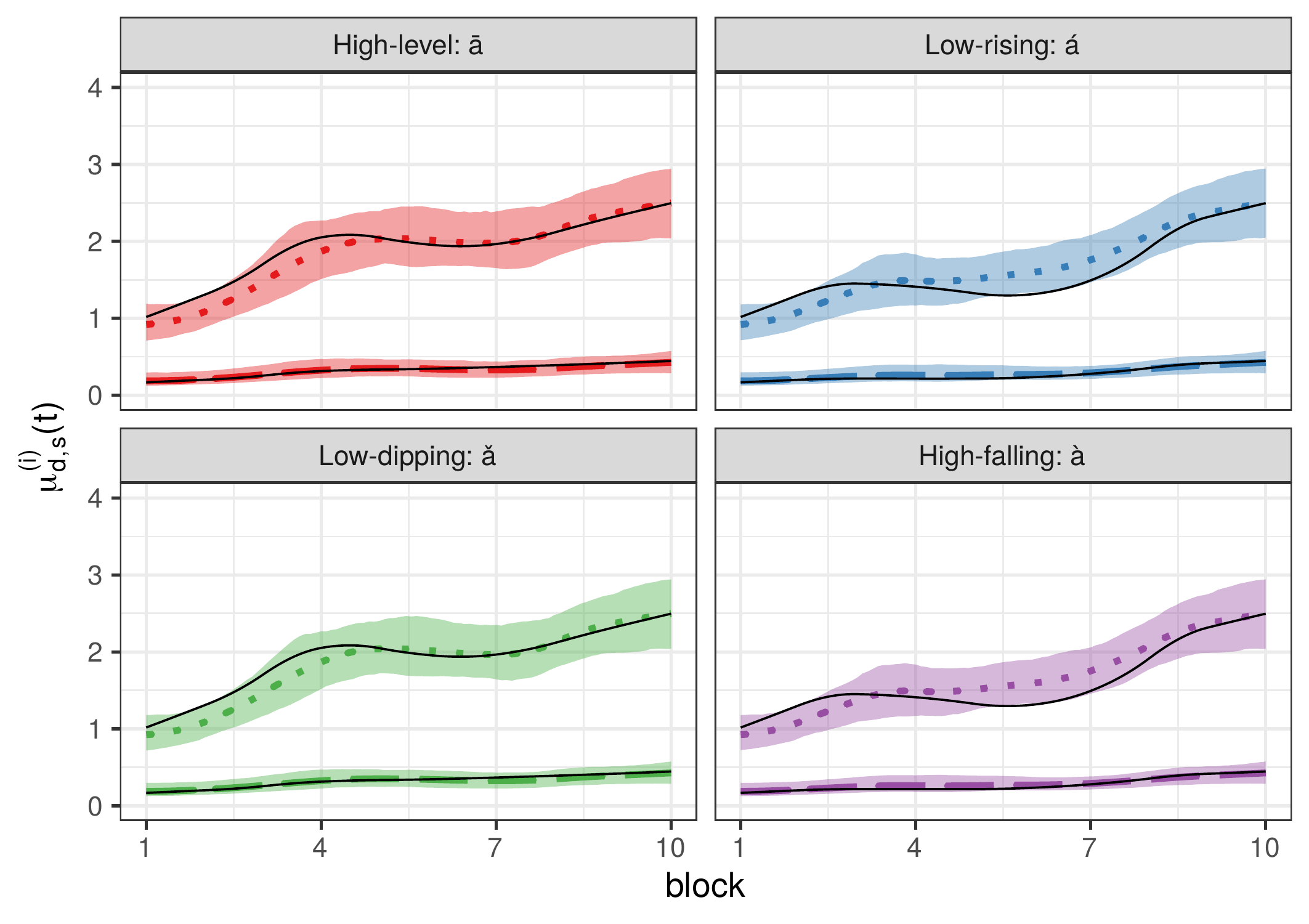} \hspace*{1cm}
	\includegraphics[width=6.5cm, trim=2cm 1.25cm 1cm 1.25cm]{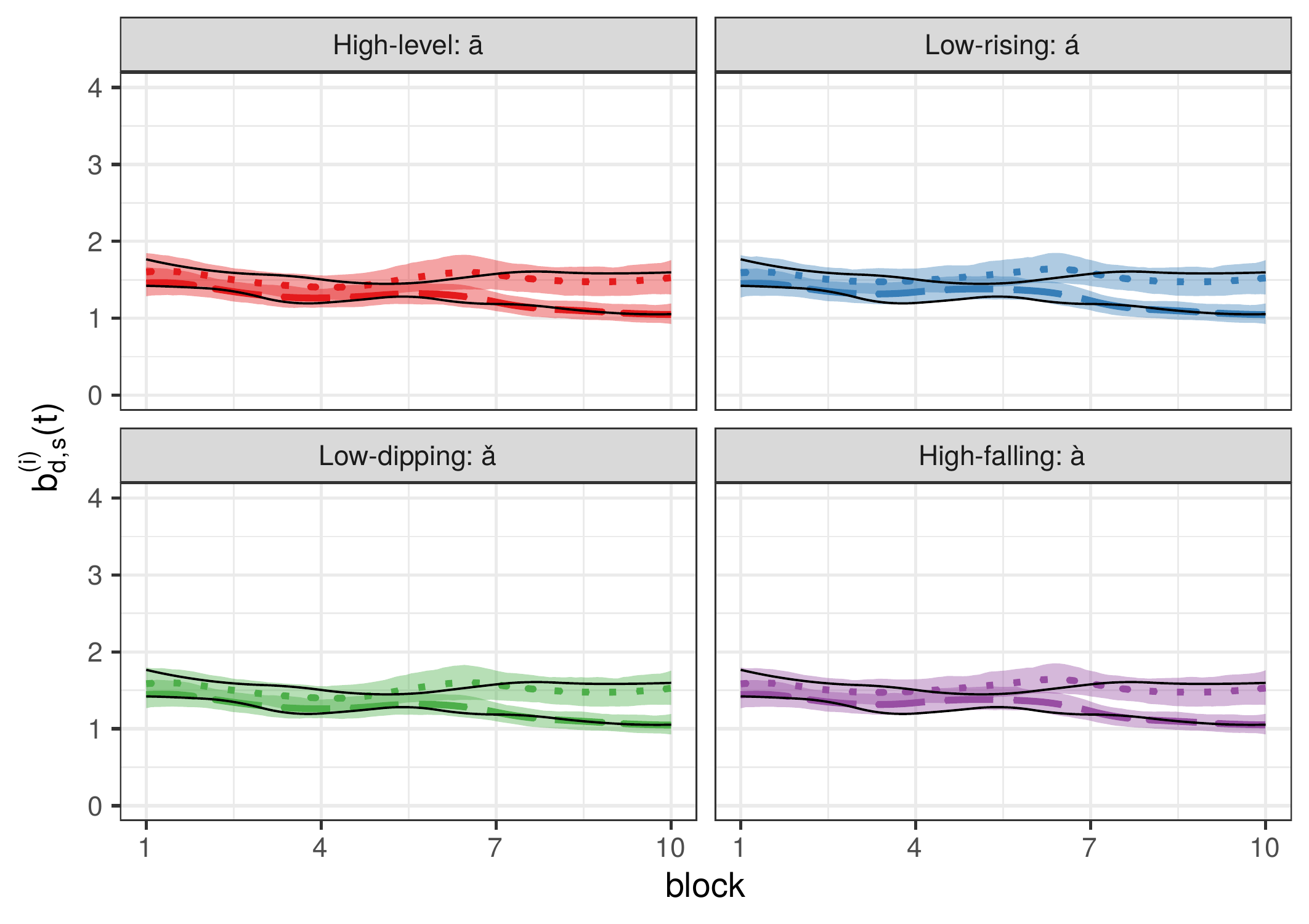}  \\[5pt]
	\caption{Results for synthetic data: Estimated posterior mean trajectories for individual specific drifts $\mu_{d,s}^{(i)}(t)$ (left panel)
	and boundaries $b_{d,s}^{(i)}(t)$ (right panel) for two different participants 
	- one performing well (dotted line) and one performing poorly (dashed line). 
	The shaded areas represent the corresponding $90\%$ point wise credible intervals.
	The solid black lines represent underlying true curves.
	Parameters for the high-level tone response category T1 are shown in red; low-rising T2 in blue; low-dipping T3 in green; and high-falling T4 in purple.}
	\label{fig: individual_eff_sim}
\end{figure}

\begin{figure}[ht!]
	\centering
	\includegraphics[width=6.5cm, trim=2cm 1.25cm 1cm 1.25cm]{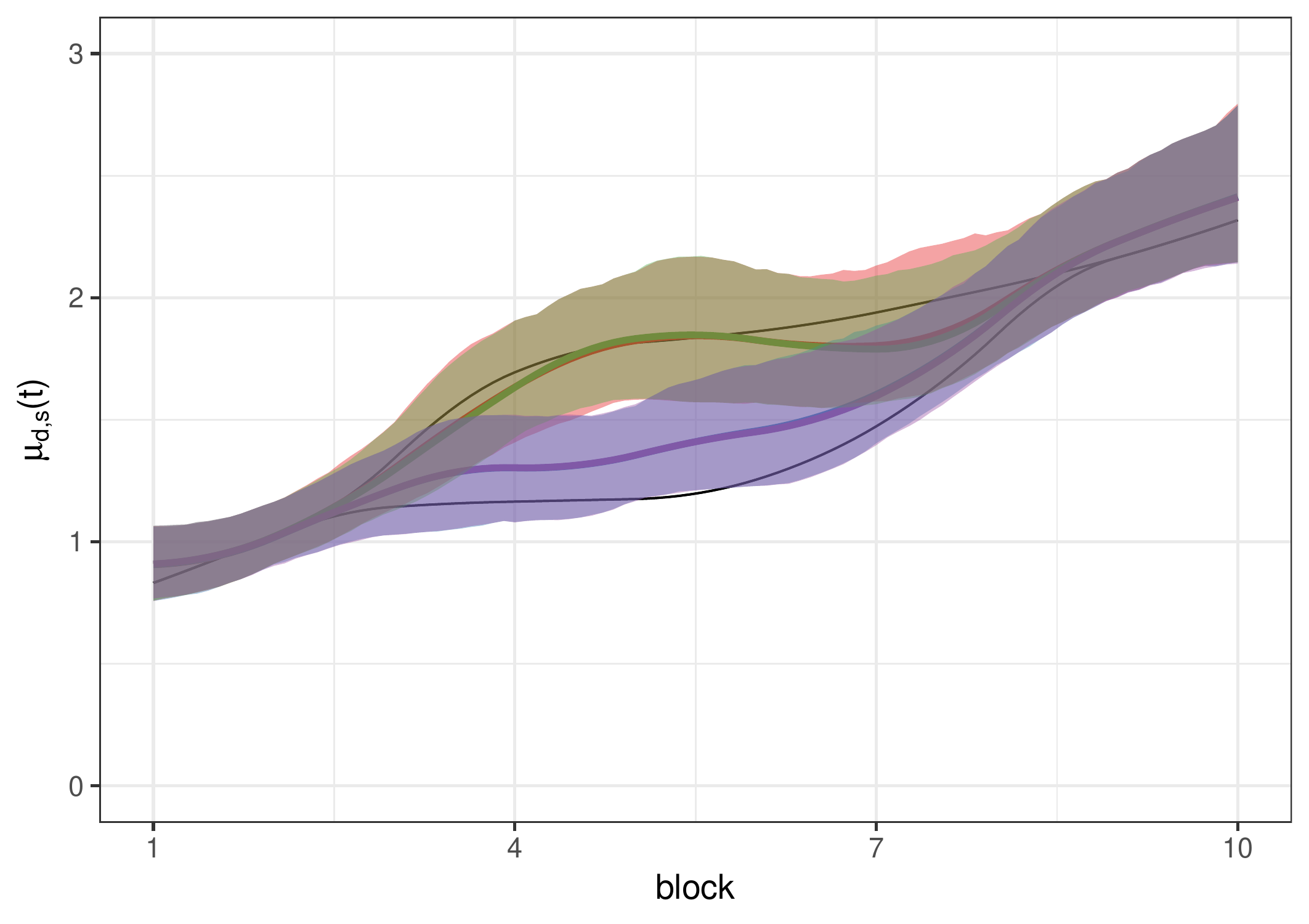} \\[5pt]
	\caption{Results for synthetic data: Estimated posterior mean trajectories of the population level drifts $\mu_{d,s}(t)$ for successful identification ($d=s$) of different input tones for the proposed longitudinal inverse Gaussian drift-diffusion mixed model. 
	The shaded areas represent the corresponding $90\%$ point wise credible intervals.
	The solid black lines represent underlying true curves.
	Parameters for the high-level tone response category T1 are shown in red; low-rising T2 in blue; low-dipping T3 in green; and high-falling T4 in purple.}
	\label{fig: local_diff_drift_sim}
\end{figure}

Figure \ref{fig: population_effect_sim} shows the posterior mean trajectories and associated $90\%$ credible intervals for the the drift rates $\mu_{d,s}(t)$ and boundaries $b_{d,s}(t)$, for every possible combination of $(d,s)$. 
Figure \ref{fig: local_diff_drift_sim} additionally presents the drift curves for successful identifications $(d=s)$ superimposed on each other.
These figures suggest that the underlying true curves are all recovered well by our method. 
In comparison, the results obtained by the LBA model, displayed in Figure \ref{fig: LBA_sim}, suffer from the same limitations discussed in Section \ref{sec: application}.
Furthermore, Figures \ref{fig: coclust_sim} and \ref{fig: individual_eff_sim} suggest that the underlying true local partition structure, as well as the individual specific parameter trajectories, are also estimated quite well by our method.

Figure \ref{fig: LBA_sim} presents the results obtained by the LBA model applied to the synthetic data set.  
There is a general agreement between the population level estimates produced by our method and the LBA. 
However, as discussed in detail in Section \ref{sec: application} in the main paper and Section \ref{sec: sm lba} in the supplementary materials, the LBA model has many serious limitations, including being incapable of producing individual level estimates, having shared boundary parameters across all input tones, not borrowing any information across adjacent time stamps etc. 
Only a very limited set of inferential questions can therefore be answered by the LBA model.

\clearpage
\newpage
\section{Additional Figures} \label{sec: add_figures}

\vskip 10pt
\begin{figure}[ht!]
	\centering
	\includegraphics[width=6.5cm, trim=2cm 1.25cm 1cm 1.25cm]{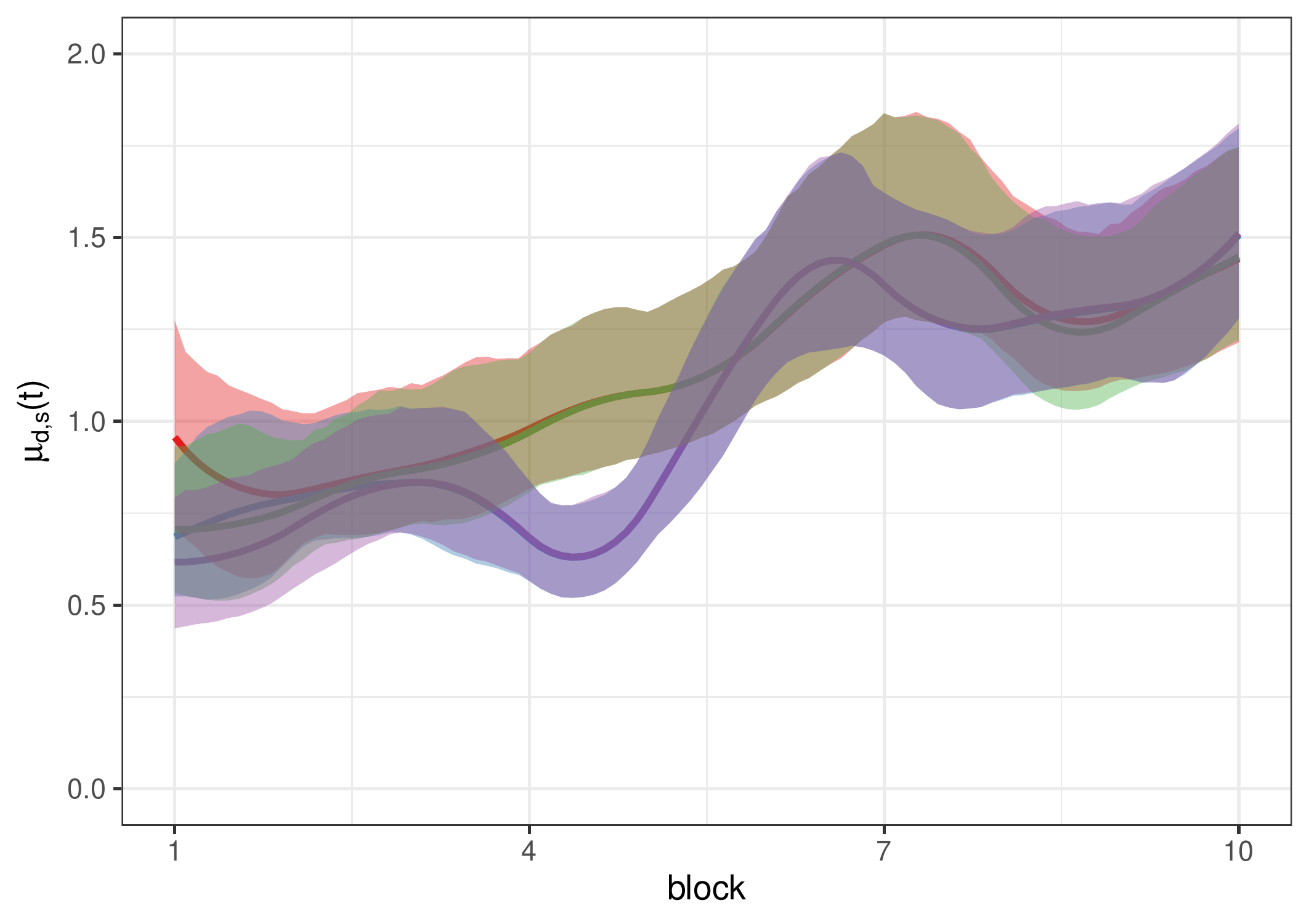} \\[5pt]
	\caption{Results for tone learning data: Estimated posterior mean trajectories of the population level drifts $\mu_{d,s}(t)$ for successful identification ($d=s$) of different input tones for the proposed longitudinal inverse Gaussian drift-diffusion mixed model. The shaded areas represent the corresponding $90\%$ point wise credible intervals.
	Parameters for the high-level tone response category T1 are shown in red; low-rising T2 in blue; low-dipping T3 in green; and high-falling T4 in purple.}
	\label{fig: local_diff_drift_real}
\end{figure}

\clearpage
\bibliographystylelatex{natbib}
\bibliographylatex{Categorical,Diffusion,FDA_LDA,HMM,HOHMM,MCMC_Latent_Var_Models}

\end{document}